\documentclass{emulateapj}

\usepackage{apjfonts}
\usepackage{verbatim}
\usepackage{natbib}
\usepackage{enumerate}
\usepackage{amsmath}
\usepackage{graphicx}

\graphicspath{{./figures/}}

\newcommand{\der}[2]{\frac{d #1}{d #2}}

\newcommand{\pder}[2]{\left(\frac{\partial #1}{\partial #2}\right)}

\newcommand{\pdernp}[2]{\frac{\partial #1}{\partial #2}}
\newcommand{\pderl}[2]{\partial #1/\partial #2}

\begin{document}

%\title{Laterally Propagating Helium Detonations on Accreting White Dwarfs: The Effects of Expansion}
\title{The Effects of Curvature and Expansion on Helium Detonations on White Dwarf Surfaces}

\author{Kevin Moore}
\affil{Department of Physics, University of California, Santa Barbara, CA}
%\authoremail{kmoore@physics.ucsb.edu}
\author{Dean M. Townsley}
\affil{Department of Physics and Astronomy, University of Alabama, Tuscaloosa, AL}
\and
\author{Lars Bildsten}
\affil{Kavli Institute for Theoretical Physics, University of California, Santa Barbara, CA}
\affil{Department of Physics, University of California, Santa Barbara, CA}

\begin{abstract}
Accreted helium layers on white dwarfs have been highlighted for many
decades as a possible site for a detonation triggered by a
thermonuclear runaway. In this paper, we find the minimum helium layer
thickness that will sustain a steady laterally propagating detonation and
show that it depends on the density and composition of the helium
layer, specifically $^{12}$C and $^{16}$O. Detonations in these thin helium 
layers have speeds slower than the Chapman-Jouget (CJ) speed from complete 
helium burning, $v_{CJ}=1.5\times 10^9$cm/s. Though gravitationally unbound, 
the ashes still have unburned
helium ($\approx 80\%$ in the thinnest cases) and only reach up to
heavy elements such as $^{40}$Ca, $^{44}$Ti, $^{48}$Cr, and
$^{52}$Fe. It is rare for these thin shells to generate large amounts
of $^{56}$Ni. We also find a new set of solutions that can propagate
in even thinner helium layers when $^{16}$O is present at a
minimum mass fraction of $\approx 0.07$. Driven by energy release from
$\alpha$ captures on $^{16}$O and subsequent elements, these slow
detonations only create ashes up to $^{28}$Si in the outer detonated
He shell. We close by discussing how the unbound helium burning ashes
may create faint and fast ``.Ia'' supernovae as well as events with
virtually no radioactivity, and speculate on how the slower helium
detonation velocities impact the off-center ignition of a carbon
detonation that could cause a Type Ia supernova in the double
detonation scenario.

\end{abstract}


%\author{
%Kevin Moore \altaffilmark{1},
%Dean M. Townsley\altaffilmark{2},
%and
%Lars Bildsten\altaffilmark{1,3}
%}

%\altaffiltext{1}{
%Department of Physics,
%University of California, Santa Barbara, CA; kmoore@physics.ucsb.edu
%}
%\altaffiltext{2}{
%Department of Physics and Astronomy,
%University of Alabama, Tuscaloosa, AL
%}
%\altaffiltext{3}{
%Kavli Institute for Theoretical Physics, Santa Barbara, CA
%}

\section{Introduction}
Recent observations indicating a large population of compact double white dwarf (WD) binaries with one member being a low mass ($M<0.2\ M_\odot$) helium WD \citep{ELM-I, ELM-II, ELM-III, ELM-IV, ELM-V}, especially those that will merge within a Hubble time, highlight the need for understanding the outcomes of helium accretion in such systems. Previous theoretical work on helium accretion in compact binaries has shown the possibility of the accreted helium layer becoming dynamically unstable \citep{Bildsten07} leading to deflagrations or detonations in the accreted layer \citep{Shen09, Shen10, Woosley11}. Such events are predicted to be characteristically `fast and faint' due to their low ejecta mass and partial burning to radioactive isotopes such as $^{56}$Ni. At the same time, rapidly-evolving transients that stand apart from traditional supernova categories were discovered such as SN 2002bj \citep{Poznanski10}, SN 2010X \citep{Kasliwal10}, and SN 2005ek \citep{Drout13}, while some historic supernovae such as SN 1885A and SN 1939B were recognized as potential members of a class of fast, low ejecta-mass events \citep{Perets11}.

Nondegenerate helium burning stars may also be relevant donors for peculiar Type Ia explosions. Some of the members of the recently proposed ``Type-Iax'' supernovae (SNe 2004cs \& 2007J) show helium lines in their spectra that may point to helium playing a role in their explosion \citep{Foley13}, though perhaps through deflagration rather than detonation \citep{Woosley11}. The discovery of helium burning stars in binaries with WDs that will initiate mass transfer before their helium burning phase ends \citep{Vennes12, Geier13} widens the spectrum of helium accretion scenarios. This also allows for the possibility of accreting material enriched with helium burning products such as $^{12}$C and $^{16}$O.

State of the art calculations of helium detonations on WDs include computationally intensive full-star 3D simulations \citep{Moll13}, models with 1D radially-propagating detonations \citep{Shen10, Waldman11, Woosley11}, and level-set methods \citep{Fink10, Sim12}, including detonation-shock-dynamics \citep{Dunkley13}. Comparison against a 1D model that can be treated with numerical techniques that have explicit error control is an essential verification test of multidimensional models.  This allows confidence that the relevant reaction scales are determined independent of resolution limitations.  Additionally, the 1D model allows a physical understanding of the mechanism of reaction freeze out.  Along with some parameterization of the WD properties, this makes it possible to efficiently survey the expected yields for the full variety of He shell thicknesses, compositions, and densities. 
%For these reasons, we present a detailed comparison between a physically motivated 1D model for laterally propagating detonations that includes the truncation of burning due to post-shock expansive effects, and multidimensional hydrodynamic simulations of such detonations. 
Motivated by the stable, laterally propagating helium detonation solutions found in the simulations of \citet{Townsley12}, we investigate the effects of post-shock expansion through both shock-front curvature and radial expansion on a 1D steady-state detonation. In this paper we present criteria for the propagation of a steady detonation wave in a uniform medium that is allowed to expand behind the shock front, and characterize the speeds and burning products of such detonations.

%Although there are difficulties fitting such events with 

%Work on both radially and laterally propagating helium detonations 

In \S \ref{sec:znd}, we outline the standard one-dimensional formalism for determining the post-shock structure of detonations. Section \ref{sec:flash_intro} discusses astrophysically motivated helium accretion scenarios and the results of multidimensional detonation simulations we wish to capture in our 1D model, while \S \ref{sec:znd_mod} explains our modifications of the ZND formalism to include the effects of radial expansion and curvature. In \S \ref{sec:results} we report how these effects modify detonation velocities and nucleosynthesis, and present comparisons to 2D hydrodynamic simulations of detonations in constant-density layers in {\ttfamily FLASH}. We extend these comparisons to detonations in finite gravity environments and map our results to WD core + envelope configurations in \S \ref{sec:finite_g} and conclude in \S \ref{sec:conclusions}.

\section{Detonations in One Dimension}
\label{sec:znd}
A detonation consists of a leading shock wave, self-sustained by net exothermic reactions which release a specific energy $\Delta q$ (erg/g) in the shocked material. In contrast to a deflagration, where heat transport by thermal conduction leads to subsonic propagation of the burning, detonations move at supersonic speeds that depend most strongly on $\Delta q$ via $v^2_{\rm det} \propto \Delta q$. In this section, we summarize the properties of 1D detonations in plane-parallel geometry and explain why modifications are necessary to model surface detonations on WDs. We first discuss the equations that describe steady detonations in \S\ref{subsec:znd}, followed by detonation velocity determination for 1D plane-parallel detonations in \S \ref{subsec:CJ}.

\subsection{Zel'dovich-von Neumann-D\"oring formalism}
\label{subsec:znd}
The one-dimensional model for a steady detonation was developed by \citet{Zeldovich40}, \citet{vonNeumann42, vonNeumann63}, and \citet{Doring43}, coupling the hydrodynamic equations to the burning equations under the steady-state assumption in the frame moving with the detonation front. The Zel'dovich-von Neumann-D\"oring (ZND) model calculates the post-shock structure, and is an excellent starting point for modeling post-shock expansion terms later in this paper. We derive the ZND equations from the hydrodynamic equations in Eulerian coordinates, neglecting viscosity and conduction:
\begin{align}
\pdernp{\rho}{t} + {\bf \nabla} \cdot \left(\rho {\bf u} \right) &= 0, \label{eq:hydro_mass} \\
\pdernp{{\bf u}}{t} + ({\bf u} \cdot {\bf \nabla}){\bf u} + \frac{1}{\rho}{\bf \nabla}P - \frac{{\bf F_{\rm ext}}}{\rho} &=0, \label{eq:hydro_mom}  \\
\pdernp{}{t} \left( \rho e \right) + {\bf \nabla} \cdot \left[(\rho e + P) {\bf u} \right] - \rho \epsilon &= 0, \label{eq:hydro_energy}.
\end{align}
Here, $\rho(t, {\bf x})$ is the density of the fluid, ${\bf u}(t, {\bf x})$ is the (Eulerian) velocity, $P(t, {\bf x})$ is the pressure, ${\bf F_{\rm ext}}(t, {\bf x})$ are any external forces on the fluid (e.g. gravity), $e(t, {\bf x})$ is the internal specific energy including the nuclear binding energy (in erg/g), and $\epsilon(t, {\bf x})$ is the local heating/cooling function (in erg g$^{-1}$ s$^{-1}$). These equations simplify under the assumptions used in the ZND formalism. For steady one-dimensional flow, we move into the shock frame ($x$ is now the distance behind the shock front) and write ${\bf u} = u \hat{x}$, ignore terms with $\pderl{}{t}$, and set ${\bf F_{\rm ext}} = 0$. The energy generation rate $\epsilon$ comes from the release of nuclear binding energy $\Delta q$, where the total nuclear binding energy is
\begin{equation}
q = N_A \sum_i Q_i Y_i.
\end{equation}
Here, $N_A$ is Avogadro's number, $Q_i$ is the nuclear binding energy of the $i^{th}$ species $(Q_i > 0)$, and $Y_i$ is the molar mass fraction, $Y_i = X_i/A_i$, with the mass fraction and atomic weight of the $i^{th}$ species being $X_i$ and $A_i$, respectively. Since there are no neutrino losses, the energy generation rate is thus
\begin{align}
\epsilon = \dot{q} &= \pdernp{q}{t} + ({\bf u} \cdot {\bf \nabla})q = u \der{q}{x},
\end{align}
where
\begin{align}
\der{q}{x} = N_A \sum_i Q_i \der{Y_i}{x},
\end{align}
so that the energy release of a fluid element throughout the detonation,
\begin{equation}
\Delta q = \int \epsilon \ dt = \int \der{q}{x} dx,
\end{equation}
can be calculated.

This leaves a concise set of equations describing one-dimensional reactive flow \citep{Fickett79} :
\begin{eqnarray}
\der{}{x}\left( \rho u \right) &=& 0, \label{eq:mass} \\
\der{}{x}\left( P + \rho u^2 \right) &=& 0, \label{eq:mom} \\
\der{}{x}\left( (e-q) + \frac{P}{\rho} + \frac{u^2}{2} \right) &=& 0, \label{eq:energy} \\
\der{}{x}\left( X_i \right) &=& \frac{R_i (\rho, T, \bf{X})}{u} \label{eq:comp}.
\end{eqnarray}

Equations (\ref{eq:mass})-(\ref{eq:energy}) express mass, momentum, and energy conservation, while equation (\ref{eq:comp}) evolves the composition due to nuclear burning. The reaction rates $R_i$ are taken from MESA's {\ttfamily net} module \citep{Paxton11, Paxton13}, using data from NACRE and JINA Reaclib \citep{Rauscher00, Cyburt10}.

In order to compare to other derivations, we take $\rho, T, u, {\bf X}$ as the independent variables for our ZND calculation, where ${\bf X}$ is the vector of mass fractions of all isotopes in the reaction network. Using the transformations
\begin{eqnarray}
\der{P}{x} &=& \pder{P}{\rho}_{{\bf X}, T} \der{\rho}{x} + \pder{P}{T}_{{\bf X}, \rho} \der{T}{x} \nonumber \\
 &+& \sum_i \pder{P}{X_i}_{X_j, \rho, T} \der{X_i}{x}, \\
\der{e}{x} &=& \pder{e}{\rho}_{{\bf X}, T} \der{\rho}{x} + \pder{e}{T}_{{\bf X}, \rho} \der{T}{x} \nonumber \\
 &+& \sum_i \pder{e}{X_i}_{X_j, \rho, T} \der{X_i}{x},
\end{eqnarray}
we derive evolution equations for $\rho, T$, and $u$ as a function of position behind the shock front. Henceforth, partial derivatives are assumed to hold all variables in the $\{\rho, T, {\bf X}\}$ set constant except the one(s) being differentiated by unless otherwise noted. In this case, equations (\ref{eq:mass})-(\ref{eq:comp}) become
\begin{eqnarray}
\der{\rho}{x} &=& \frac{1}{c_s^2 - u^2} \sum_i \der{X_i}{x} \left[ \pder{P}{e}_\rho \pder{e}{X_i} - \pder{P}{X_i} \right] \nonumber \\
 &=& - \frac{\rho c_s^2 \sigma_i R_i}{u(c_s^2 - u^2)}, \label{eq:znd_rho}\\
\der{T}{x} &=& \pder{P}{T}^{-1}\left[\der{\rho}{x}\left(u^2 - \pder{P}{\rho}\right) - \sum_i\pder{P}{X_i}\der{X_i}{x} \right], \label{eq:znd_t} \\
\der{u}{x} &=& -\frac{u}{\rho} \der{\rho}{x}, \label{eq:znd_ux} \\
\der{X_i}{x} &=& \frac{R_i (\rho, T, \bf{X})}{u} \label{eq:znd_rxn}.
\end{eqnarray}
where
\begin{equation}
\label{eq:af2}
c_s^2 = \pder{P}{\rho}_{{\bf X}, S} = \pder{P}{\rho} + \pder{P}{e}_\rho\left(\frac{P}{\rho^2} - \pder{e}{\rho}\right),
\end{equation}
evaluated at fixed entropy, $S$, and composition, is the frozen sound speed, and consistent with the detonation literature \citep{Fickett79, Sharpe99}, the thermicity, $\sum_i \sigma_i R_i$, is defined as the rate of conversion of nuclear energy into thermal energy, where
\begin{equation}
%\sigma_i = \frac{1}{\rho c_s^2} \left\{ \pder{P}{Y_i} - \left[ \pder{U}{Y_i} - \pder{q}{Y_i} \right] \pder{P}{T} \right\},
\sigma_i = \frac{1}{\rho c_s^2} \left[ \pder{P}{Y_i} - \pder{e}{Y_i} \pder{P}{T} \right],
\end{equation}
and the equation governing reaction rates remains equation (\ref{eq:comp}).
%Summation over the indices in the right hand side of equation (\ref{eq:znd_rho}) is implied.

Numerical integration of equations (\ref{eq:znd_rho}) - (\ref{eq:znd_rxn}) can be made far more efficient if we have analytic expressions for their derivatives with respect to each of the independent variables, i.e. the Jacobian terms. In order to compute these, we rewrite the compositional derivatives in terms of the independent variables that the MESA equation of state (EOS) uses, $\{\rho, T, \bar{A}, \bar{Z}\}$, where $\bar{A}$ is the mean atomic weight of the mixture,
\begin{equation}
\bar{A} = \frac{\sum_i n_i A_i}{\sum_i n_i},
\end{equation}
and $\bar{Z}$ is the mean charge of the mixture,
\begin{equation}
\bar{Z} = \frac{\sum_i n_i Z_i}{\sum_i n_i},
\end{equation}
where $n_i = \rho N_A Y_i$ is the particle density of the $i^{th}$ species and $Z_i$ the respective charge. The ZND equations, (\ref{eq:znd_rho}) - (\ref{eq:znd_rxn}), are then rewritten with $\pderl{}{X_i}$ in terms of $\pderl{}{\bar{A}}$ and $\pderl{}{\bar{Z}}$. The appropriate transformations are
\begin{equation}
\pder{}{X_i}_{X_j} = \pder{\bar{A}}{X_i}_{X_j}\pder{}{\bar{A}}_{\bar{Z}} + \pder{\bar{Z}}{X_i}_{X_j}\pder{}{\bar{Z}}_{\bar{A}},
\end{equation}
where
\begin{align}
\pder{\bar{A}}{X_i}_{X_j} &= \frac{\bar{A}(A_i - \bar{A})}{A_i} \\ %- \frac{\bar{A}^2}{A_i}, \\
\pder{\bar{Z}}{X_i}_{X_j} &= \frac{\bar{A}(Z_i - \bar{Z})}{A_i}.
\end{align}
Thus, for example
\begin{equation}
\pder{P}{X_i}_{X_j} = \frac{\bar{A}(A_i - \bar{A})}{A_i} \pder{P}{\bar{A}}_{\bar{Z}} + \frac{\bar{A}(Z_i - \bar{Z})}{A_i} \pder{P}{\bar{Z}}_{\bar{A}}.
\end{equation}
These relations will be used in constructing the Jacobian for the ZND equations presented here, and beyond as we explore modifications to the ZND equations.

\subsection{Chapman-Jouget Detonations}
\label{subsec:CJ}
A Chapman-Jouget (CJ) detonation is one in which the burning ceases (due to exhaustion of fuel) when the post-shock flow becomes sonic relative to the shock front \citep{Fickett79}. Physically, this means that all the nuclear energy released can be used to propagate the detonation front, since the entire reaction zone is in sonic contact with it. Such a detonation occurs at a unique velocity, given the initial conditions ($\rho_0$, $T_0$, ${\bf X}_0$). The CJ velocity for helium detonations has a weak dependence on initial density \citep{Timmes00}, and is $v_{\rm CJ} = 1.52\times 10^9$ cm/s at an initial density of $\rho_0 = 5\times 10^5$ g/cm$^3$. Calculating this velocity requires a knowledge of the final state of the ashes, and in general requires a calculation of nuclear statistical equilibrium (NSE) near the end of the reaction zone. At these low densities, the final state is virtually pure $^{56}$Ni, while at higher densities $(\rho_0 \gtrsim 10^6$ g/cm$^3)$ the high post-shock temperatures can lead to photo-disintegration and electron capture, which significantly alters the final state. In order to avoid a lengthy digression on the final states of CJ detonations in pure helium at high densities, we employ a simple 13-isotope alpha chain for our CJ calculations here \citep{Timmes99}. In a $\gamma$-law equation of state, $P = (\gamma - 1)\rho e$, the detonation velocity is related to the post-shock energy release $\Delta q$ as \citep{Fickett79}
\begin{equation}
\label{eq:vq}
v_{\rm det}^2 = 2\left(\gamma^2 -1 \right) \Delta q.
\end{equation}
Although our equation of state is more detailed, the requirement that the burning used to propel the detonation remains in sonic contact with the shock front guides us to expect the scaling $v_{\rm det}^2 \propto \Delta q$ in general.

We first calculate $v_{\rm CJ}$ via the standard Hugoniot construction (see \citet{Fickett79}) assuming a final state of pure $^{56}$Ni and then use the post-shock conditions immediately behind the shock front to start integrating the ZND equations (\ref{eq:znd_rho}) - (\ref{eq:znd_rxn}). We integrate the reaction network using a stiff ODE integrator - a linearly implicit Runge-Kutta method included with MESA \citep{Lang01}. Since determination of $v_{\rm CJ}$ is independent of the reaction network used in integrating the ZND equations, the resultant nucleosynthesis may differ from the assumed final state in the CJ calculation. 
%For example, if we assume a final state of pure $^{44}$Ti, we could calculate a corresponding CJ velocity, but that final state would not be realized when integrating an alpha-chain reaction network with that velocity. 
We therefore check the final NSE state and iterate this procedure - choosing a new smaller value for $v_{\rm CJ}$ and thus a slightly less fully burned CJ state if the NSE state corresponds to a smaller $\Delta q$ than was assumed, otherwise choosing a larger value for $v_{\rm CJ}$ - until the composition used in the CJ state conforms to the NSE state found by network integration.

\begin{figure}
  %\centering
  %\includegraphics[width=\textwidth]{figures/Abundances_plot_cj_r5d5_approx13}
  \plotone{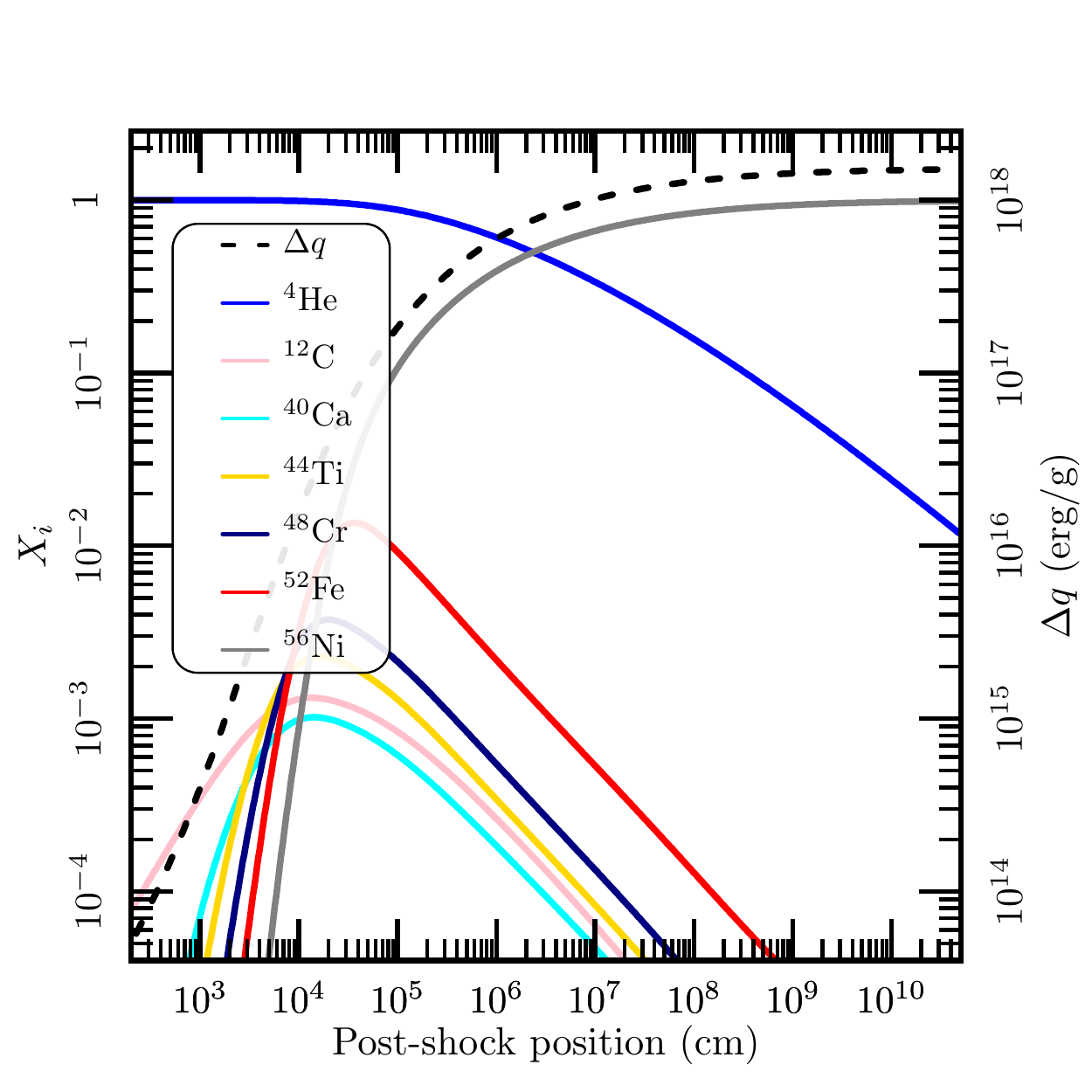}
  \caption{Nucleosynthesis profile of a CJ detonation in pure helium with $\rho_0 = 5\times 10^5$ g/cm$^3$, $T_0 = 10^8$ K. The detonation velocity is $v_{\rm CJ} = 1.523\times 10^9$ cm/s. Solid lines are nuclear abundances and correspond to the left axis, while the dashed line is the cumulative energy release - corresponding to the right axis. It takes a distance of $1.8 \times 10^9$ cm to achieve $95\%$ of the total energy release. \label{fig:nuc_cj_r5d5}}
\end{figure}

\begin{figure}
  %\centering
  %\includegraphics[width=\textwidth]{figures/Abundances_plot_cj_r2d6_approx13}
  \plotone{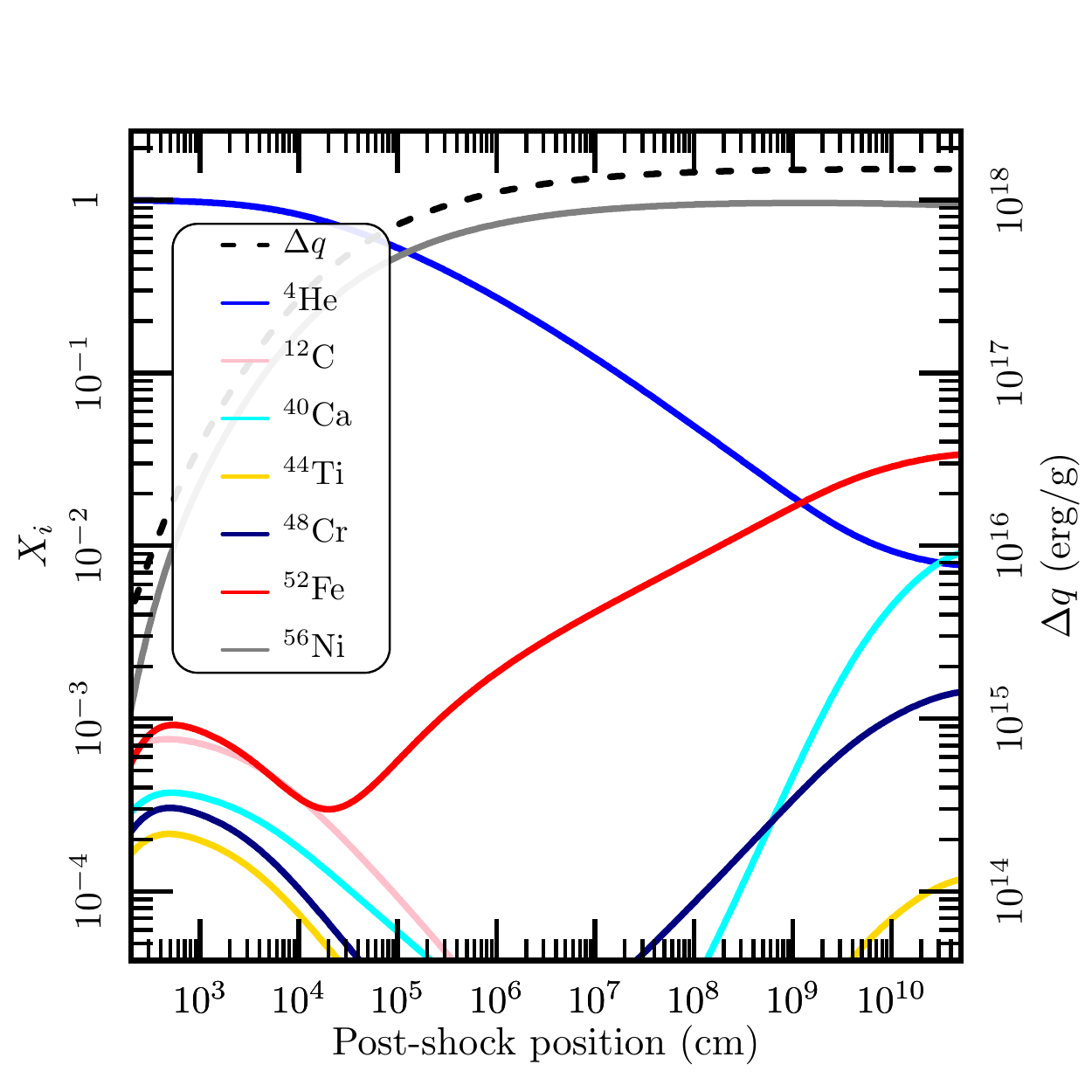}
  \caption{Nucleosynthesis profile of a CJ detonation in pure helium with $\rho_0 = 2\times 10^6$ g/cm$^3$, $T_0 = 10^8$ K. The detonation velocity is $v_{\rm CJ} = 1.566\times 10^9$ cm/s. 
  %$v_{\rm CJ} = 1.607\times 10^9$ cm/s. 
  Solid lines are nuclear abundances and correspond to the left axis, while the dashed line is the cumulative energy release - corresponding to the right axis. It takes a distance of $6.3\times 10^7$ cm to achieve $95\%$ of the total energy release. \label{fig:nuc_cj_r2d6}}
\end{figure}

\begin{figure}
  %\centering
  %\includegraphics[width=\textwidth]{figures/thermo_pane_cj_both}
  \plotone{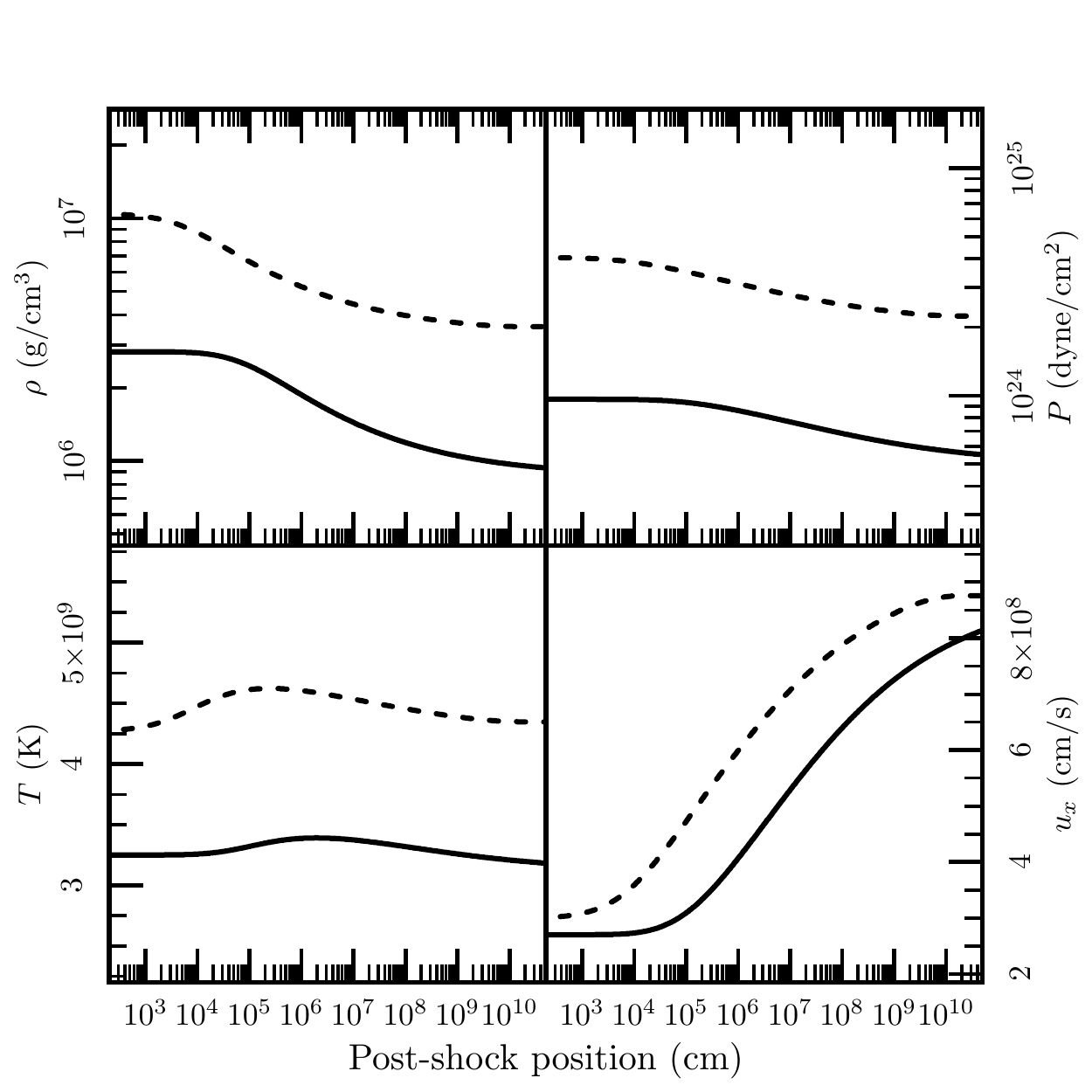}
  \caption{Thermodynamic profile of CJ detonations in pure helium with the same starting conditions as Figures \ref{fig:nuc_cj_r5d5} ($\rho_0 = 5\times 10^5$ g/cm$^3$ - solid lines) and \ref{fig:nuc_cj_r2d6} ($\rho_0 = 2\times 10^6$ g/cm$^3$ - dashed lines). \label{fig:thermo_cj_both}}
\end{figure}

\begin{figure}
  %\centering
  %\includegraphics[width=\textwidth]{figures/rho_sweep_lengths_95}
  \plotone{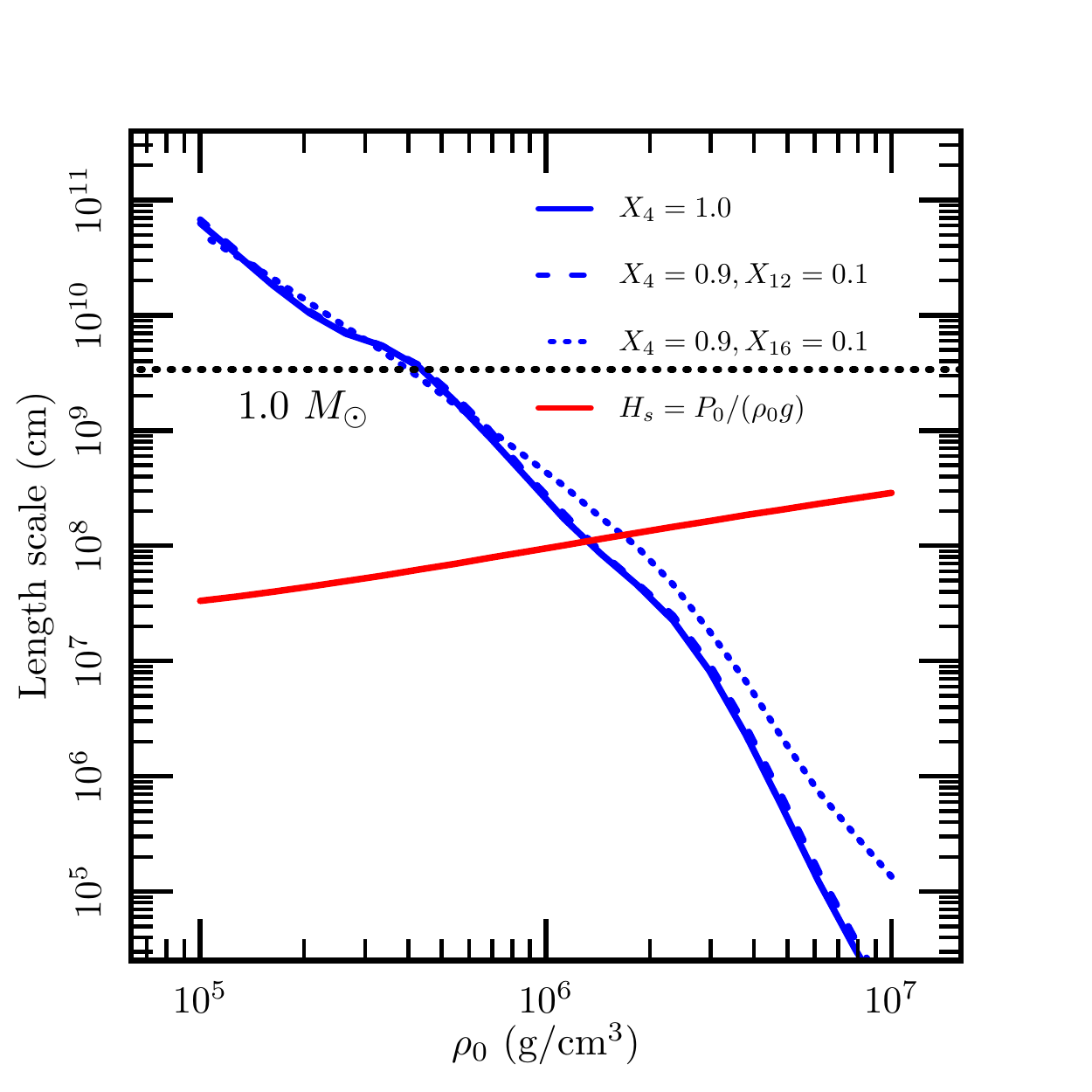}
  \caption{Length scales to release $95\%$ of the total energy, $l_{95}$, as a function of initial density for helium layers with initial $^{12}$C and $^{16}$O mass fractions and $T_0 = 10^8$ K (blue lines). The scale height of the atmosphere (red line) is computed using $g=3\times 10^8$ cm/s$^2$ and is smaller than $l_{95}$ for lower initial densities, showing that complete burning and CJ detonations are not formally possible in such envelopes, while higher-density envelopes will be affected less by the finite scale height of the material. The horizontal black dotted line shows the circumference of a $1.0\ M_\odot$ WD with core temperature $10^7$ K. \label{fig:rho_sweep_lengths_95}}
\end{figure}

%Once we calculate $v_{\rm CJ}$ (description of procedure involves digression into Hugoniots and Rayleigh lines, so put in appendix perhaps?), we find the post-shock conditions immediately behind the shock front start integrating the ZND equations [(\ref{eq:znd_rho}) - (\ref{eq:znd_rxn})] from those values. 
We show a spatial profile of the thermodynamic variables and composition in the shock frame in Figures \ref{fig:nuc_cj_r5d5} - \ref{fig:thermo_cj_both} for CJ detonations in pure helium at initial densities of $\rho_0 = 5\times 10^5$ g/cm$^3$ and $\rho_0 = 2\times 10^6$ g/cm$^3$, respectively. The detonation with $\rho_0 = 5\times 10^5$ g/cm$^3$ burns almost completely to $^{56}$Ni, while the detonation with $\rho_0 = 2\times 10^6$ g/cm$^3$ shows signs of photo-disintegration at late times. Low-density CJ detonations in pure helium take a very long distance to achieve near-total energy release, typically much larger than the circumference of a C/O WD as shown in Figure \ref{fig:rho_sweep_lengths_95}. Formally the material will never become fully burned, so length scales of ever increasing energy release percentage will grow until NSE is achieved. 

%Regardless of other physics at work, a complete burn to $^{56}$Ni is not possible for a laterally propagating detonation on the surface of a WD, and the detonation velocity must be lower than the corresponding CJ velocity. 

A comparison that highlights when the finite thickness of a helium layer will affect the burning is the ratio of scale height to the burning length. We define a measure of the energy release length scale, $l_{95}$, as the distance behind the shock front (in the shock frame) where $95\%$ of the total energy is released. We compare this to the pressure scale height on a star, $H_s = P_0/(\rho_0 g)$, as a function of ambient density $\rho_0$ in Figure \ref{fig:rho_sweep_lengths_95} (cf. \citet{Timmes00}). For lower densities, $l_{95}$ is much larger than $H_s$, while at higher initial densities, the energy release length scale is much smaller than the scale height of the helium layer. We therefore expect the finite thickness of the helium layer to have the largest impact on nucleosynthesis for the thinner cases with lower base densities. A complete burn to $^{56}$Ni is not possible for a laterally propagating detonation on the surface of a WD. Low-density detonations do not have enough space to burn to completion, while high-density cases experience significant photo-disintegration of synthesized isotopes.

%The initial conditions ($\rho$, $T$, ${\bf X}$) and assumed final state determine a unique detonation velocity in the following manner. The equations for Momentum and mass conservation combine into 

%\begin{figure}
%  \centering
%  \includegraphics[width=\textwidth]{figures/rho_sweep_lengths_50}
%  \caption{Length scales to release $50\%$ of the total energy, $l_{50}$, as a function of initial base density for helium layers with a range of initial $^{12}$C mass fractions and $T_0 = 10^8$ K. The scale height of the atmosphere is computed using $g=3\times 10^8$ g/cm$^2$ and is comparable to $l_{50}$ for lower base densities, showing that those will likely be the cases most modified due to the finite helium layer thickness. \label{fig:rho_sweep_lengths_50}}
%\end{figure}

\begin{figure}
  %\centering
  %\includegraphics[width=\textwidth]{figures/thermo_pane_CJ}
  \plotone{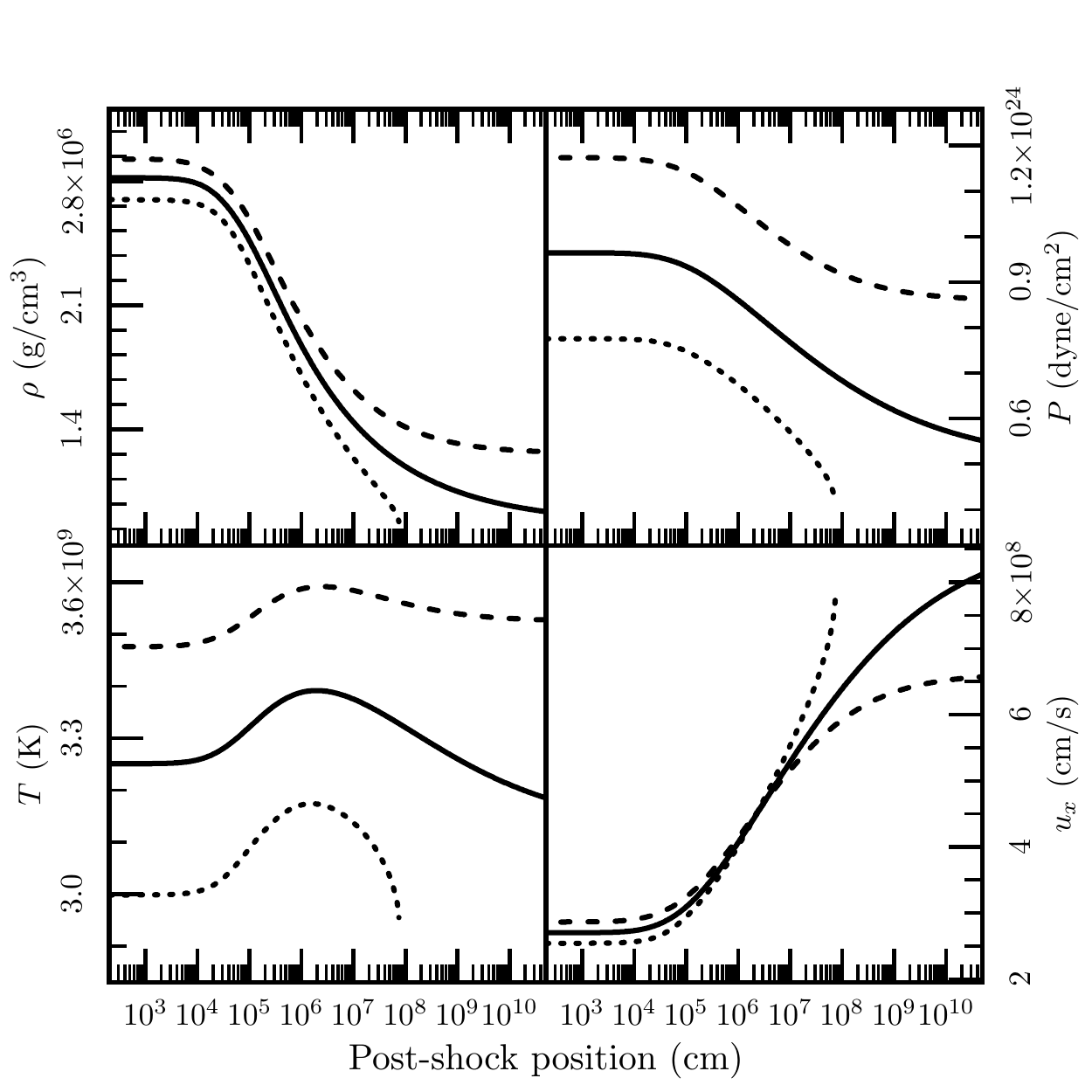}
  \caption{Thermodynamic profile of detonations in pure helium with the same starting conditions as Figure \ref{fig:nuc_cj_r5d5} - $\rho_0 = 5\times 10^5$ g/cm$^3$. The line types correspond to the velocity used to integrate the ZND equations: $v_{\rm det} = v_{\rm CJ}$ (solid lines), $v_{\rm det} = 1.1\ v_{\rm CJ}$ (dashed lines), and $v_{\rm det} = 0.9\ v_{\rm CJ}$ (dotted lines). The CJ detonation velocity is $v_{\rm CJ} = 1.523\times 10^9$ cm/s. As discussed in the text, only the detonation with $v_{\rm det} < v_{\rm CJ}$ hits the sonic point, which occurs at $\approx 10^{8}$ cm behind the shock front. \label{fig:thermo_cj}}
\end{figure}

Understanding what happens to the CJ detonation velocity when post-shock energy losses due to curvature and radial expansion occur requires understanding what happens in the post-shock flow at detonation velocities near the CJ velocity. In the absence of endothermic reactions in a simple alpha chain nuclear network, the CJ velocity is the unique velocity at which a freely propagating planar detonation will travel. However, we can integrate the ZND equations with any detonation velocity that we choose. If we pick $v_{\rm det} < v_{\rm CJ}$, then the integration encounters a singularity (the sonic point) when the post-shock flow relative to the shock front reaches the local frozen sound speed, $c_s^2 = (\pderl{P}{\rho})_{{\bf X}, S}$ - see equation (\ref{eq:znd_rho}). These detonations are referred to as underdriven detonations, and numerical hydrodynamic simulations show that such planar detonations in helium will strengthen to propagate at $v_{\rm CJ}$, given enough time \citep{Townsley12}. If we use $v_{\rm det} > v_{\rm CJ}$ then the integration will not hit a singularity, but a supporting pressure is required to sustain such a detonation, called an overdriven detonation. In the absence of such a sustaining pressure (eg. via a piston in a shock tube), this type of detonation will weaken until it propagates at $v_{\rm CJ}$. Figure \ref{fig:thermo_cj} shows the evolution of the thermodynamic variables behind the shock front for a CJ detonation in pure helium, as well as an overdriven case at a $10\%$ higher detonation velocity, and an underdriven case at $10\%$ lower detonation velocity. Intuition about what happens in planar detonations for velocities near $v_{\rm CJ}$ will guide our analysis of detonation velocities when we include the effects of curvature and expansion.

\section{Simulations of Detonations in Single Hydrostatic Helium Layers}
\label{sec:flash_intro}
Here we present the astrophysically-motivated configurations and
detonation structures for which, in the following section, we will develop
methods for directly computing the basic detonation properties via a ZND-like
formalism.  A WD with a thin ($\sim$few$\times 10^{-2}M_\odot$) He shell
undergoing a thermonuclear shell flash is the environment we will study for
the propagation of a lateral detonation.  Since we will
develop ZND-like steady-state calculations of this detonation
structure, we utilize a plane parallel calculation under constant (spatially
uniform) gravity rather than working in a full star.  This is as done in
\citet{Townsley12}, and allows us to propagate a detonation into steady
state for a given helium layer.
We make the further simplification that there is only a
$10^8$~K He layer, no hot overlying convective zone as was used in
\citet{Townsley12}.  This isolates expansive
characteristics of a single layer and facilitates more direct comparison with
our ZND-like calculations.

The computational setup used here is otherwise the same as that used
in Townsley et al. (2012).  All physics is included in the public
{\ttfamily FLASH 4} code release, and the Simulation Units, which define the
initial condition and mesh refinement, for this and the strip
detonation configuration described below are available for download
via the web \footnote{\url{http://astronomy.ua.edu/townsley/code/}}. 
While for high densities and thick layers NSE will be reached,
requiring careful treatment of the coupling between energy release and
hydrodynamics, we limit our hydrodynamic cases to those that do not burn to
completion, having peak temperatures around 2-3$\times10^9$~K.  The
hydrodynamic step is taken as 0.8 of the CFL, and the nuclear reactions are
substep-integrated assuming constant temperature in the usual operator-split
fashion used in the {\ttfamily FLASH} code.  With the low peak temperatures, this is
numerically stable. 
Nuclear reactions are suppressed in the vicinity of shocks, since
the computations presented here are not over-resolved.  Even so, we see,
given sufficient resolution, minimal resolution dependence for integrated
quantities like yields, energy release, and detonation speed.  The
small-scale cellular structure near the detonation front does vary with
resolution, in a way similar to that seen for resolved detonations with
reactions suppressed in shocks \citep{Papathedore13}.  However, the
effect on the burning products well behind the detonation front appear to be
mild.
The lower boundary of the hydrostatic layer is treated as discussed
in \citet{Zingale02}, with local hydrostatic gradient and
a reflecting velocity, as implemented in {\ttfamily FLASH 4}.
The top boundary, located at $8.5\times 10^8$~cm above the base
of the He layer, is zero-gradient outflow, and the side boundaries near the
ignition and at the far end of the domain, where the detonation does not
reach in the time of the simulation, are reflecting.

\begin{figure}
  %\centering
  %\includegraphics[width=0.8\textwidth]{figures/singlelayer_sonic_2e8_4e8_combine.png}
  \plotone{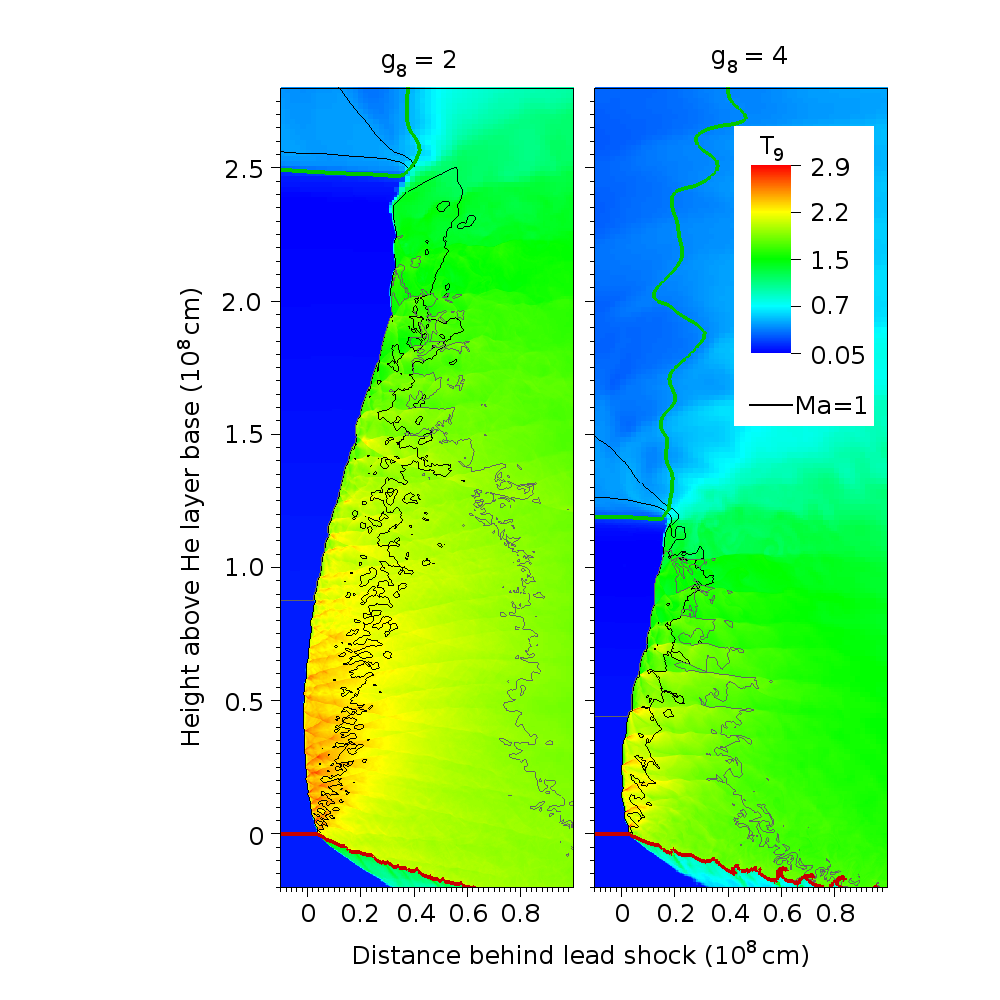}
  \caption{Steady-state detonation structures in cold helium layers,
  $T=10^8$~K, under two different gravities, $g_8=2$ (left) and $4$ (right),
  in units of $10^8$~cm~s$^{-2}$.  Both calculations are performed in plane
  parallel geometry with constant gravity.  The detonation is propagating to
  the left.  Material within sonic contact of the shock front is bounded by the
  black contour. Specifically, it indicates the sonic locus, the location that the 
  post-shock flow becomes supersonic with respect to the shock front.
  %is shown (along with the shock front) by the thin black contour.  
  The thick green contour shows a density of $10^4$~g~cm$^{-3}$,
  near the surface of the star prior to blowout.  The grey contour is at
  $\rho_5=2.5$, half the density at the base of the He layer.  The red
  contour shows the division between the He layer and the underlying core
  material.
  \label{fig:surface_2grav}}
\end{figure}

A higher gravity will lead to a geometrically thinner
shell for the same base pressure (base density).  We expect that the
vertical expansion of this shell, i.e. blowout, will have a more significant
effect on this thinner shell, causing differences in detonation speed and
products despite the fuel density being the same.  Figure
\ref{fig:surface_2grav} shows the steady-state temperature and density
structure of detonations propagating in helium layers at two different
gravities, $g_8=2$ (left) and $4$ (right), where $g_8$ is gravity in units of
$10^8$~cm~s$^{-2}$.  The density at the base of the He layer is $5\times
10^5$~g~cm$^{-3}$ in both cases, giving scale heights $H_s= P_0/\rho_0 g$ of 1
and 0.5$\times 10^8$~cm respectively.  The surface of the star is located at
a height of approximately 2.5 and 1.2$\times 10^8$~cm respectively.  The
steady-state detonation speeds obtained for these two cases are 0.98 and
0.89$\times 10^9$~cm~s$^{-1}$ respectively. The more prompt lateral expansion
of the thinner layer leads to the burning being truncated sooner and a lower
propagation speed of the detonation front. 
These simulations were performed at a resolution of 1km, and the
detonation speed varies less than 2\% for a factor of 4 coarser resolution.
Throughout work in this paper, resolution was adjusted to achieve this level
of consistency.

An important contrast created by the difference in shell thickness is the
width of the subsonic region behind the shock front.  In the shock frame,
material entering from the left at the detonation speed, $v_{\rm det}$, is slowed to
subsonic speeds (and compressed) by the shock and then accelerated to the
right by expansion driven by energy release.  This is the driving region for
the propagating detonation and, as will be seen below, nearly all the
reaction and energy release occurs in this zone.  In
Figure~\ref{fig:surface_2grav} this region is indicated by the sonic locus in
the shock-attached frame, i.e. the Mach number $=1$ contour shown as a thin
black
line.  While the cellular structure of the detonation makes the sonic locus
fairly irregular on scales of around $10^7$~cm, its average position is
stable as the detonation propagates.  The distance between the shock and the
sonic locus is smaller near the base of the He layer, and the sonic locus in
fact meets the shock front at the bottom edge of the He layer.  
%This is due to the rarefaction wave reflected back up into the shell at the shell-core boundary as the shock crossing this boundary continues into the interior ofthe WD.  
This is due to the rarefaction wave in the fuel layer arising from the interaction with the inert underlying layer as the shock crossing the shell-core boundary locally continues into the interior of the WD rather than being reflected.This corresponds to the low-impedance case shown in Figure 7.26(a) in \citet{Bdzil12}.
The behavior of the extent of the subsonic region with increasing
height is more subtle but qualitatively similar.  At lower
densities, a propagating detonation will have a larger subsonic
region and longer burning time.  However after a maximum width at around one
scale height above the base of the helium layer, the sonic region becomes
slowly smaller, closing with the shock front in a somewhat less clean way
near the surface of the star.  The decrease in the peak temperature of the
burning with height is also evident.

Comparing the two gravities shown in Figure~\ref{fig:surface_2grav}, we find
that the subsonic region is larger at the lower gravity with the thicker
shell.  At $g_8=2$ it spans $\approx0.2\times 10^8$~cm at its
thickest in the central reaction region (height of about $0.5\times 10^8$~cm),
while at $g_8=4$, the subsonic region spans $\approx0.11\times 10^8$~cm.
The curvature of the detonation front also shows variation, with the detonation
in higher gravity exhibiting a more curved front.
As we show in the next sections, this is consistent with truncation of the
detonation structure by the upward expansion of the shell transverse to the
direction of propagation.  In this case the expansion is vertical behind the
shock front due to the boundaries above and below.  In both cases it
appears that the \emph{entire shell} is in the region affected by the
rarefaction originating from the boundary.  From the ZND analysis in
the previous section (structure shown in Figure \ref{fig:nuc_cj_r5d5}),
$>1$\% He by mass remains out to distances as large as $10^{10}$~cm, the 50\%
reacted length is $\approx 2\times 10^6$cm.  As a result the burning is
quite incomplete on the reaction widths $\sim 10^7$~cm observed here.

\begin{figure}
  %\centering
  %\includegraphics[width=0.8\textwidth]{figures/strip_diagram.pdf}
  \plotone{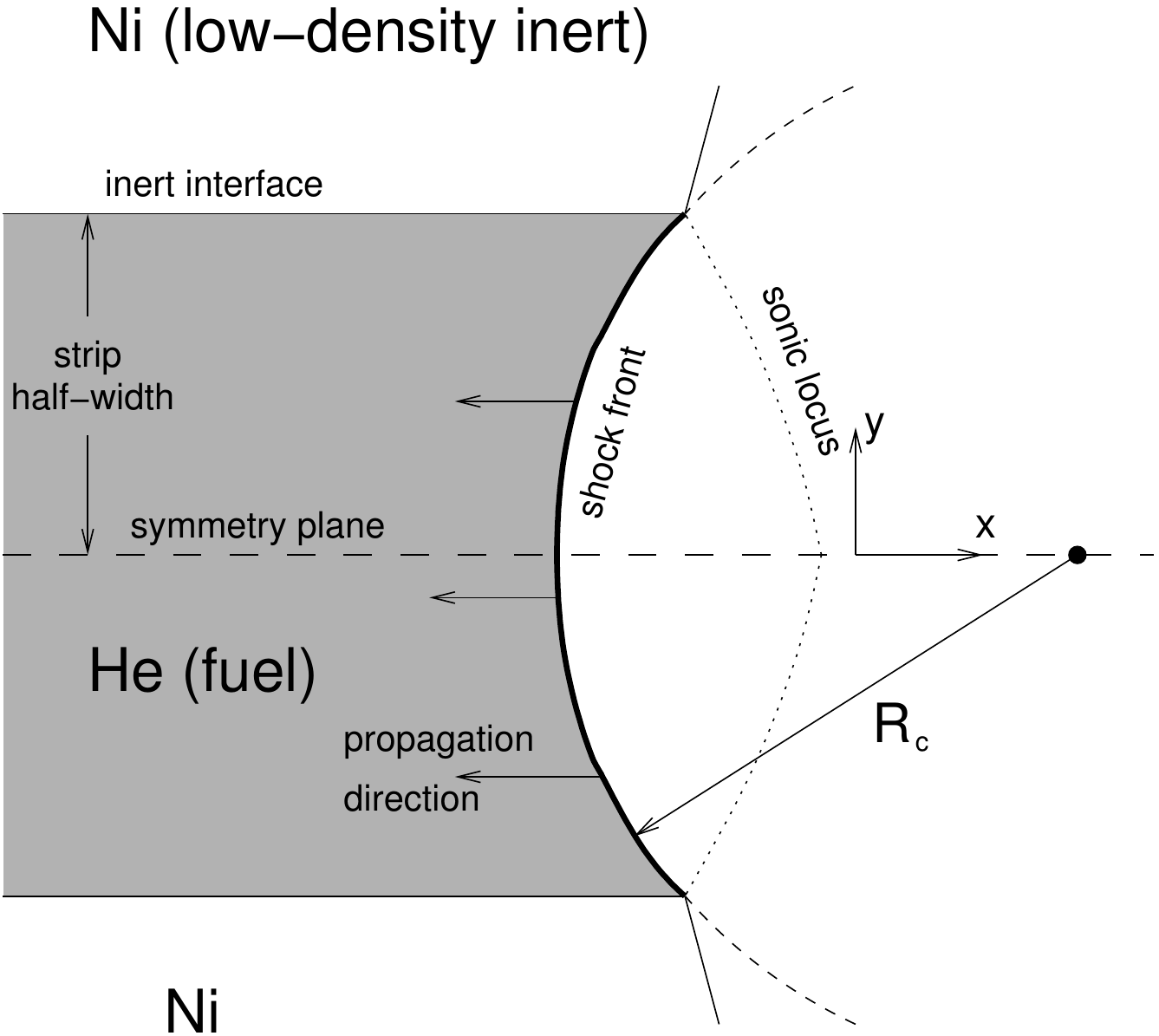}
  \caption{Configuration for uniform-density strip detonation in 2 spatial
dimensions.  The strip is confined in pressure equilibrium by a hot,
low-density gas of Ni, to minimize reactivity.
  \label{fig:strip_diagram}}
\end{figure}

Even this simplified configuration of a
single helium layer in plane parallel presents challenges for comparison to a
direct computation of the steady-state structure.  The two most
significant difficulties are the stratification of the envelope, i.e. the
variation of density across the shock front, and the interaction with the
mildly-reactive underlying layer, where any minor (numerical) mixing leads to a
small amount of extra $\alpha$-capture.  
%The interaction with the lower boundary also makes the detonation cells more prominent, and therefore the sonic surface more irregular.  
This interaction at the shell-core boundary also makes the detonation cells more prominent
For these reasons, in sections
below we will resort to an even more simplified configuration in which a
uniform density strip of He is confined by a low-density non-reactive gas.
This configuration is shown in Figure~\ref{fig:strip_diagram}.  The strip is
confined in pressure equilibrium by a hot field of Ni gas with a density of
$100$~g~cm$^{-3}$.  In this geometry, the detonation shock will also be
curved, with a radius of $R_c$, denoted in
Figure~\ref{fig:strip_diagram}.  The curvature is mostly determined by the
shock interaction with the edge of the fuel strip \citep{Bdzil12}.
This further simplified geometry is comparable to that in the star, but has
the added advantage of symmetric expansion and a uniform fuel density.
There are still rarefaction
waves entering the helium strip from above and below.  In the next section we
will develop a direct computation to be compared to the properties of the
detonation along the center line of this strip configuration.

\begin{figure}
  %\centering
  %\includegraphics[width=0.8\textwidth]{figures/strip_det_example_w25e6.png}
  \plotone{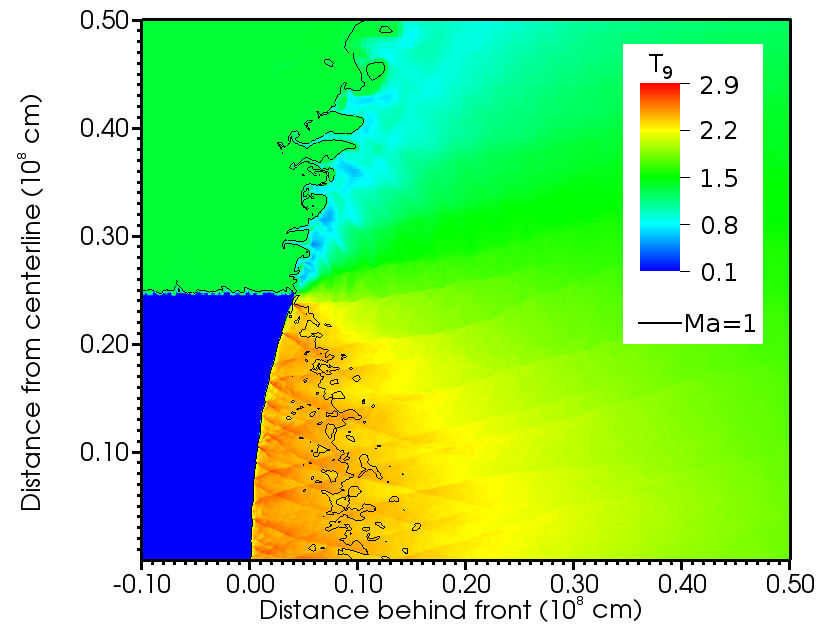}
  \caption{Temperature distribution of a detonation propagating through a strip
  of helium at a uniform density of $5\times 10^5$~g~cm$^{-3}$.  The helium
  extends up to a half-width of $0.25\times10^8$~cm.  The detonation was
  ignited far to the right and at the time shown, $1.0$~s, has propagated
  nearly $10^9$~cm.  A black countour at Mach number of 1 indicates the sonic
  locus, the boundary of the subsonic region behind the shock front.  The
  extent of the subsonic region shows no region of uniform distance behind
  the front, and therefore is clearly everywhere effected by the proximity of
  the blowout at the edge of the strip.
  \label{fig:strip_det_example}}
\end{figure}

Figure~\ref{fig:strip_det_example} shows an example of a detonation
propagating in the strip geometry with a fuel density of $\rho_0 = 5\times
10^5$~g~cm$^{-3}$ and a lateral strip half-width of $0.25\times 10^8$~cm,
half the scale height of the $g_8=4$ case shown in
Figure~\ref{fig:surface_2grav}.  The structure shown is in steady-state
according to a measurement of shock position as a function of time, which
gives a very stable speed.  It has propagated for 1.0~s since ignition,
covering nearly $10^9$~cm. The detonation was initiated by setting a high 
initial temperature, typically $T_9=3$, in a circular region the same diameter 
as the full width of the strip.  This initially leads to an overdriven detonation 
that then weakens to the steady-state self-propagating state as the initial 
extra pressure support behind is lost to lateral expansion. This simulation is 
performed with a resolution 
of $3\times 10^6$~cm.  The detonation speed the same to within less than 
$1$\% compared to a factor of 2 coarser resolution of $6\times 10^6$~cm, and spatial 
features are very similar at the two resolutions.
More features of this calculation will be discussed in later sections.

A critical similarity with the stratified atmosphere is the non-uniform width
of the subsonic region due to the expansion from the unconfined edge.  This
demonstrates that this feature is not related to the density stratification.
The distance from the shock to the average sonic locus is also not uniform
for any region across the detonation front.  This is in contrast to cases
shown in \citet{Bdzil12} in which the reaction length scale is much less than
the lateral width of the strip of fuel.  In the limit of a short reaction
length, the shock front can still be curved in much the same way, but away
from the edge of the strip, near the centerline, the sonic locus is parallel
to the shock.  In our case the slowness of the late-time He consumption
causes the driving region to extend until it is quenched by the blowout,
which depends on the distance from the edge all the way to the centerline.
That is, the entire detonation front is in the edge boundary region.

\section{Generalized ZND formalism}
\label{sec:znd_mod}
Surface detonations in accreted layers on WDs are inherently multidimensional, and the standard ZND equations do not capture important effects such as the post-shock radial expansion or the curved detonation front. These effects have a dramatic impact on the total energy released in a detonation, and thus its propagation speed and burning products. 

The modifications to the hydrodynamic equations that simulate post-shock expansion, either due to the curvature of the detonation front or radial expansion, can be written in the form of source terms in the standard 1D hydrodynamic equations (\ref{eq:mass}) - (\ref{eq:energy}):
\begin{align}
\der{}{x}\left( \rho u_x \right) &= f_1, \label{eq:mass_gen} \\
\der{}{x}\left( P + \rho u_x^2 \right) &= f_2, \label{eq:mom_gen} \\
\der{}{x}\left( e + \frac{P}{\rho} + \frac{u_x^2}{2} \right) &= f_3 \label{eq:energy_gen}.
\end{align}
The standard ZND equations have $f_{1-3} = 0$, whereas here they can be arbitrary functions. Deriving the ZND equations from equations (\ref{eq:mass_gen})-(\ref{eq:energy_gen}) and (\ref{eq:comp}) is straightforward, following the derivation of the standard ZND equations - (\ref{eq:znd_rho})-(\ref{eq:znd_ux}), we obtain
\begin{align}
\der{\rho}{x} &= \frac{1}{c_s^2 - u_x^2} \left( \sum_i \der{X_i}{x} \left[ \pder{P}{e}_\rho \pder{E}{X_i} - \pder{P}{X_i} \right] \right. \nonumber \\
&+ \left. f_2 - 2u_xf_1 - \pder{P}{e}_\rho\left[f_3 - \frac{f_2}{\rho} + \frac{u_x f_1}{\rho}\right] \right), \label{eq:znd_gen_rho}\\
\der{T}{x} &= \pder{P}{T}^{-1}\left[\der{\rho}{x}\left(u_x^2 - \pder{P}{\rho}\right)\right. \nonumber \\
&- \left. \sum_i\pder{P}{X_i}\der{X_i}{x}  + f_2 - 2u_xf_1\right], \label{eq:znd_gen_t} \\
\der{u_x}{x} &= \frac{f_1}{\rho}-\frac{u_x}{\rho} \der{\rho}{x} \label{eq:znd_gen_ux}.
\end{align}
These equations, along with equation (\ref{eq:znd_rxn}), govern the post-shock structure of steady, one-dimensional detonation waves with arbitrary source terms in the hydrodynamic equations. We use them here to examine expansive effects (now through radial expansion and curvature) on the detonation, but they are quite general at this point. We derive specific expressions for $f_{1-3}$ due to curvature and expansion in sections \ref{app:curvature} and \ref{app:blowout} of the appendix.

\subsection{Determining the detonation velocity from initial geometry}
\label{subsec:velocities}
\begin{figure}
	%\plotone{figures/max_mach_plot_he100.pdf}
	\plotone{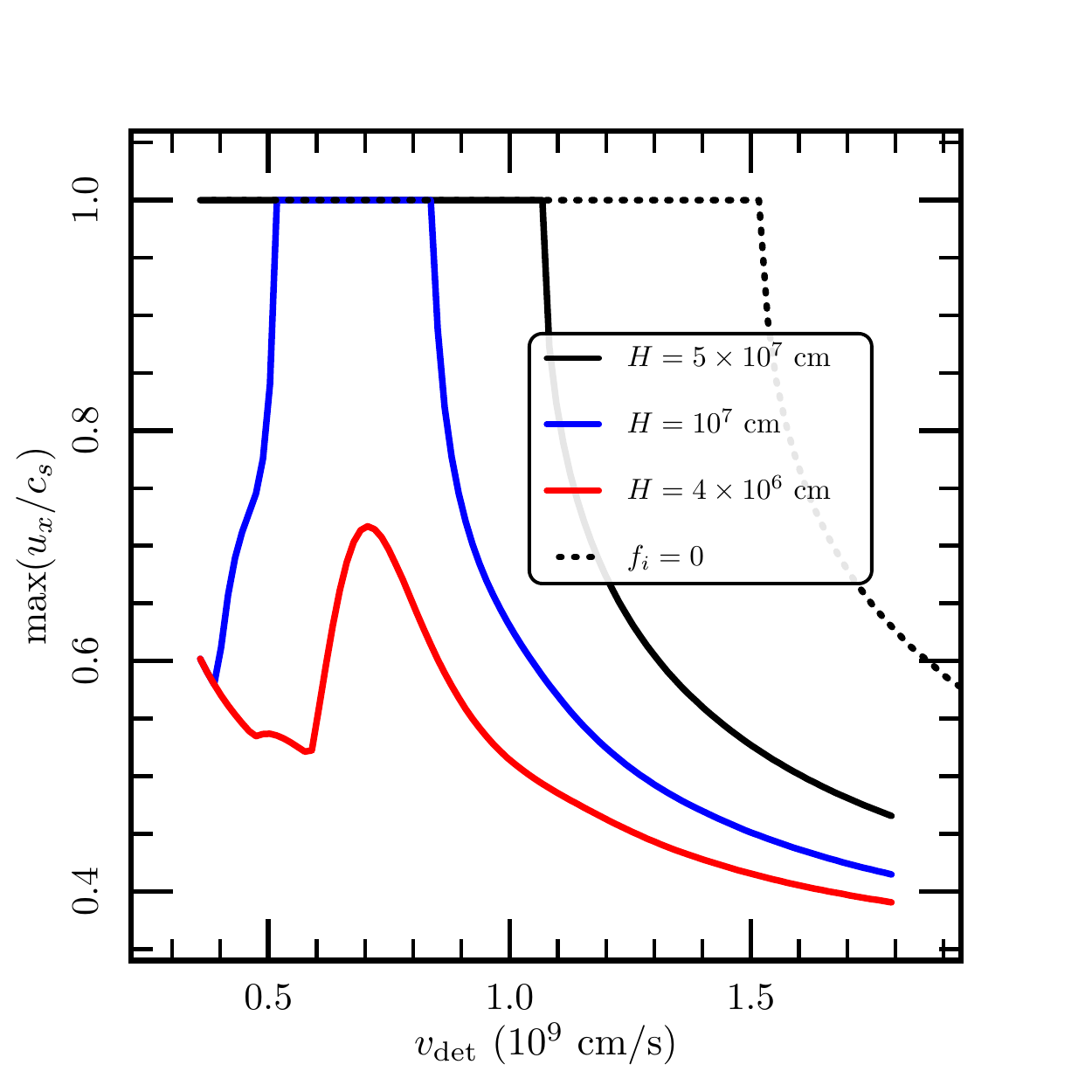}
	\caption{\label{fig:maxmach} Maximum mach number ($u_x/c_s$) achieved in the post-shock flow for pure helium detonations with $\rho_0 = 5\times 10^5$ g/cm$^3$ and $T_0 = 10^8$ K as a function of detonation velocity, $v_{\rm det}$. Since there is a singularity in the equations at $u_x/c_s = 1$, the integration terminates if there is a sonic point in the post-shock flow at that given $v_{\rm det}$. The dotted line shows the results of a ZND integration with no source terms, while the colored lines show the results when source terms due to blowout and curvature are included for different layer thicknesses parameters, $H$, under the condition $R_c = 3H$. Detonations that hit a sonic point during burning are possible for all the cases except the $H = 4\times 10^6$ cm case, since expansion occurs too rapidly for the sonic point to be hit.}
\end{figure}

The standard ZND equations (\ref{eq:znd_rho})-(\ref{eq:znd_ux}), along with our generalized ZND equations (\ref{eq:znd_gen_rho})-(\ref{eq:znd_gen_ux}), encounter a singularity if the flow in the shock frame becomes sonic ($u_x = c_s$) - called the sonic point. Recall from section \ref{subsec:CJ} that the CJ detonation velocity separated solutions that hit the sonic point ($v_{\rm det} < v_{\rm CJ}$) from solutions that did not hit a sonic point ($v_{\rm det} > v_{\rm CJ}$). In a detonation at $v_{\rm det} = v_{\rm CJ}$ the sonic point appears when burning is complete, but this may formally take infinite length and time to achieve depending on the reaction network used. When we find the CJ velocity using the generalized ZND equations (\ref{eq:znd_gen_rho})-(\ref{eq:znd_gen_ux}) with blowout and/or curvature effects, we find that the velocity that separates solutions that encounter the sonic point from those that do not is always less than the $v_{\rm CJ}$ we found without source terms. For example, in the case without source terms shown in Figure \ref{fig:nuc_cj_r5d5}, the separating velocity is $v_{\rm CJ}=1.52\times10^9$ cm/s, whereas for $H=R_c/3=5\times10^7$ cm we get a separating velocity of $1.08\times10^9$ cm/s (the relation $R_c=3H$ used here will be discussed in Section 5 when comparing to simulations). This general effect on detonation velocity is apparent in Figure \ref{fig:maxmach} which shows the separation between underdriven and overdriven solutions in cases with and without source terms. Similar to \citet{He94}, we refer to this CJ-like velocity when source terms are included as the \textit{generalized CJ velocity}, $v_{\rm gcj}$. While the underdriven solution is unphysical because of the singularity, the overdriven detonation requires a supporting pressure in the following flow, and thus a small $u_x$ at large distance.  The unsupported, freely propagating solution can be obtained by integrating through the point where both the numerator and denominator in equation (\ref{eq:znd_gen_rho}) are simultaneously zero, thus avoiding the singularity and giving a consistent solution that is sonically disconnected from the following flow.  This can only occur at $v_{\rm gcj}$, also called the eigenvalue detonation speed \citep{Fickett79}, and gives a following flow in which the pressure and temperature monotonically fall to cessation of burning.

The main difference from a normal CJ detonation is that the reactions will freeze out due to expansive effects caused by blowout and/or curvature.
%From figure \ref{fig:maxmach}, we see that for a certain $H_s$ parameters there are regions with $v_{\rm det} < v_{\rm CJ}$ where detonations hit the sonic point. 
Another important difference between a standard CJ detonation and a generalized CJ detonation with source terms is that the burning no longer stops at the singularity for detonations at $v_{\rm gcj}$. Since we are interested in the nucleosynthesis of such detonations we must calculate the additional burning past this singularity, even though the energy released there does not propel the shock front. The ignorance of the final state requires us to guess detonation velocities and use a bisection search to find $v_{\rm gcj}$ for a given set parameters controlling the blowout and curvature source terms, $H$ and $R_c$. The situation is that of an eigenvalue problem as described in \citet{Sharpe99}. The sonic point in such a detonation at $v_{\rm gcj}$ is called the pathological point, and is a saddle point in the sense that any integration path with $v_{\rm det}$ either too high or too low is repelled from it. At the pathological point, the post-shock structure bifurcates into either a frozen subsonic solution ($u_x < c_s$) or a frozen supersonic solution ($u_x > c_s$) for flow beyond the pathological point.

\begin{figure}
	\plotone{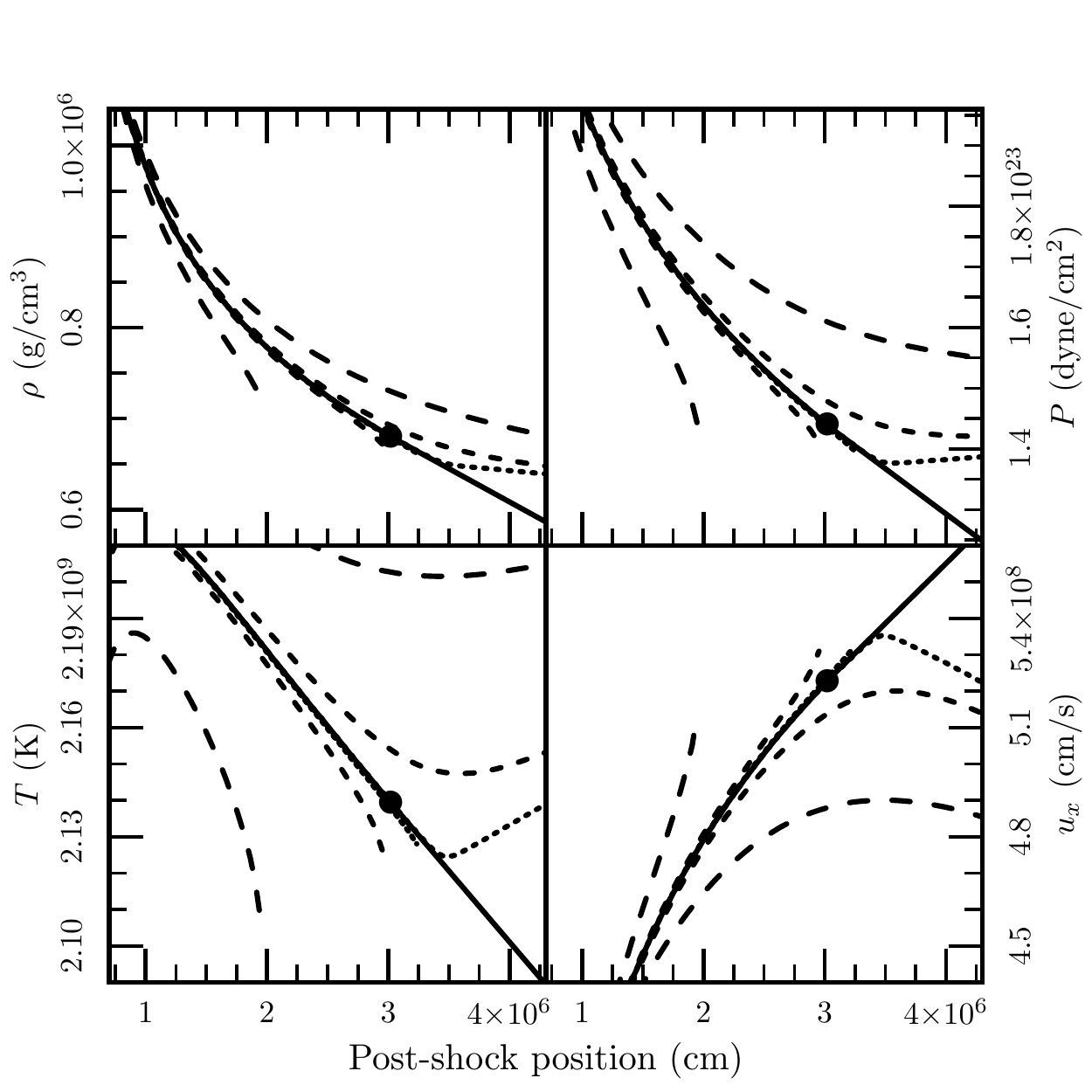}
	\caption{\label{fig:thermo_pane_pathological_point} Thermodynamic variables near the pathological point for a pure helium detonation with initial conditions $\rho_0 = 5\times 10^5$ g/cm$^3$, $T_0 = 10^8$ K and source term parameters $H = 10^7$ cm, $R_c = 3\times 10^7$ cm - resulting in a detonation velocity of $v_{\rm gcj} = 8.418\times 10^8$ cm/s. The broken lines correspond to integrations using detonation velocities given by the bracketing velocities in a bisection search for $v_{\rm gcj}$. Integrations for $v_{\rm det}$ greater/less than $v_{\rm gcj}$ are shown using tolerances of $10^{-2}$ (long dashed lines), $10^{-3}$ (medium dashed lines), and $10^{-4}$ (short dashed lines). The solid line is the eigenvalue solution described in the text using a velocity within $10^{-5}$ of $v_{\rm gcj}$ and linearized when the mach number reached $0.98$, indicated by the solid dot. Detonation velocities less than $v_{\rm gcj}$ hit a sonic point, while those greater than $v_{\rm gcj}$ lie on the frozen subsonic branch.}
\end{figure}

For a fixed set of initial thermodynamic and compositional conditions, each value of $v_{\rm det}$ will correspond to a different set of post-shock initial conditions for the generalized ZND equations. The source terms require additional information, namely the ambient thickness of the medium, $H$, and the radius of curvature of the detonation front, $R_c$. We show the behavior of solutions around the pathological point with $v_{\rm det}$ near $v_{\rm gcj}$ graphically in Figure \ref{fig:thermo_pane_pathological_point}. Each line style corresponds to a pair of $v_{\rm det}$ values that are within a certain tolerance of $v_{\rm gcj}$ $(10^{-2} - 10^{-4})$ as determined by a bisection search. Numerical solutions will never reach the pathological point since it is a saddle point, but we can get arbitrarily close by picking a sufficiently stringent tolerance. Solutions with $v_{\rm det} < v_{\rm gcj}$ hit a sonic point and terminate, while solutions with $v_{\rm det} > v_{\rm gcj}$ always remain subsonic. In order to reach the branch where the flow is supersonic with respect to the shock front (the freely propagating solution), we need to traverse the pathological point.

%Each line is a separate integration of the generalized ZND equations with blowout and curvature source terms using a different value of $v_{\rm det}$ either slightly above or below where we show how tightening the bounds on $v_{\rm gcj}$ allows us to get very close to the pathological point. 

Numerical integration of a system of ODEs containing a coordinate singularity is a challenge. Before we go on, it is useful to examine another way of writing the conditions of a CJ (and generalized CJ) detonation. For a normal CJ detonation, the condition that the flow hits the sonic point when burning is complete means that the numerator and denominator of equation (\ref{eq:znd_rho}) both go to zero at the same time. When we move to the generalized CJ velocity, we are effectively requiring the numerator and denominator of equation (\ref{eq:znd_gen_rho}) to go to zero at the same place - the pathological point. This implies that the thermodynamic derivatives are defined near the pathological point. Although we can never reach the pathological point in a numerical integration, we can get arbitrarily close, and use the fact that the thermodynamic derivatives are well-behaved around the pathological point to linearize our solution past the pathological point. Figure \ref{fig:thermo_pane_pathological_point} shows how choosing tighter bounds on the integration velocity allows us to linearize the equations closer to the pathological point.

%This procedure is shown graphically in figure \ref{fig:thermo_pane_pathological_point}, where we show how tightening the bounds on $v_{\rm gcj}$ allows us to get very close to the pathological point. 

\begin{figure}
	\plotone{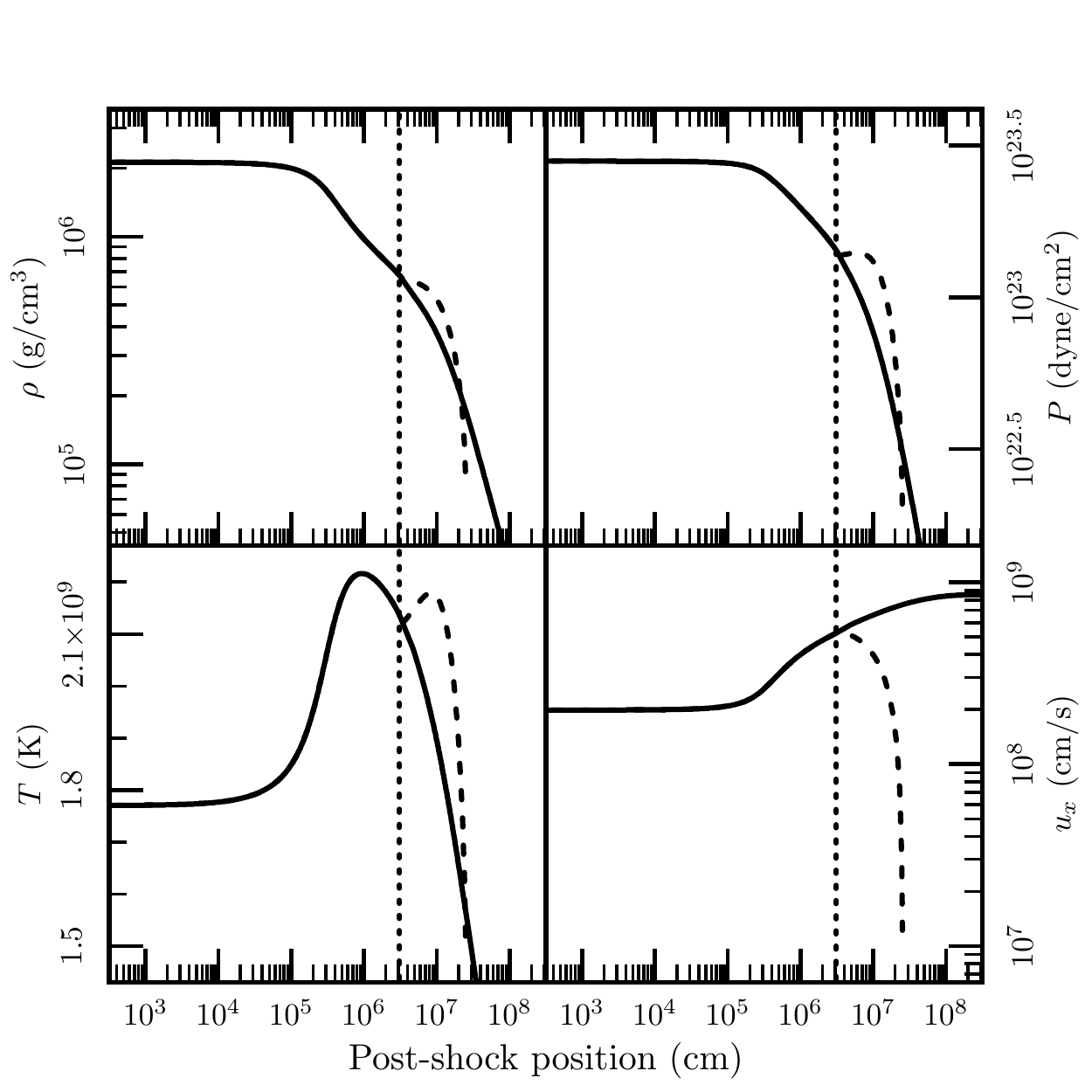}
	\caption{\label{fig:thermo_pane_pathological} Thermodynamic variables for a pure helium detonation with initial conditions $\rho_0 = 5\times 10^5$ g/cm$^3$, $T_0 = 10^8$ K and source term parameters $H = 10^7$ cm, $R_c = 3\times 10^7$ cm - resulting in a detonation velocity of $v_{\rm gcj} = 8.418\times 10^8$ cm/s. Solid lines correspond to the solution that traverses the pathological point and becomes frozen supersonic beyond it, while dashed lines correspond to the frozen subsonic solution. The vertical dotted lines indicate where the solution was linearized to jump over the pathological point.}
\end{figure}

\begin{figure}
	\plotone{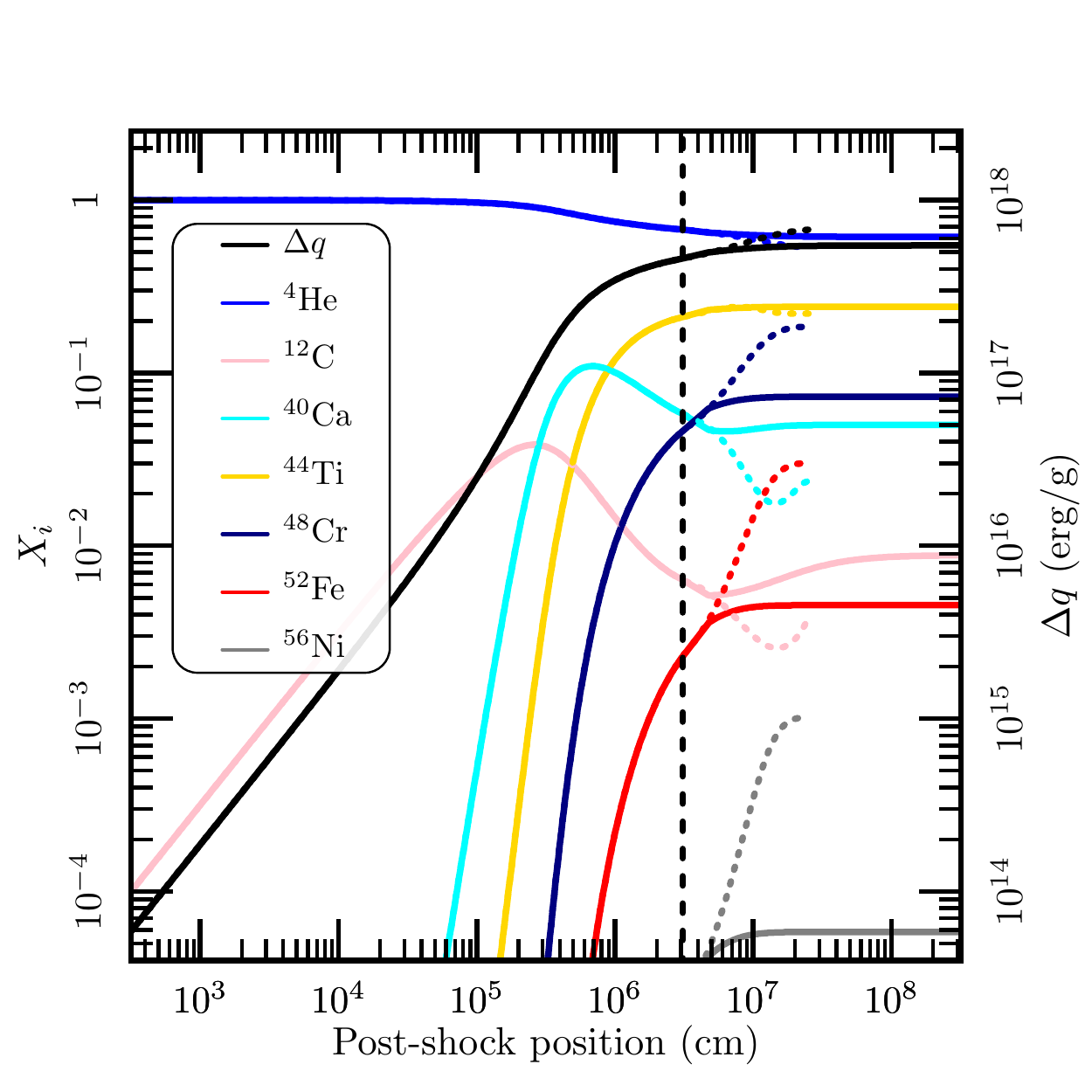}
	\caption{\label{fig:abundances_pathological} Nucleosynthesis for a pure helium detonation with initial conditions $\rho_0 = 5\times 10^5$ g/cm$^3$, $T_0 = 10^8$ K and source term parameters $H = 10^7$ cm, $R_c = 3\times 10^7$ cm - resulting in a detonation velocity of $v_{\rm gcj} = 8.418\times 10^8$ cm/s. Colored lines are abundances (left axis) and the black line is the cumulative energy release, $\Delta q$ (right axis). Solid lines correspond to the solution that traverses the pathological point and becomes frozen supersonic beyond it, while dotted lines correspond to the frozen subsonic solution. The vertical dashed line indicates where the equations were linearized to jump over the pathological point.}
\end{figure}

We traverse the pathological point via the following linearization method, similar to that of \citet{Sharpe99}. Given a set of blowout and curvature parameters $(H, R_c)$ we find the eigenvalue detonation velocity, $v_{\rm gcj}$, to a tolerance of a part in $10^{-5}$. We then use the lower bound velocity (which hits a sonic point slightly before the true pathological point), and integrate the generalized ZND equations until we get close to the sonic point (typically a mach number limit of $\sim 0.99$ is used). We then linearize the generalized ZND equations over a length given by $10-50\%$ of the current $x$ coordinate and resume integration, now on the frozen supersonic branch. The difference between the frozen subsonic and frozen supersonic solutions can be dramatic, requiring us to traverse the pathological point and find the frozen supersonic solution since self-sustaining detonations are frozen supersonic. The resulting solutions for a specific case are shown in Figures \ref{fig:thermo_pane_pathological} \& \ref{fig:abundances_pathological}. The frozen subsonic solutions shown in dashed lines require supporting pressure behind the pathological point to keep the flow subsonic relative to the shock front. The flow velocity $u_x$ goes to zero, indicating that the burned material is traveling at the same speed as the shock front. The thermodynamic derivatives are seen to be discontinuous across the pathological point for the frozen subsonic solution, while they are continuous along the frozen supersonic branch.

\begin{figure}
	\plotone{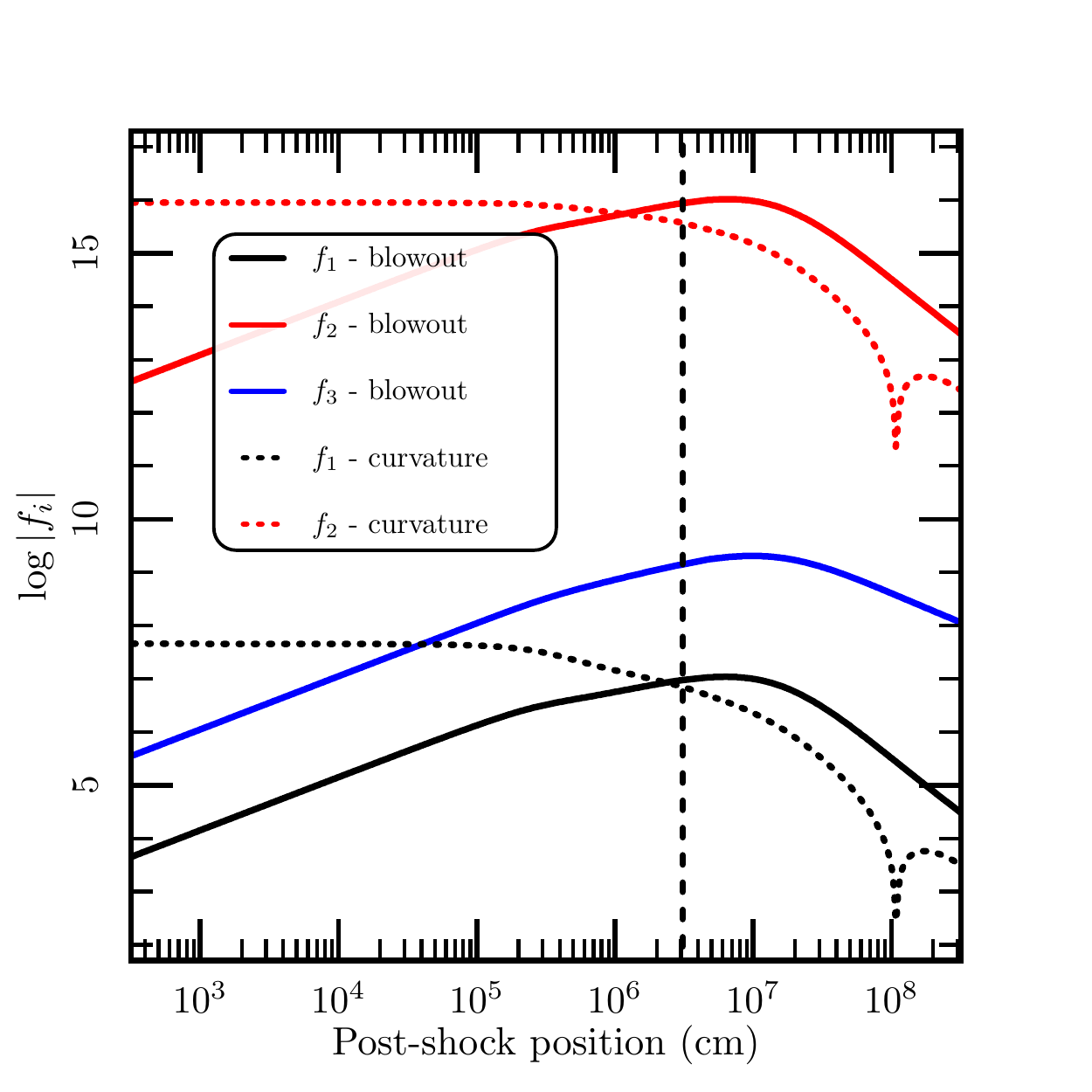}
	\caption{\label{fig:source_term_pathological} Absolute values of source terms $f_1 - f_3$ due to blowout (solid lines) and curvature (dotted lines) as a function of distance behind the shock front for a pure helium detonation with initial conditions $\rho_0 = 5\times 10^5$ g/cm$^3$, $T_0 = 10^8$ K and source term parameters $H = 10^7$ cm, $R_c = 3\times 10^7$ cm. Note that $f_3 = 0$ for curvature. The vertical dashed line indicates where the equations were linearized to jump over the pathological point. All terms start negative, but the curvature terms switch sign when $u_x = v_{\rm det}$ at $x\approx 10^8$ cm. At early times, the velocity divergence due to curvature is the important modification to the ZND equations, while at late times blowout dominates.}
\end{figure}

Once we have a full frozen supersonic solution, knowing the relative strengths of the source terms due to blowout and curvature along the solution will tell us where each physical effect is important. Recall that our derivation of the curvature source terms assumed that we were in the limit of $x \ll R_c$, while the blowout source terms had no such restriction. Figure \ref{fig:source_term_pathological} shows that the blowout source terms become dominant when we are out of the range of validity for the curvature source terms, making the curvature source terms inconsequential. This allows us to integrate the generalized ZND equations beyond the $x \ll R_c$ limit and be confident in our results. We leave the curvature terms turned on for calculational simplicity, but they could be turned off around $x \approx R_c$ without affecting the integration.
%If we do not include the effects of blowout, then we are not able to find the final state of the material using the curvature source terms derived in section \ref{subsec:curvature}.

The generalized CJ velocity corresponds to the steady-state detonation velocity when the expansive effects of  blowout and curvature are considered. We need to check and calibrate our 1D prescription using multidimensional hydrodynamic simulations of steady laterally propagating detonations in order to make sure our treatment of blowout is reasonable. The next section details our comparisons between the 1D integrations and {\ttfamily FLASH} models.

\section{Comparisons at constant density}
\label{sec:results}

Our goal is to use our 1D ZND model with curvature and blowout to predict detonation speeds and nucleosynthesis. Although the 1D model is physically motivated, it remains a prescription for simulating multidimensional expansive effects. Hence, in this section we compare its predictions to those from reactive compressible hydrodynamic simulations in {\ttfamily FLASH}. Analysis of {\ttfamily FLASH} simulations in finite gravity atmospheres led us to investigate a simpler constant density scenario, as described in \S \ref{sec:flash_intro}, so our comparisons begin with those models. {\ttfamily FLASH} simulations use the aprox13 reaction network, while our 1D model uses aprox19 \citep{Timmes99}. These networks are effectively identical for the regimes of helium burning we investigate (aprox19 has additional reactions for H burning and photodisintegration, both of which are unimportant in the detonations we consider due to reaction freeze out). We first describe how we relate the curvature of the detonations front, $R_c$, to the layer thickness, $H$, reducing the number of free parameters by one. We then compare detonation velocities, nucleosynthesis, and thermodynamic profiles resulting from the 1D generalized ZND equations of section \S \ref{sec:znd_mod} to those from 2D plane-parallel {\ttfamily FLASH} simulations with helium layers of various densities, thickness, and compositions. We finally summarize the regions of $(\rho_0, H)$ space where steady detonations can propagate for various compositions.

%The simplest 2D model has a layer of helium at uniform ambient density $\rho_{\rm He}$ and thickness $H_s$ with no gravity. We choose the helium layer to have a density and temperature agreeing with out accretion models and keep it contained by putting a layer of hot, low density $^{56}$Ni above it. Detonations initiated in the helium will rapidly expand into the $^{56}$Ni region, but without the added complications of gradients in the variables due to gravity.

%\subsection{Curvature of the detonation front in FLASH}
We developed a treatment of curvature and blowout in the previous sections, with each effect adding an additional length scale - $R_c$ for curvature, and $H$ for blowout.
%but do not yet have a way of relating the curvature of the detonation front to the thickness of the helium layer. 
Steady detonations in constant density layers have shock fronts with well-defined curvature, as evident in Figure %\ref{fig:flash_curvature}. 
\ref{fig:strip_det_example}. Empirically, we found that the detonation fronts have a radius of curvature that relates to the layer thickness as $R_c \approx 3 H$. This was found through a large set of constant density {\ttfamily FLASH} simulations. From here on, we use this relation so that there is only one independent parameter controlling the rate of post-shock expansion, $H$. The parameters left to vary are the initial density, $\rho_0$, composition, ${\bf X_0}$, and layer thickness, $H$.

%\begin{figure}
%	\centering
%  	\includegraphics[width=1.0\textwidth]{figures/Hestripdet_d5e5_w1e8_he4_mach_0094.png}
%	\caption{\label{fig:flash_curvature} Snapshot of a steady detonation propagating in a pure helium layer of thickness $10^8$ cm with $\rho_0 = 5\times 10^5$ cm and $T_0 = 10^8$ K.}
%\end{figure}

\subsection{Comparisons between 2D simulations and 1D analytics}
The first and simplest point of comparison is the steady-state detonation velocity in a helium layer of varying thickness, $H$, and initial density, $\rho_0$. We performed {\ttfamily FLASH} simulations with helium layers with various initial abundances of $^{12}$C and $^{16}$O having $\rho_0 = 5\times 10^5$ g/cm$^{3}$ and $H = 10^6-10^9$ cm, finding steady-state detonation velocities of $v_{\rm gcj} = 0.59 - 1.37 \times 10^9$ cm/s. Each layer thickness yields a unique detonation velocity, $v_{\rm gcj}$, as described in \S \ref{subsec:velocities}. This generalized CJ velocity continuously connects to the standard CJ detonation velocity in the $H \rightarrow \infty$ limit. Resolution studies of the {\ttfamily FLASH} simulations indicate that a spatial resolution of $0.6$ km is sufficient for determining the steady-state detonation speed, with slightly smaller detonation velocities found for coarser resolutions.

\begin{figure}
  %\centering
  %\includegraphics[width=1.0\textwidth]{figures/vcj_hs_plot-curvature_blowout}
  \plotone{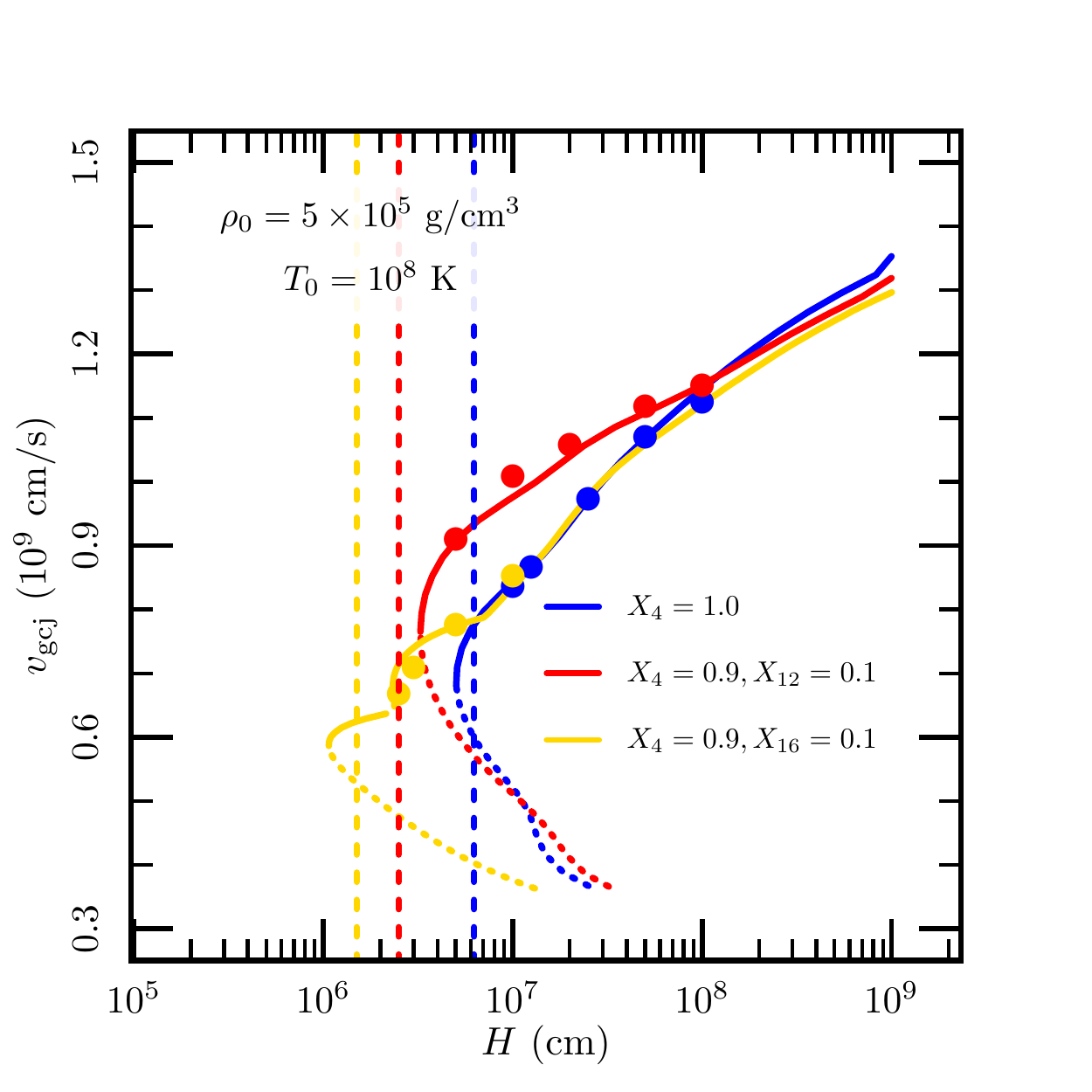}
  \caption{Generalized CJ velocities as a function of initial layer thickness $H$, using $R_c = 3H$, for various initial compositions. The initial thermodynamic conditions were $\rho_0 = 5\times 10^5$ g/cm$^3$ and $T_0 = 10^8$ K for all cases. Solid line segments indicate $v_{\rm gcj}$ values, while the dotted sections correspond to the unphysical solutions at lower detonation velocities (corresponding to the low-velocity side of the cliff in Figure \ref{fig:maxmach}). Note the small secondary forbidden region of $v_{\rm gcj}$ for $X_{16} = 0.1$ at $H\approx 2.5\times 10^7$ cm. Colored dots correspond to individual realizations in {\ttfamily FLASH}, showing good agreement between our 1D model and direct numerical simulation. The vertical dashed lines indicate the thickest layer in which a detonation would not propagate in {\ttfamily FLASH}. \label{fig:vcj_hs}}
\end{figure}

The lines in Figure \ref{fig:vcj_hs} show generalized CJ velocities using $\rho_0 = 5\times 10^5$ g/cm$^3$ as a function of $H$ for a few characteristic initial compositions, along with points corresponding to individual 2D {\ttfamily FLASH} simulations. 
%Scaling $H_s$ up by a factor of $5$ produces good agreement between our 1D model and FLASH simulations (shown by the dashed line), indicating the physical thickness of the helium layer, $d$, is connected to the scale height parameter via $H_s \approx d/5$. 
The minimum in $H$ for each composition shows that steady detonations cannot exist for layers that are too thin - the post-shock expansion occurs so quickly that the pathological point cannot be reached and the flow always remains frozen subsonic. This means that the post-shock rarefaction can communicate back to the shock front and quench the detonation, stopping it from propagating. This effect can also be seen in Figure \ref{fig:maxmach}, where decreasing $H$ (also under the assumption $R_c = 3H$) eventually prevented solutions from hitting a sonic point in the post-shock flow. We observe a similar effect in the {\ttfamily FLASH} simulations, where reducing the layer thickness beyond a point does not allow us to ignite steadily propagating detonations. We report the detonation limits on $H$ found in {\ttfamily FLASH} with vertical dashed lines in Figure \ref{fig:vcj_hs}, indicating the thickest layer that failed to yield a propagating detonation for each composition.
The slower detonations in $^{4}$He + $^{16}$O mixtures having speeds around $0.6\times 10^9$ cm/s are difficult to realize in hydrodynamic simulations at this density due to stability considerations, but we found them to be realizable at lower densities. 
Such detonations only burn up to $^{28}$Si and can be unstable 
in the sense that small fluctuations in post-shock temperature 
can allow burning further up the alpha chain 
(typically to $^{40}$Ca), thereby increasing the detonation velocity.
A steady-state model will not capture this effect, 
however, and predicts such detonations to occur.
This is consistent with the broader extent at lower densities of the regime in which burning terminates at $^{28}$Si before dominance of $^{40}$Ca (see \S \ref{subsec:prob_limits_theory}).
%While these minimum thicknesses do not exactly agree, the disagreement is going in a less severe direction. It would be worse to have seen propagating detonations in FLASH in our predicted ``forbidden region'', rather than FLASH detonations that 

%\begin{figure}[h!]
 % \centering
  %\includegraphics[width=1.0\textwidth]{figures/vcj_hs}
 % \caption{Generalized CJ velocities as a function of scale height parameter $H_s$ for various initial compositions. The initial thermodynamic conditions were $\rho_0 = 5\times 10^5$ g/cm$^3$ and $T_0 = 10^8$ K for all cases. Solid lines indicate solutions that continuously connect to the standard CJ detonations in the $H_s \rightarrow \infty$ limit, while the dotted sections correspond to the unphysical solutions at lower detonation velocities (corresponding to the other side of the cliff in figure \ref{fig:maxmach}). The dashed black line shows the predicted detonation velocities when $d = 5H_s$ is used, and the black dots are specific FLASH realizations. The vertical dotted line indicates the thickest helium layer in which we did not see a detonation propagate (we ran no simulations with thicknesses in between this and the minimum predicted thickness). \label{fig:vcj_hs}}
%\end{figure}

\begin{figure}
  %\centering
  %\includegraphics[width=1.0\textwidth]{figures/final_abundances_r5d5_he100_vdet_pathological}
  \plotone{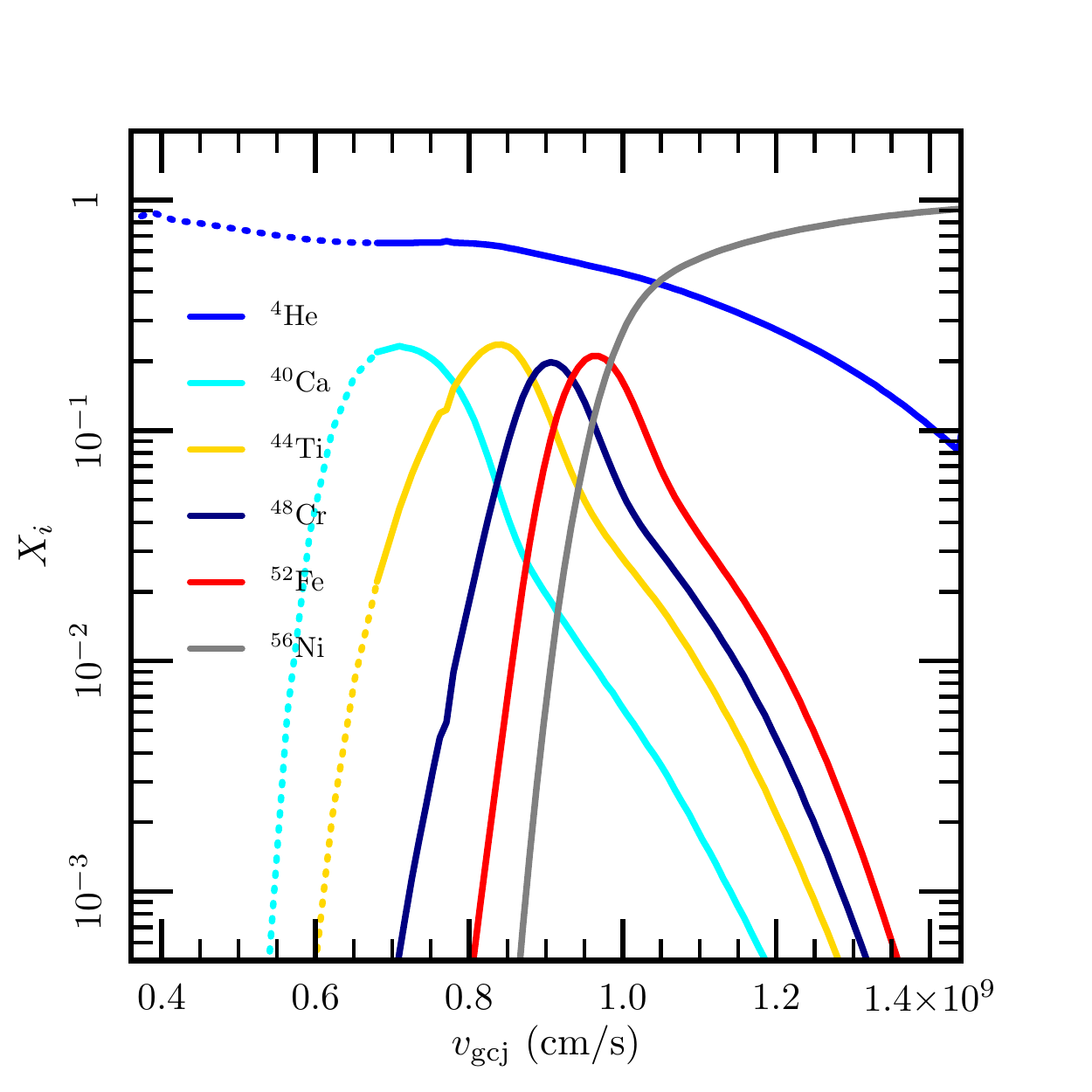}
  \caption{Final abundances of a pure helium detonation with $\rho_0 = 5\times 10^5$ g/cm$^3$ and $T_0 = 10^8$ K in the generalized 1D ZND model. As shown in Figure \ref{fig:vcj_hs}, each value of $v_{\rm gcj}$ corresponds to a different $H$, necessary to make that detonation velocity the generalized CJ velocity.\label{fig:final_abundances_r5d5_he100_vdet}}
\end{figure}

\begin{figure}
  %\centering
  %\includegraphics[width=\textwidth]{figures/Abundances_plot_flash}
  \plotone{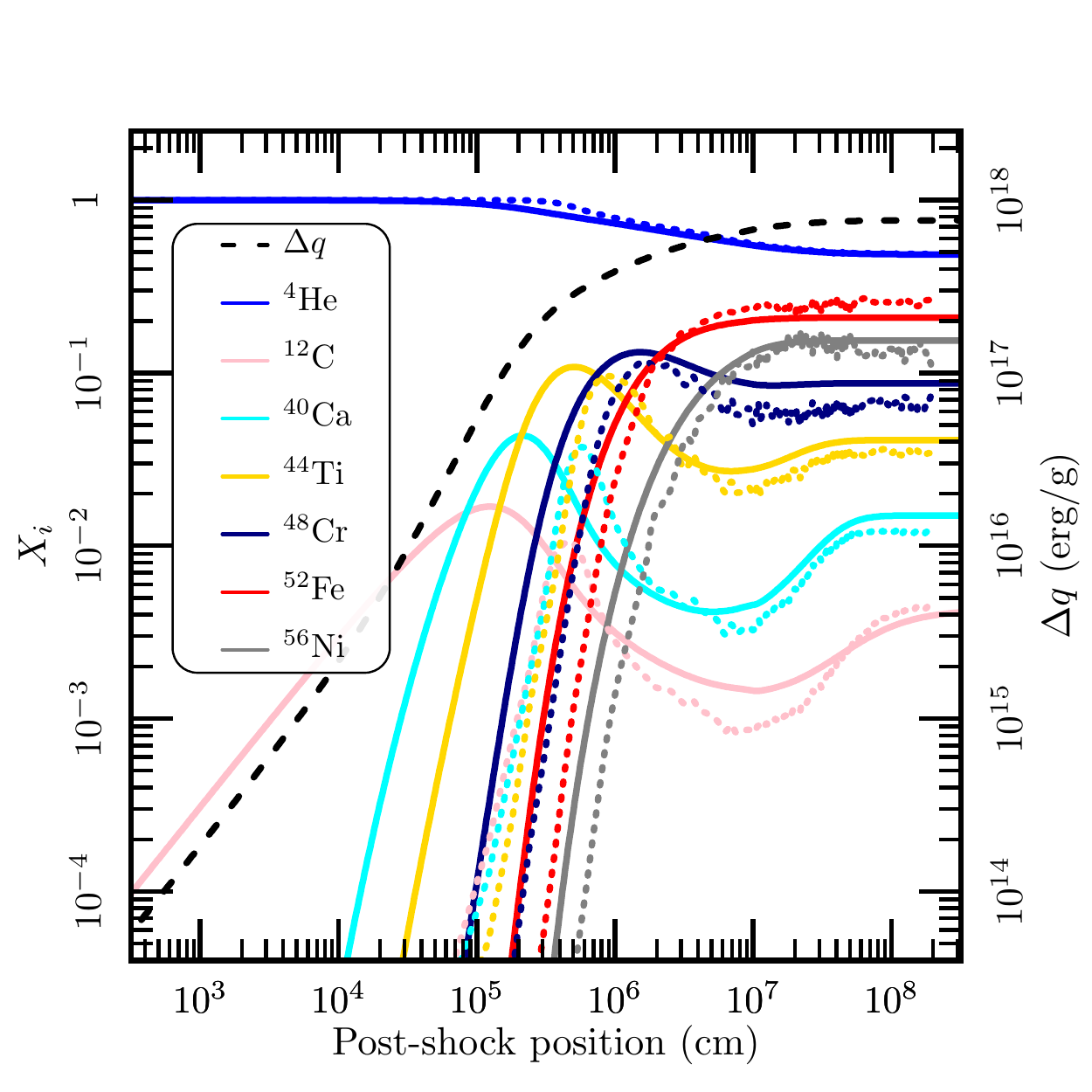}
  \caption{Nucleosynthesis comparisons between a {\ttfamily FLASH} model with $\rho_0 = 5\times 10^5$ g/cm$^3$, thickness $H=2.5\times 10^7$ cm, and steady-state detonation velocity $v_{\rm FLASH} = 0.97\times 10^9$ cm/s (dotted lines) and a 1D generalized ZND integration using the same initial conditions, also finding $v_{\rm gcj} = 0.97\times 10^9$ cm/s (solid lines).
  %The scale height parameter that reproduces the actual FLASH velocity is $H_s = 5\times 10^6$ cm, while we need $H_s = 6\times 10^6$ to get the $v_{\rm gcj}$ value used here, in line with the displaced dashed curve in figure \ref{fig:vcj_hs}. 
The dashed black line is the cumulative energy release, to be read off the right axis. Since the {\ttfamily FLASH} simulation has a resolution of $0.6$ km, the early-time nucleosynthesis is slightly different from the 1D model, although they reach very nearly the same final state. \label{fig:1d_to_flash_nucleosynthesis}}
\end{figure}

\begin{figure}
  %\centering
  %\includegraphics[width=1.0\textwidth]{figures/thermo_pane_flash}
  \plotone{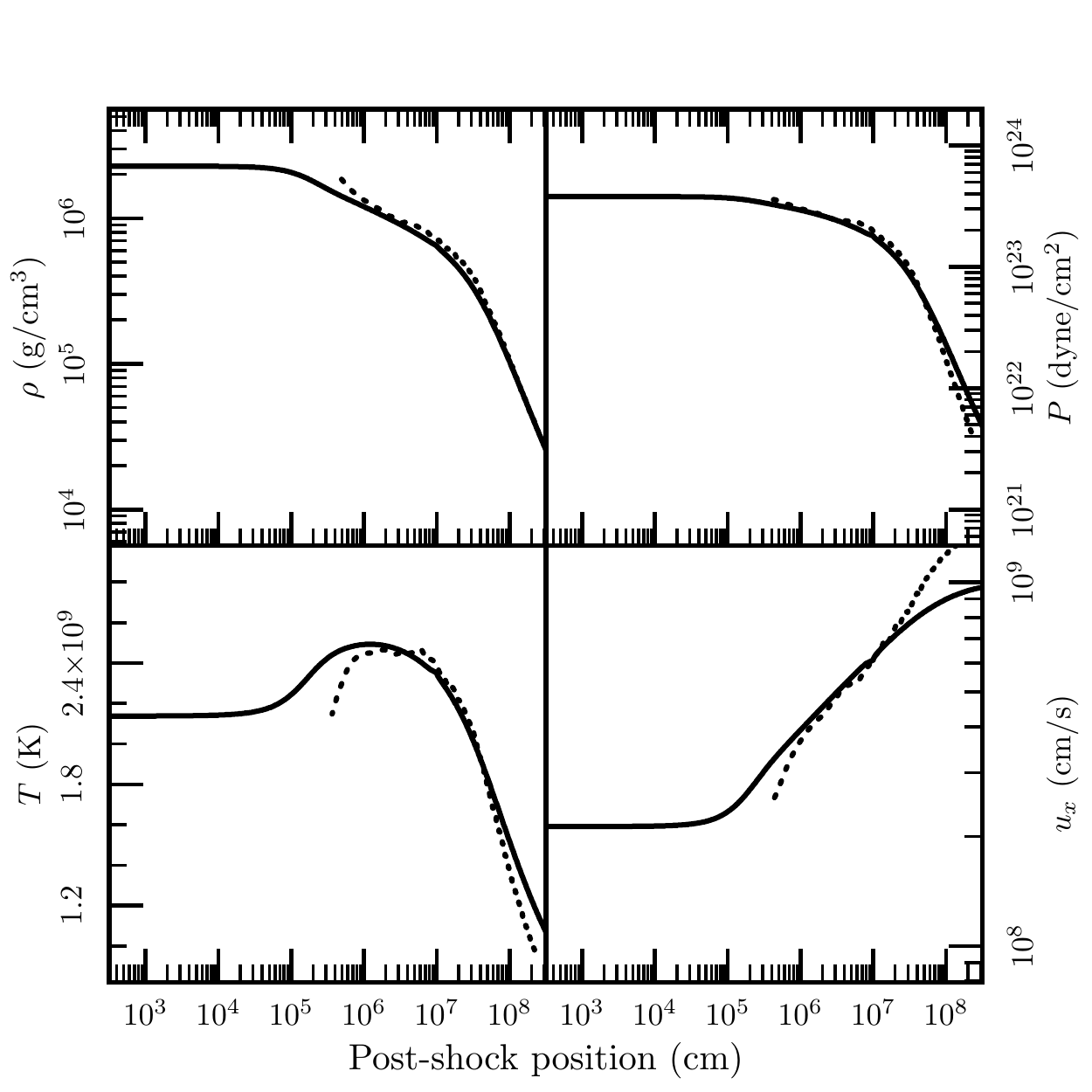}
  \caption{Thermodynamic comparisons between a {\ttfamily FLASH} model with $\rho_0 = 5\times 10^5$ g/cm$^3$, thickness $H=2.5\times 10^7$ cm, and steady-state detonation velocity $v_{\rm FLASH} = 0.97\times 10^9$ cm/s (dotted lines) and a 1D generalized ZND integration using the same initial conditions, also with $v_{\rm gcj} = 0.97\times 10^9$ cm/s (solid lines). The early-time variations are mainly due to the finite resolution ($0.6$ km) of the {\ttfamily FLASH} simulations. \label{fig:thermo_pane_flash}}
\end{figure}

Since the detonation velocities are in good agreement with our 1D models, we now move to post-shock structure comparisons. We first note the strong sensitivity of nucleosynthesis on detonation velocity via the initial post-shock conditions - mainly temperature. Small changes in $v_{\rm det}$ ($\sim 1\%$) can produce very large changes in the final product mass fractions ($\sim 100\%$), as Figure \ref{fig:final_abundances_r5d5_he100_vdet} shows, so we do not expect to see very close agreement between {\ttfamily FLASH} models and our 1D models, a priori. To show that our spectrum of solutions is reasonable, we compare the {\ttfamily FLASH} nucleosynthesis of a pure helium layer with a 1D model using the same thickness in Figure \ref{fig:1d_to_flash_nucleosynthesis}. In order to obtain smoother quantities for comparison to 1D, abundances and thermodynamic quantities from {\ttfamily FLASH} are averaged perpendicular to the direction of propagation over a region within $4\times 10^6$~cm of the symmetry plane shown in Figure \ref{fig:strip_diagram}.  This is approximately the width of the cellular detonation structures observed in Figure \ref{fig:strip_det_example}. The close agreement shows that our equations describing the post-shock expansive effects are capturing the relevant physics. Thermodynamic comparisons are shown in Figure \ref{fig:thermo_pane_flash}, again showing that the post-shock evolution is well described by our 1D model.

%Although the final states are somewhat different, particularly with regards to the $^{56}$Ni abundances, we note that the composition of the ashes is extremely sensitive to $v_{\rm gcj}$, so we don't expect as strong agreement with nucleosynthesis. 

\subsection{Impact of carbon and oxygen in the fuel}
We now investigate the effects of adding $^{12}$C and $^{16}$O to the initial fuel. There are a number of ways in which these isotopes may be produced. First, in the convective burning period prior to the helium shell going dynamical ($t_{\rm burn} < t_{\rm dyn}$), significant amounts of $^{12}$C + $^{16}$O (5-20\% by mass) may be synthesized before a detonation is ignited \citep{Shen09}. Additionally, interaction of the base of the convective zone with either the core of the WD or the layer of C/O ashes from previous shell flashes may mix carbon/oxygen into the reaction zone. Similar mixing could occur along the base of the laterally propagating detonation wave as well. Finally, accretion scenarios from helium burning stars allow the composition of accreted material to contain $^{12}$C and $^{16}$O \citep{Yungelson08}.

\begin{figure}
	\plotone{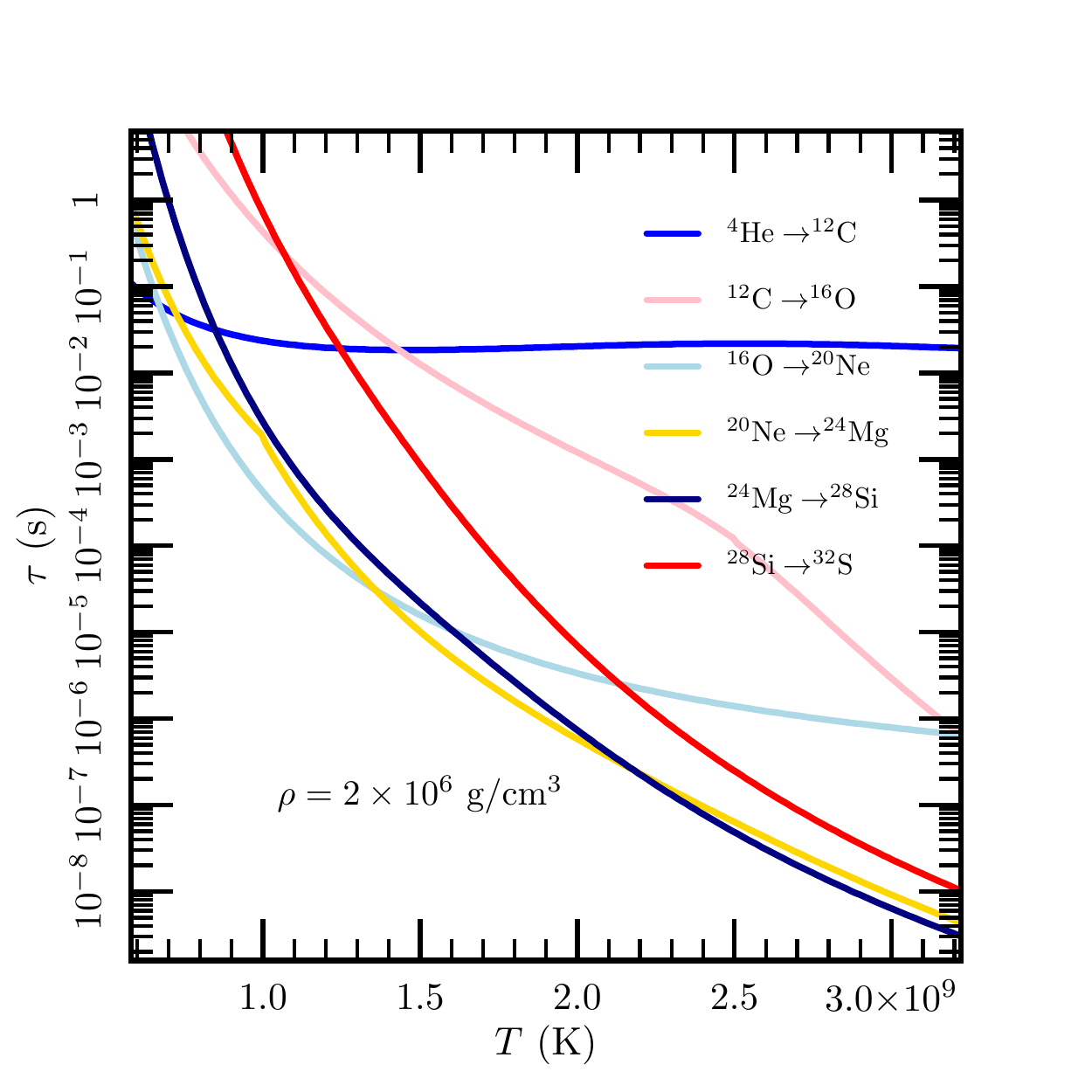}
	\caption{\label{fig:t_alpha} Lifetimes of various $\alpha$-chain nuclei against $\alpha$-capture, including both $(\alpha, \gamma)$ and $(\alpha, p)(p, \gamma)$ channels, computed using pure $^4$He at a density of  $\rho =  2\times10^6$ g/cm$^3$. The lifetimes are defined as $\tau_i = (n_\alpha \left<\sigma v \right>_i)^{-1}$, where $n_\alpha$ is the number density of $\alpha$ particles, and $\left<\sigma v \right>_i$ is the net velocity-averaged $\alpha$-capture cross section onto the $i^{\rm th}$ species.}
\end{figure}

\begin{figure}
	\plotone{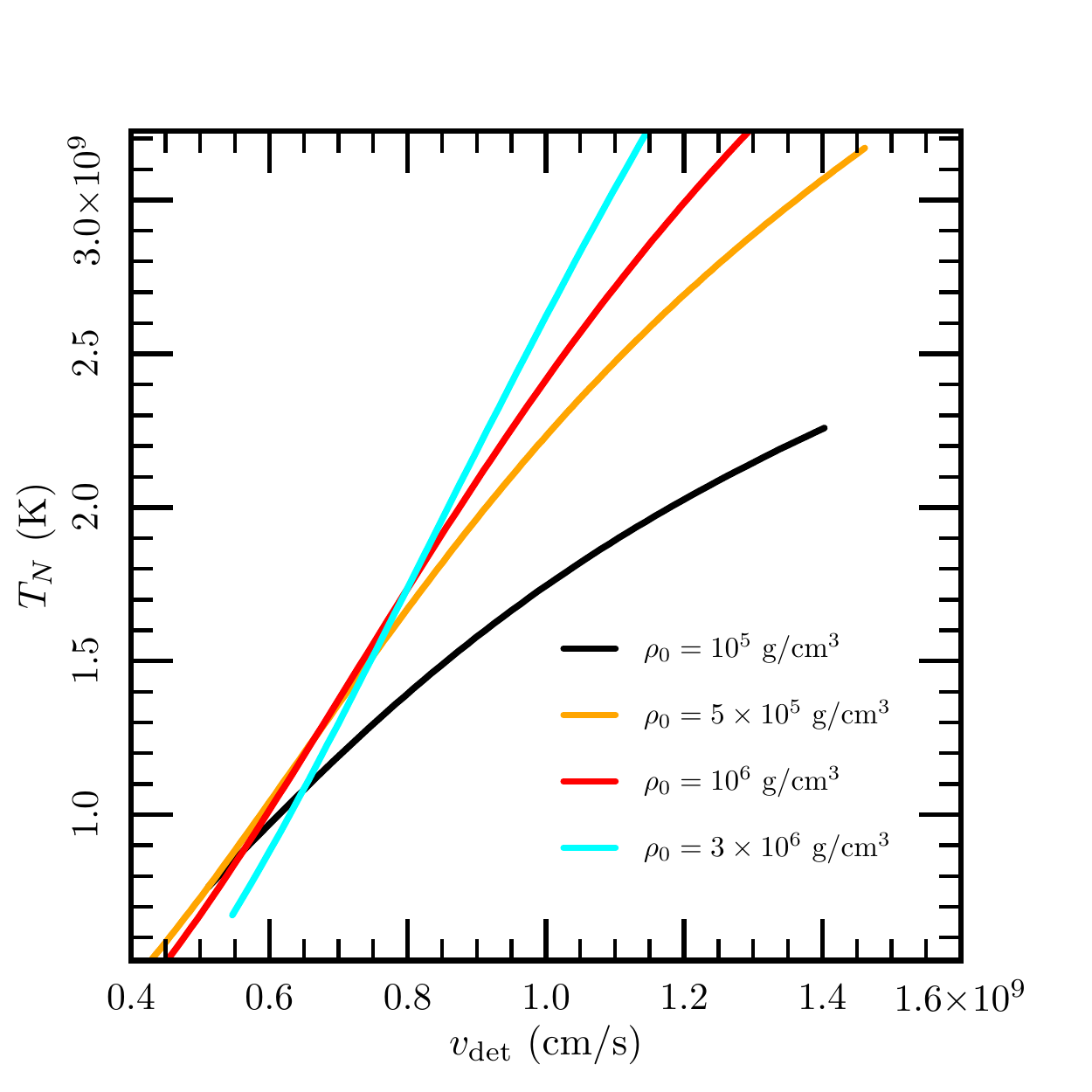}
	\caption{\label{fig:vt_plot} Post shock temperature ($T_N$) as a function of detonation velocity for a variety of initial densities. $T_N$ only depends on the shock jump conditions, so we label the $x$-axis as $v_{\rm det}$ since these jump conditions do not always correspond to generalized CJ detonations.}
\end{figure}

\begin{figure}
	\plotone{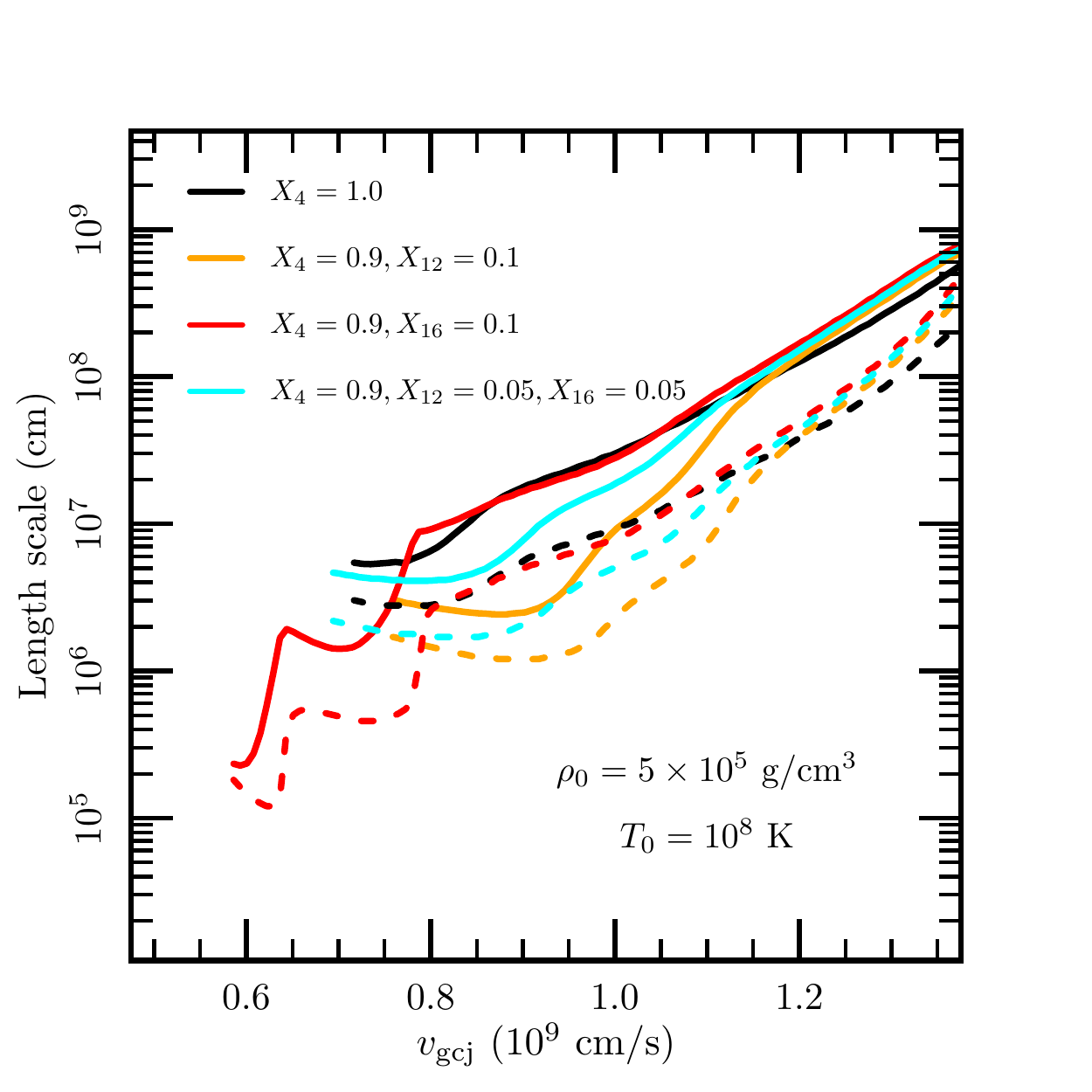}
	\caption{\label{fig:lznd_plot_he100-90} Various length scales for generalized CJ detonations in layers having $\rho_0 = 5\times 10^5$ g/cm$^3$ and various initial compositions down to $90\%$ helium. Solid lines are $l_{95}$, the length scale over which $95\%$ of the nuclear burning energy is released. Dashed lines correspond to the distance behind the shock front to the pathological point.}
\end{figure}

\begin{figure}
	\plotone{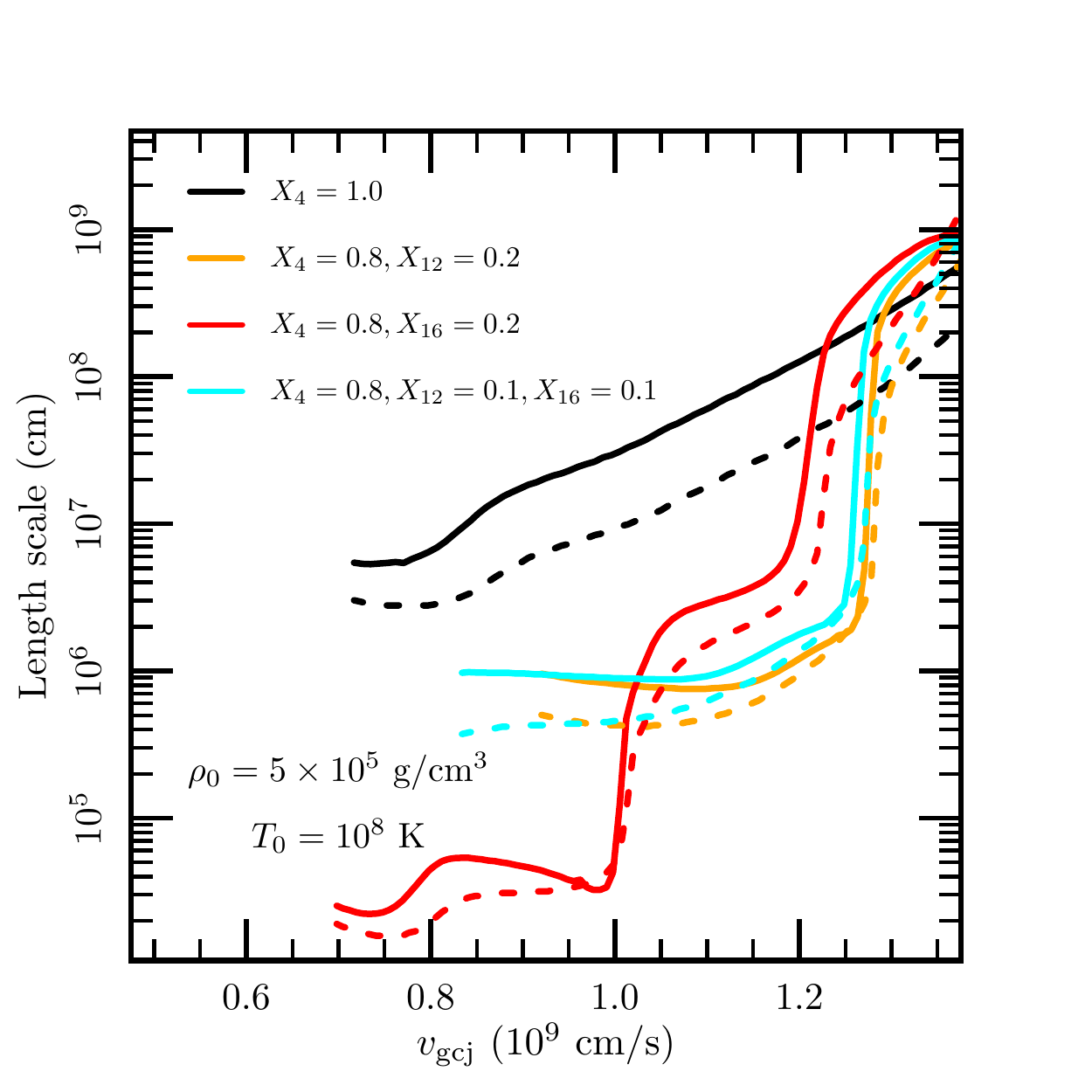}
	\caption{\label{fig:lznd_plot_he100-80} Same as Figure \ref{fig:lznd_plot_he100-90}, but with compositions down to $80\%$ helium by mass.}
\end{figure}

\begin{figure}
	\plotone{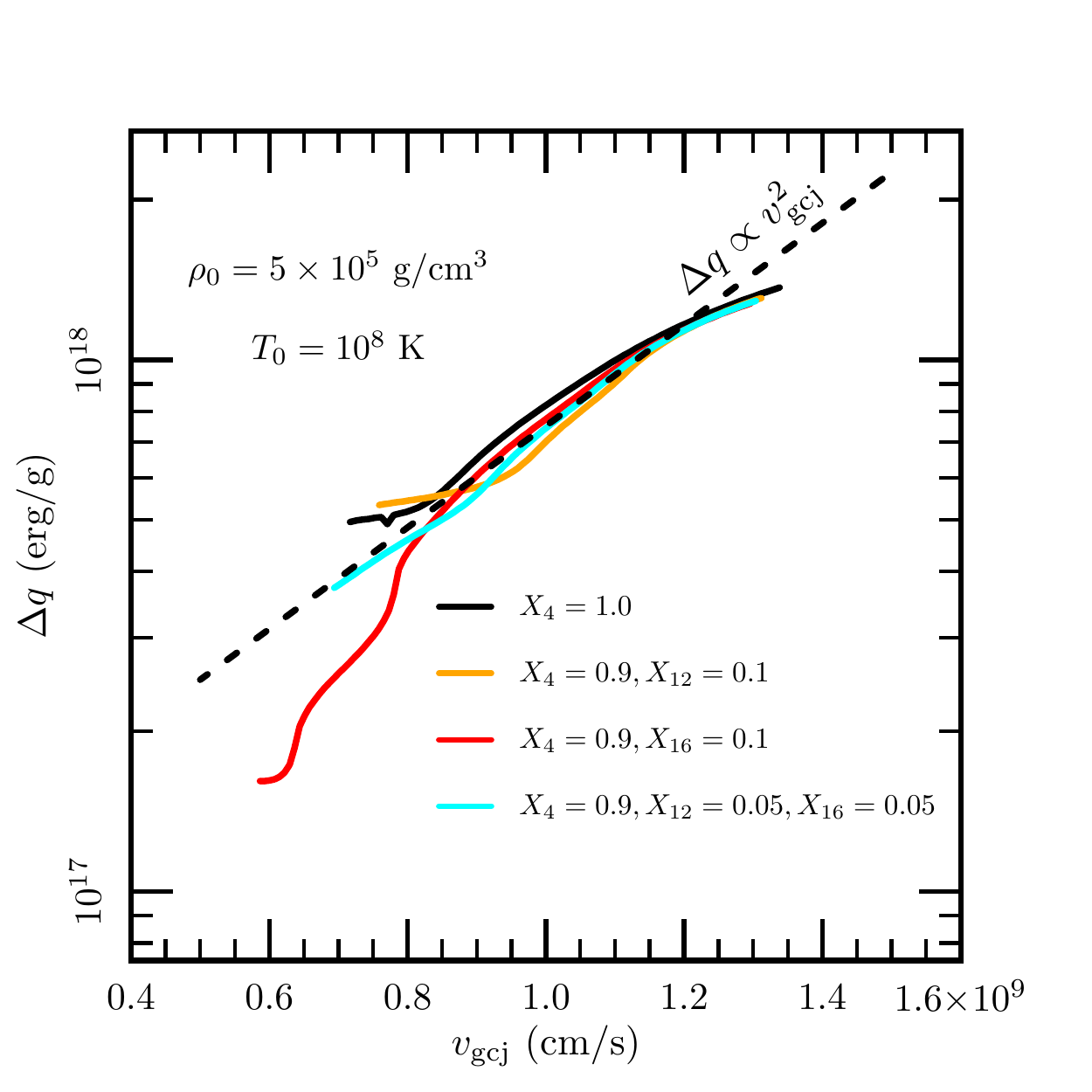}
	\caption{\label{fig:vq_he100-90} Detonation energy release $\Delta q$ as a function of generalized CJ velocity, $v_{\rm gcj}$, for detonations in layers having $\rho_0 = 5\times 10^5$ g/cm$^3$, $T_0 = 10^8$ K, and various initial compositions down to $90\%$ helium by mass. }
\end{figure}

\begin{figure}
	\plotone{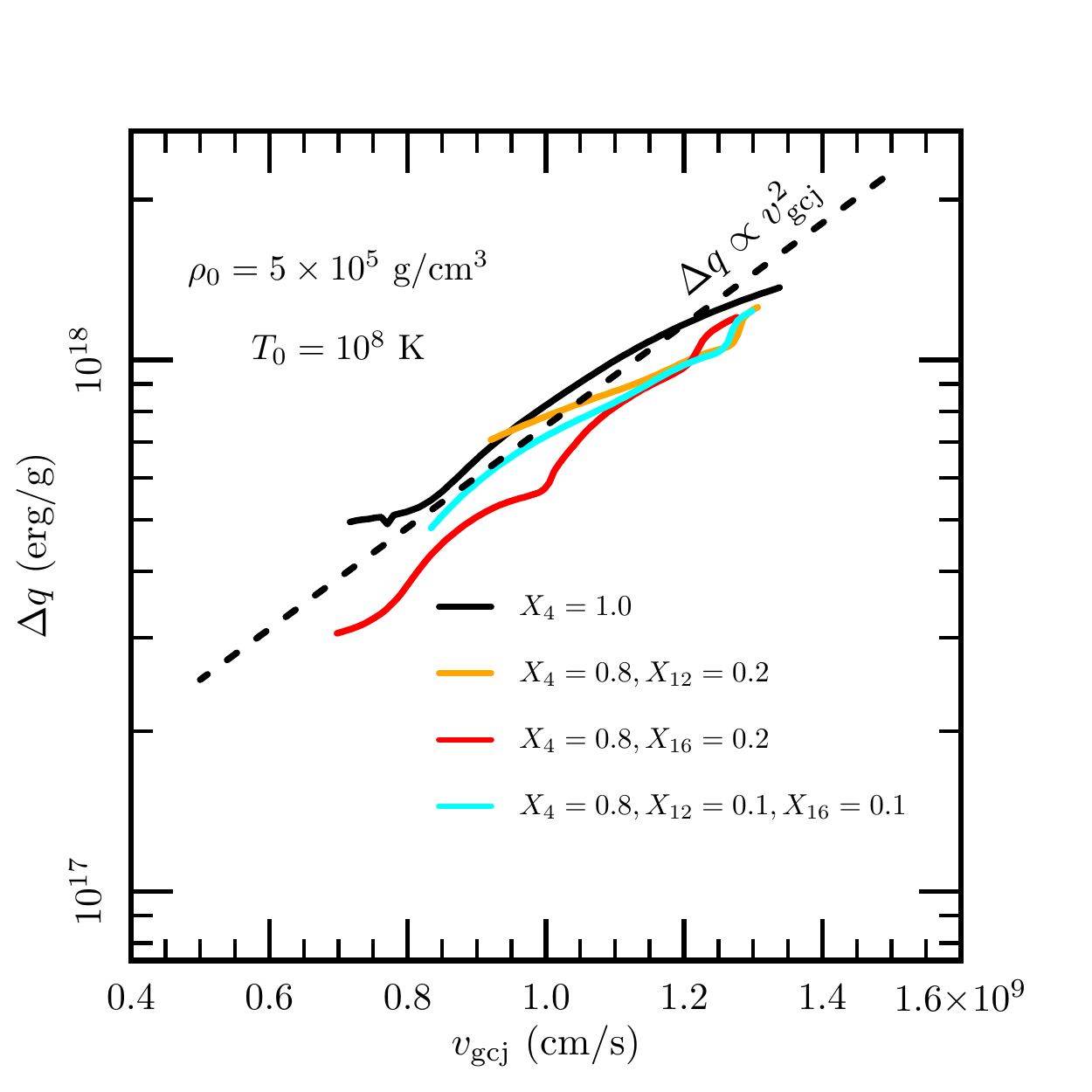}
	\caption{\label{fig:vq_he100-80} Same as Figure \ref{fig:vq_he100-90}, but with compositions down to $80\%$ helium by mass.}
\end{figure}

We have already seen in Figure \ref{fig:vcj_hs} that the range of allowable detonation velocities and layer thicknesses varies greatly with initial composition. The overall effect of adding $^{12}$C and $^{16}$O to the fuel can be seen in Figure \ref{fig:t_alpha}, which shows the lifetimes of several nuclei on the $\alpha$-chain as a function of temperature. While the triple-$\alpha$ rate is relatively flat over this temperature range, $\alpha$-captures on nuclei going up the $\alpha$-chain are quite sensitive to temperature. Typical post-shock temperatures for generalized CJ detonations with characteristic layer thicknesses are $1-3 \times 10^9$ K as shown in Figure \ref{fig:vt_plot}. For these temperatures, $\alpha$-captures on $^{16}$O up to $^{28}$Si are much faster than both the triple-$\alpha$ process and $\alpha$-capture onto $^{12}$C. We therefore expect any $^{16}$O in the initial fuel to burn to $^{28}$Si very quickly. Rapid $\alpha$ captures onto $^{16}$O can dramatically decrease the burning length scale, $l_{95}$, and therefore reduce the minimum layer thicknesses that allows for steady detonations as well as the speeds of such detonations. Figures \ref{fig:lznd_plot_he100-90} and \ref{fig:lznd_plot_he100-80} show the reaction length scale, $l_{95}$, and distance to the pathological point. Our intuition from Equation (\ref{eq:vq}) suggests that the small energy release allows for extremely low detonation velocities. We check this scaling in Figures \ref{fig:vq_he100-90} and \ref{fig:vq_he100-80}, where we plot the net energy release, $\Delta q$, against the generalized CJ velocity for various compositions. We find that $\Delta q$ scales as expected, albeit with some deviation at low velocities for cases with $10\%$ $^{16}$O in the fuel. 

From Figures \ref{fig:vq_he100-90} and \ref{fig:vq_he100-80}, we see that while a $^{16}$O mass fraction of $X_{16} = 0.05$ does not produce a large reduction in the minimum $v_{\rm gcj}$, a mass fraction of $X_{16} = 0.1$ does. We can estimate the minimum mass fraction of $^{16}$O required to dramatically change the detonation structure as follows. The post-shock temperature range where $^{16}$O can burn to $^{28}$Si before any other reactions take place is (from Figure \ref{fig:t_alpha}) $T_N \gtrsim 0.9 \times 10^9$ K. This temperature corresponds to a shock strength given by a detonation with $v_{\rm det} \approx 0.6\times 10^9$ cm/s, from Figure \ref{fig:vt_plot}. We can estimate the energy release from a detonation at that velocity using Equation (\ref{eq:vq}). We use $\gamma = 5/3$ since this is a relatively weak shock, and the post-shock conditions will not be radiation-dominated ($P_{\rm rad}/P_{\rm gas} \approx 10^{-2}$). This yields a minimum $\Delta q \approx 10^{17}$ erg/g for $^{16}$O to $^{28}$Si burning to dominate the early-time energy release. The energy released from only burning $^{16}$O to $^{28}$Si in terms of the initial mass fraction of $^{16}$O is 
\begin{align}
\Delta q \left(^{16}{\rm O} \rightarrow ^{28}{\rm Si}\right) = &\left[\left(1-\frac{7}{4}X_{16}\right) \frac{Q_4}{A_4} + \frac{7}{4}X_{16}\frac{Q_{28}}{A_{28}} \right]  - \nonumber \\
&\left[\left(1-X_{16}\right) \frac{Q_4}{A_4} + X_{16}\frac{Q_{16}}{A_{16}}\right].
\end{align}
We can find the mass fraction $X_{16}$ that allows for a detonation strong enough that the post-shock temperature allows for burning of $^{16}$O to $^{28}$Si before anything else, giving
\begin{equation}
X_{16} = \frac{\Delta q}{\frac{7}{4}Q_{28} - Q_{16} - \frac{3}{4}Q_{28}} \approx 0.07,
\end{equation}
in good agreement with the behavior observed at lowest values of $v_{\rm gcj}$. Based on the $\alpha$-capture timescales in Figure \ref{fig:t_alpha}, we expect to see similar effects if $^{20}$Ne or $^{24}$Mg are present in the fuel, as they are nearly interchangeable with $^{16}$O in terms of burning speed for the low temperatures relevant here, but release less energy when burning to $^{28}$Si.

\subsection{Larger reaction networks}
An advantage of our 1D model is that it allows us to use much larger reaction networks than those feasible in {\ttfamily FLASH}. Recall {\ttfamily FLASH} is using a 13-isotope alpha chain (aprox13), and for our comparisons to {\ttfamily FLASH} we use a 19-isotope alpha chain with the addition of $^{1}$H, $^{3}$He, $^{14}$N, $^{54}$Fe, along with special neutrons and protons for photo-disintegration. Both helium white dwarf donors and non-degenerate helium burning star donors are expected to have significant enrichments (up to few percent) of $^{14}$N due to CNO burning earlier in the star's life, which can be important in explosive helium burning through production of $^{14}$C \citep{Hashimoto86}. In this section we examine the predictions made with a larger $136$-isotope reaction network, as well as the effects of adding initial $^{14}$C to the helium. As discussed in \citet{Shen09}, the neutron/proton ratio at the onset of dynamical burning in a helium envelope is fixed at the beginning of the convective burning phases since there is not enough time for electron captures to happen before the burning becomes dynamical. We therefore expect any $^{14}$N or $^{14}$C excess from CNO burning to be relevant to a detonation if one develops.

\begin{figure}
	\plotone{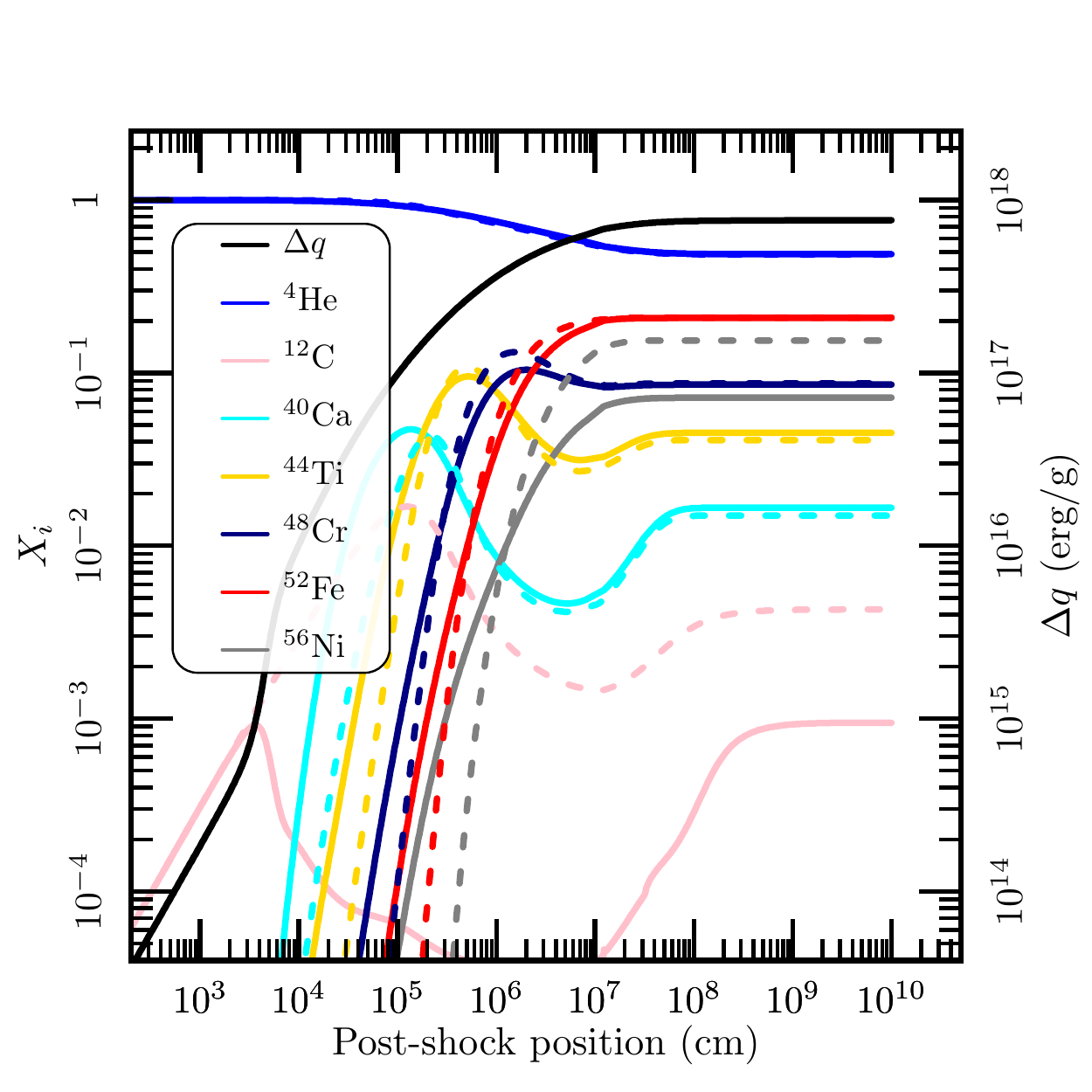}
	\caption{\label{fig:abundances_he100_136iso} Solid lines correspond to the post-shock nucleosynthesis and energy release computed with a 136-isotope network for a generalized CJ detonation in pure He and ambient conditions $\rho_0 = 5\times 10^5$ g/cm$^3$, thickness $H=2.5\times 10^7$ cm, yielding a detonation velocity $v_{\rm gcj} = 0.97\times 10^9$ cm/s. Dashed lines represent a detonation with the same initial conditions, but in pure Helium using the 19-isotope network as in Figure \ref{fig:1d_to_flash_nucleosynthesis}. The decrease in $^{56}$Ni production in the 136-isotope case is balanced by the production of isotopes such as $^{55}$Co (unstable, $t_{1/2} = 18$h) and $^{57}$Ni (unstable, $t_{1/2} = 36$h).}
\end{figure}

We first compare the nucleosynthesis of the $136$-isotope network with that of the $19$-isotope network for pure helium detonations with identical initial conditions ($\rho_0 = 5\times 10^5$ g/cm$^3$, $H=2.5\times 10^7$ cm) in Figure \ref{fig:abundances_he100_136iso}. In general, different reaction networks will give different generalized CJ velocities because the energy release rate plays a critical role in determining the detonation structure. However, both of these networks give the same detonation velocity, $v_{\rm gcj} = 0.97\times 10^9$ cm/s. The nucleosynthetic profile is slightly different between the two reaction networks, with the main difference being less $^{56}$Ni production in the $136$-isotope network (offset by production of isotopes such as $^{55}$Co and $^{57}$Ni). Both reaction networks finding the same $v_{\rm gcj}$ value means they agree very closely in total energy release, so we can be confident in the detonation velocity predictions using the $19$-isotope network.

\begin{figure}
	\plotone{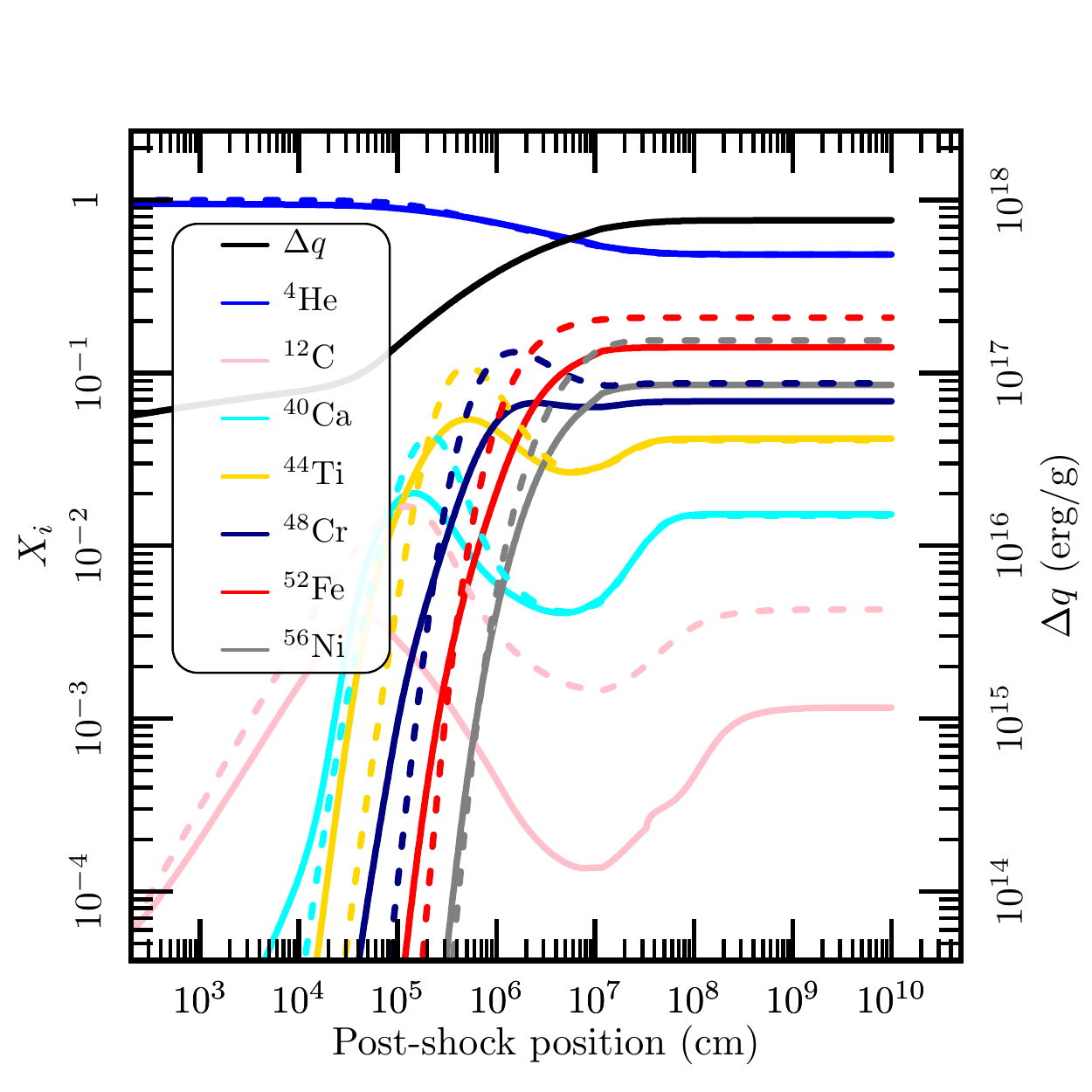}
	\caption{\label{fig:abundances_he98_c1402_136iso} Solid lines correspond to the post-shock nucleosynthesis and energy release computed with a 136-isotope network for a generalized CJ detonation with a fuel composition of $2\%$ $^{14}$C by mass and $98\%$ $^4$He, and ambient conditions $\rho_0 = 5\times 10^5$ g/cm$^3$, thickness $H=2.5\times 10^7$ cm, yielding a detonation velocity $v_{\rm gcj} = 0.97\times 10^9$ cm/s. Dashed lines represent a detonation with the same initial conditions, but in pure Helium using the 19-isotope network as in Figure \ref{fig:1d_to_flash_nucleosynthesis}. The case with the slight neutron excess leads to less production of $\alpha$-chain isotopes, but more neutron-rich isotopes such as $^{53}$Fe (unstable, $t_{1/2} = 8$m), $^{57}$Ni (unstable, $t_{1/2} = 36$h), and $^{58}$Ni (stable).}
\end{figure}

We 
%defer a more detailed study of the effects of initial metallicity to future work, as it requires analysis of an updated set of progenitor models, but 
now briefly examine the effects of adding neutron-rich isotopes to the fuel. As discussed in \citet{Timmes03}, a small amount of neutron-rich material in the fuel can have a noticeable impact on the final yields of radioactive isotopes such as $^{56}$Ni. Figure \ref{fig:abundances_he98_c1402_136iso} shows the nucleosynthesis of a detonation with the same starting conditions as Figures \ref{fig:1d_to_flash_nucleosynthesis} and \ref{fig:abundances_he100_136iso}, but with $2\%$ $^{14}$C by mass added to the fuel. The net effect is production of neutron-rich isotopes such as $^{53}$Fe, $^{57}$Ni, and $^{58}$Ni. We observe qualitatively similar effects when adding other neutron-rich isotopes such as $^{18}$O or $^{22}$Ne to the fuel. 

\subsection{Detonation propagation limits}
\label{subsec:prob_limits_theory}
\begin{figure}
	\plotone{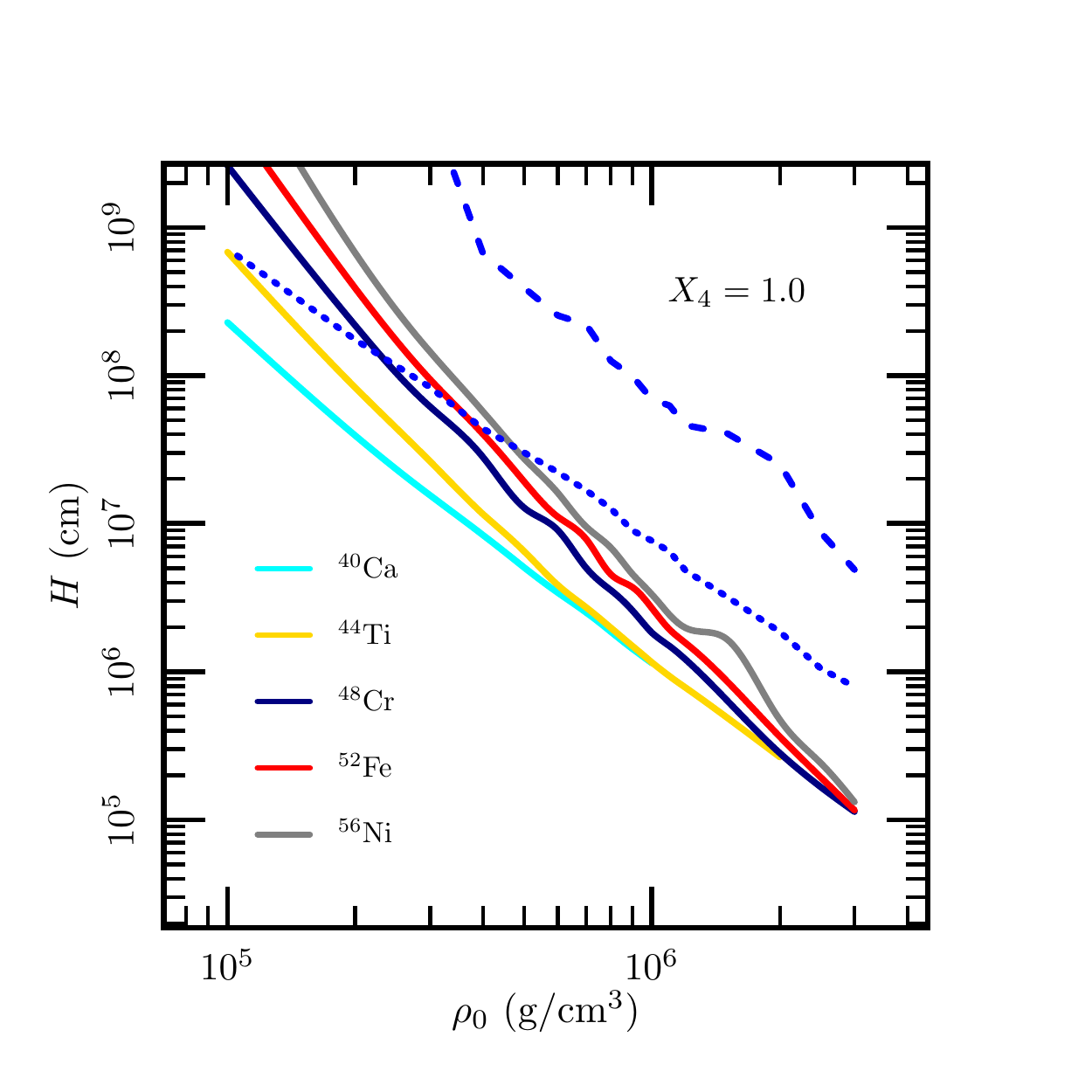}
	\caption{\label{fig:final_abundance_regions_he100} Regions where laterally propagating detonations in pure helium are possible as a function of initial layer thickness $H$ and density $\rho_0$. Regions above the lowest colored line can support generalized CJ detonations, with each point corresponding to a detonation in a layer with different initial conditions. Colors indicate where, as $H$ increases at fixed $\rho_0$, each nuclide first  becomes the dominant burning product (see Figure \ref{fig:final_abundances_r5d5_he100_vdet}). Significant amounts of initial helium remain so the dotted blue line shows where the mass fraction of unburned helium is $50\%$ and the dashed blue line shows where the mass fraction of unburned helium is $25\%$. Steady detonations cannot propagate in the region below the lowest isotope lines.}
\end{figure}

\begin{figure}
	\plotone{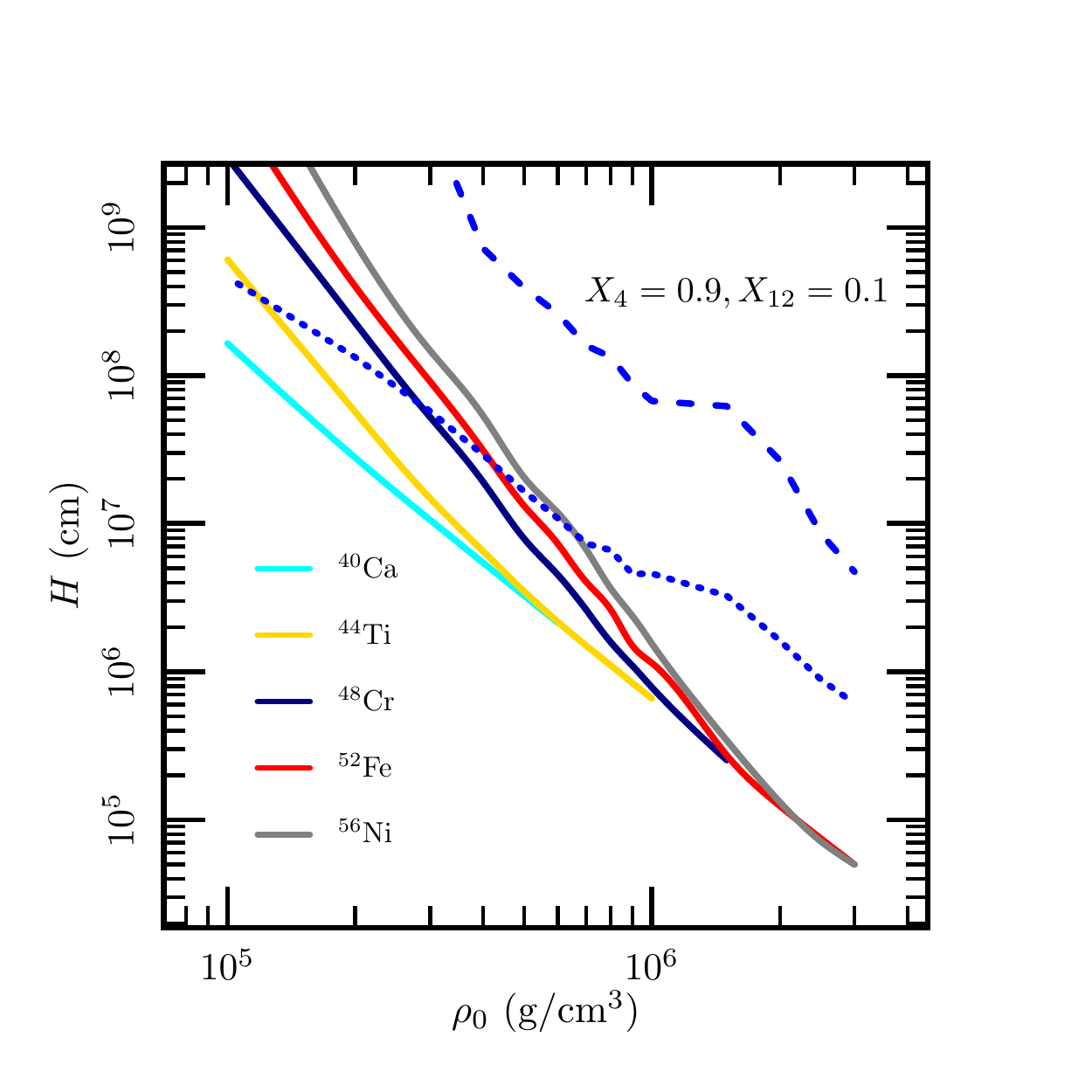}
	\caption{\label{fig:final_abundance_regions_he90_c10} Same as Figure \ref{fig:final_abundance_regions_he100}, but for detonations in material composed of $90\%$ $^4$He and $10\%$ $^{12}$C by mass.}
\end{figure}

\begin{figure}
	\plotone{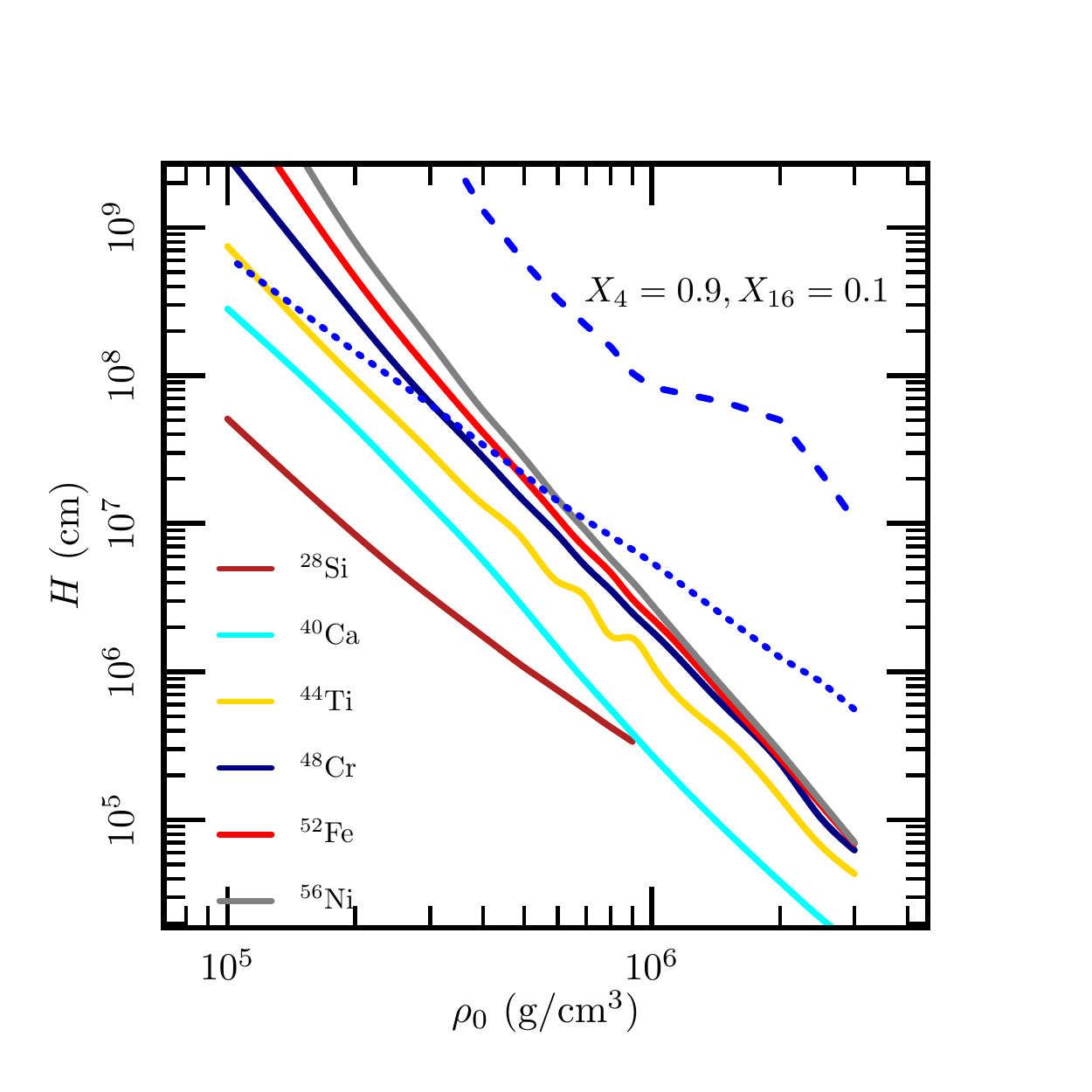}
	\caption{\label{fig:final_abundance_regions_he90_o10} Same as Figure \ref{fig:final_abundance_regions_he100}, but for detonations in material composed of $90\%$ $^4$He and $10\%$ $^{16}$O by mass. For this initial composition, the slowest detonations produce more $^{28}$Si than other products for initial densities below $10^6$ g/cm$^3$.}
\end{figure}

We showed in Figure \ref{fig:maxmach} that there is a lower limit on the layer thickness that can support a steady detonation for a given $\rho_0$ and ${\bf X}_0$. We therefore summarize the regions of the $(\rho_0, H)$ plane where laterally propagating detonations are allowed for a given composition. Figures \ref{fig:final_abundance_regions_he100}, \ref{fig:final_abundance_regions_he90_c10}, and \ref{fig:final_abundance_regions_he90_o10} show the regions of parameter space where detonations are allowed for initial compositions of $(X_4 = 1.0)$, $(X_4 = 0.9,\ X_{12} = 0.1)$, and $(X_4 = 0.9,\ X_{16} = 0.1)$, respectively. Colored lines denote where each isotope becomes the dominant burning product.
%Colored lines indicate the dominant burning product (as shown in Figure \ref{fig:final_abundances_r5d5_he100_vdet}) for a generalized CJ detonation at that initial density $\rho_0$ and layer thickness $H$. 
%The shaded regions show where laterally propagating detonations can exist, with the colors indicating the dominant burning product. 
Recall from Figure \ref{fig:final_abundances_r5d5_he100_vdet} that the burning products are typically $10-20\%$ by mass until the detonation is strong enough to produce significant amounts of $^{56}$Ni, so the majority of the material remains unburned helium. Dashed and dotted lines indicate the mass fraction of helium remaining in the ashes. The region below the lowest isotope line does not allow detonation propagation because expansion happens too quickly for a post-shock sonic locus to form. The effects of adding small amounts of $^{12}$C or $^{16}$O to the fuel are to decrease the minimum thickness $H$ that allows for detonation propagation at any given $\rho_0$. In the next section we will show how to translate detonations in $(\rho_0, H)$ space into finite-gravity helium layers on WDs.

\section{Detonations in finite gravity environments}
\label{sec:finite_g}
The results and comparisons in the previous section were presented in terms of the 1D model parameters - $H$ and $\rho_0$. We now map these variables to a set of WD core and envelope masses, connecting to the astrophysical scenarios. In this section, we describe a way to map our constant density analytics onto cases with finite gravity, and hence vertical density gradients. We again compare to 2D {\ttfamily FLASH} models in finite gravity and then present detonation limits in the form of a minimum envelope mass as a function of WD core mass and envelope composition.

\subsection{{\ttfamily FLASH} simulations}
The assumptions we make when mapping 1D constant density detonations to detonations in finite gravity arise from analyzing {\ttfamily FLASH} simulations of detonations in plane-parallel geometry with gravity. Such {\ttfamily FLASH} simulations show that the leading part of the detonation front lies above the base of the layer (see also Figure \ref{fig:surface_2grav}). Our strategy is to interpret the leading point of the detonation as the leading point of the constant-density detonation - the centerline in Figure \ref{fig:strip_diagram}. We therefore use the layer density and shock curvature at this height as the initial conditions for our 1D model. Empirically, the forwardmost point on the shock front in {\ttfamily FLASH} occurs at a height roughly $H_s/2$ above the base of the layer, where $H_s = P_b/(\rho_b g_b)$ is the scale height evaluated at the base conditions. We assume a polytrope model for the atmosphere,
\begin{equation}
\frac{P}{P_b} = \left(\frac{\rho}{\rho_b}\right)^\gamma,
\end{equation}
where $P_b$ and $\rho_b$ are the pressure and density at the base of the layer, respectively. Combining this with hydrostatic balance, $dP/dz = -\rho g$ ($z$ being the coordinate in the vertical direction), yields
\begin{equation}
\frac{\rho}{\rho_b} = \left[1 - \left(\frac{\gamma - 1}{\gamma}\right) \frac{z}{H_s}\right]^{1/(\gamma-1)}.
\end{equation}
For a convective or degenerate (non-relativistic) atmosphere, $\gamma = 5/3$, and we find $\rho(H_s/2) \approx 0.7 \rho_b$. The curvature of the shock front in the {\ttfamily FLASH} simulations relates to the layer thickness via $R_c \approx 3H \approx 3H_s/2$.

We map our calculation parameters $(\rho_0, H)$ to WD core and envelope masses ($M_c$ and $M_{\rm env}$, respectively) as follows. We construct a WD core by integrating the stellar structure equations using the MESA EOS and assuming a fixed temperature ($T_c = 10^7$ K) and composition for the core. The core composition is taken to be equal parts $^{12}$C and $^{16}$O by mass.
%for $0.6\ M_\odot \le M_c \le 1.1\ M_\odot$ or equal parts  $^{16}$O,  $^{20}$Ne, and  $^{24}$Mg by mass (?) for $1.1\ M_\odot < M_c < 1.4\ M_\odot$. 
Since we assume a constant temperature and composition, we only need the equations of hydrostatic balance and mass conservation along with an equation of state.
%\begin{align}
%\der{r}{P} &= -\frac{1}{\rho g}, \\
%\der{M}{P} &= -\frac{4\pi r^2}{g}.
%\end{align}
We use $P$ as our independent variable since the integration bounds are more easily expressed in terms of pressure than radius. The initial conditions are $r(P_c) = 0$ and $M(P_c) = 0$, where $P_c$ is the central pressure, a parameter that fixes the structure of the WD core. If we just want to characterize a WD core, we can take an outer boundary at a very low pressure (eg. $P/P_c = 10^{-6}$) to define the surface of the star. In order to add an envelope, we integrate to a higher pressure ratio $P_b/P_c > 10^{-6}$ corresponding to the base of the envelope, and then switch to an isentropic atmosphere model where the temperature varies as
\begin{equation}
T = T_b \left(\frac{\rho}{\rho_b}\right)^{\gamma-1},
\end{equation}
where $T_b$ is the temperature at the base of the convective envelope. Although the pressure is continuous, there is a small jump in density across the core-envelope boundary due to the composition and temperature change. We then integrate the envelope structure out to the outer pressure boundary (eg. $P/P_c = 10^{-6}$) to calculate the envelope mass. Each WD model is therefore determined by two parameters - the central pressure $P_c$, and the pressure ratio at the core-envelope boundary $P_b/P_c$. We can then extract the density $\rho_0$ and thickness $H = P_b/(2\rho_b g_b)$ to use in the generalized ZND integrations. 
%We finally extend our previous assumption that the radius of curvature and envelope thickness are still related via $R_c = 3H = 3H_s/2$, as in the constant density slab case. 
Along with our assumption that $R_c = 3H$, This allows us to map our previous points in $(H, \rho_0)$ space into $(M_c, M_{\rm env})$ space and vice-versa.

We end this section by noting that the $T_b$ varies depending on the progenitor scenario, with both larger envelope and core masses producing higher $T_b$ values before dynamical runaway \citep{Shen09}. We take $T_b = 10^8$ K as our fiducial for constructing hydrostatic WD envelopes, with higher temperatures corresponding to lower values of $\rho_b$ given the same $M_c$ and $M_{\rm env}$. Such envelopes that are hot and massive relative to those in the AM CVn accretion scenario may be relevant in the low accretion rates of a WD + He burning star donor \citep{Yungelson08} or unstable mass transfer during a WD + He WD merger \citep{Guillochon10, Schwab12, Shen12, Pakmor13}. These hot envelopes are different enough from our hydrostatic WD models that they are better placed in $(H, \rho_0)$ space, as in Figures \ref{fig:final_abundance_regions_he100} - \ref{fig:final_abundance_regions_he90_o10}.

%We end this section by noting that for many of the core + envelope models we consider, the envelope is not in the thin-shell limit so a stellar model using a realistic EOS as described above is necessary. Figure () shows the relative differences between the thin-shell approximation of $P_b$,
%\begin{equation}
%P_{b, {\rm thin}} = \frac{4\pi R_{\rm WD}^2 M_{\rm env}}{g},
%\end{equation}
%and the value of $P_b$ from our WD structure integrations in $(M_c, M_{\rm env})$ space, showing a significant difference between the two calculation methods.

%We first assume that the envelope is geometrically thin so we can relate the envelope mass, $M_{\rm env}$, to $P_b$ via
%\begin{equation}
%M_{\rm env} = \frac{4\pi R_{\rm WD}^2 P_b}{g},
%\end{equation}
%where $R_{\rm WD}$ is the WD radius. Lastly, we use a white mass-radius relationship computed from MESA models with core temperatures of $3\times 10^7$ K (B. Wolf, private communication),
%\begin{equation}
%R_{\rm WD} = 7.6 \times 10^{-3} R_\odot \left(\frac{M_{\rm WD}}{M_\odot}\right)^{-1.62}.
%\end{equation}
%We finally extend our previous assumption that the radius of curvature and envelope thickness are still related via $R_c = 3H = 3H_P/2$, as in the constant density slab case. This allows us to map our previous points in $(H, \rho_0)$ space into $(M_{\rm WD}, M_{\rm env})$ space.

\subsection{Detonation propagation limits on WDs}
\begin{figure}
	\plotone{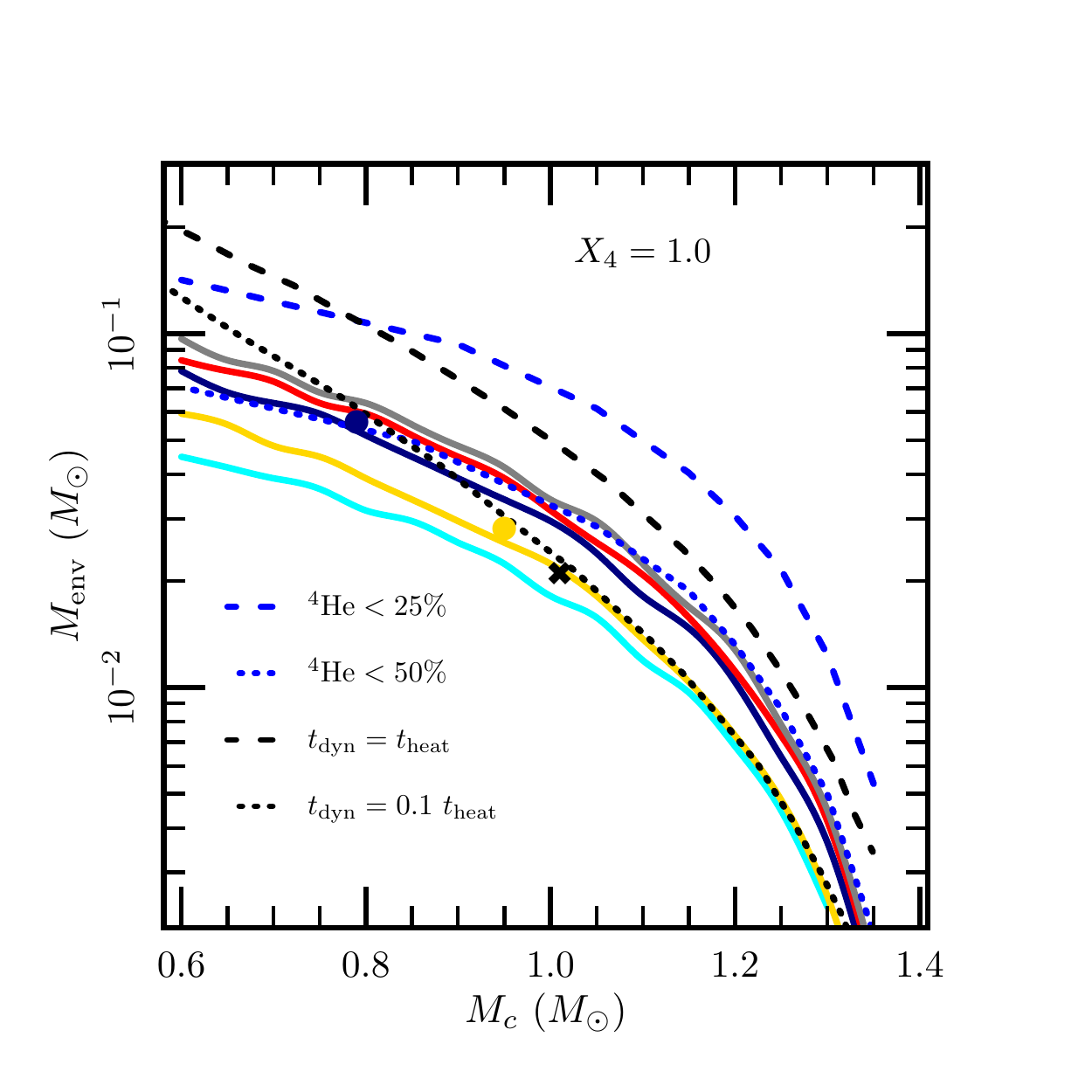}
	\caption{\label{fig:final_abundance_regions_wd_he100} Regions where laterally propagating detonations in pure helium are allowed as a function of core and envelope mass. Solid lines indicate transitions in dominant burning products as in Figures \ref{fig:final_abundance_regions_he100} - \ref{fig:final_abundance_regions_he90_o10} in the order (bottom to top): $^{40}$Ca, $^{44}$Ti, $^{48}$Cr, $^{52}$Fe, and $^{56}$Ni. The dotted blue line shows where the mass fraction of unburned helium is $50\%$ and the dashed blue line shows where the mass fraction of unburned helium is $25\%$. Regions below the $^{40}$Ca line do not allow detonation propagation because expansion happens too quickly for a post-shock sonic locus to form. Black dashed and dotted lines show calculations guiding where dynamical shells are expected by comparing the local heating timescale $t_{\rm heat}$ to the dynamical timescale $t_{\rm dyn}$ from \citet{Shen09}. Dots indicate where we performed 2D {\ttfamily FLASH} runs with finite gravity and found propagating detonations, with their colors corresponding to the most dominant isotope produced along the centerline of the detonation, while X's indicates runs where no propagating detonations were found.}
\end{figure}

\begin{figure}
	\plotone{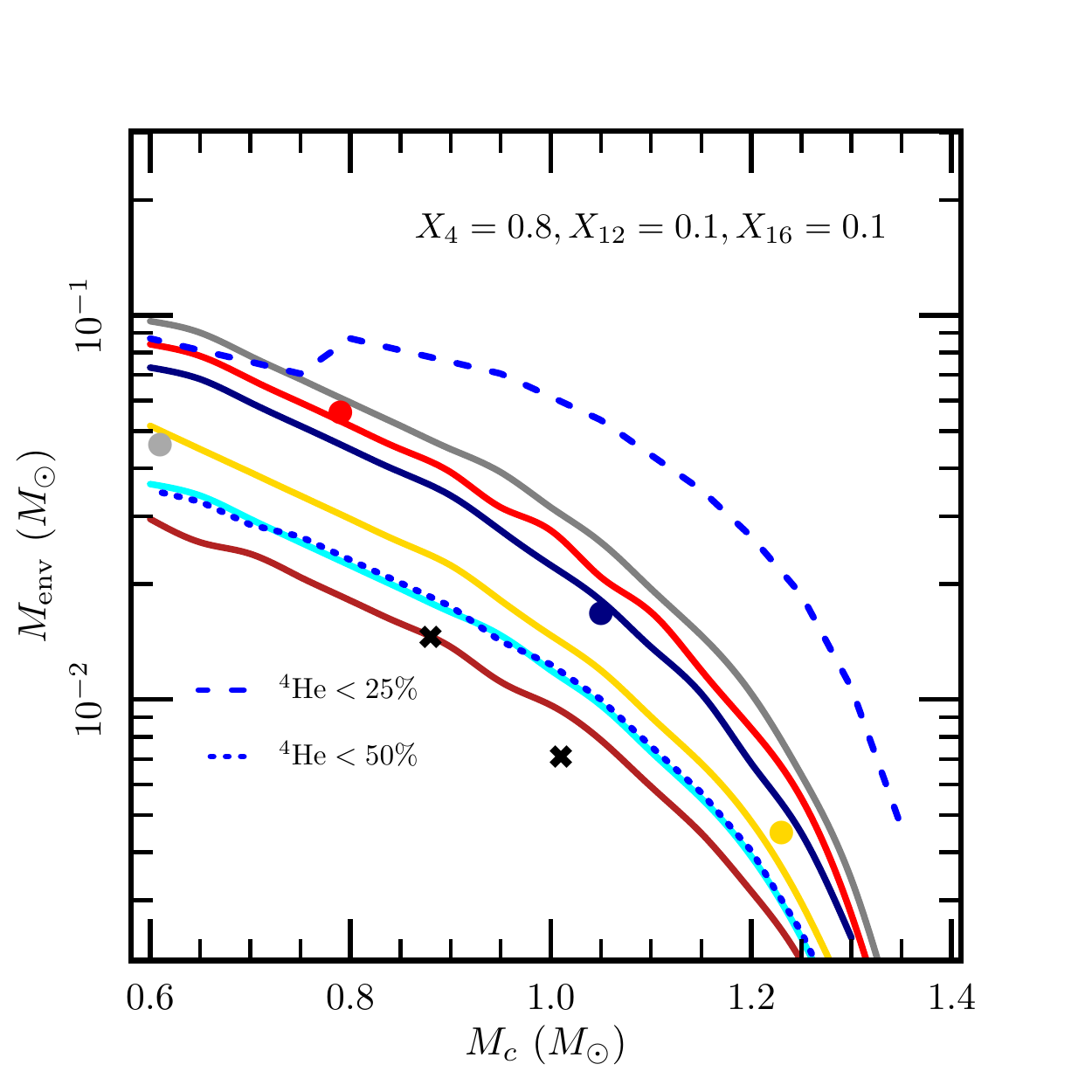}
	\caption{\label{fig:final_abundance_regions_wd_he80_c10_o10} Same as Figure \ref{fig:final_abundance_regions_wd_he100}, but with an envelope composed of $80\%$ $^4$He,  $10\%$ $^{12}$C, and $10\%$ $^{16}$O by mass. The lowest solid line here corresponds to $^{28}$Si, from the rapid $^{16}{\rm O} \rightarrow ^{28}$Si reactions. The dot representing the {\ttfamily FLASH} simulation at $(M_c = 0.61\ M_\odot, M_{\rm env} = 4.6\times 10^{-2}\ M_\odot)$ is colored for $^{32}$S production.}
\end{figure}

Figures \ref{fig:final_abundance_regions_wd_he100} and \ref{fig:final_abundance_regions_wd_he80_c10_o10} 
%and \ref{fig:final_abundance_regions_wd_he50_c25_o25} 
show the regions of allowed detonation propagation to WDs. These figures correspond to Figures \ref{fig:final_abundance_regions_he100} - \ref{fig:final_abundance_regions_he90_o10}, but in terms of WD parameters. We again compare to {\ttfamily FLASH} simulations, this time in finite gravity, with the colored dots corresponding to the most abundant nuclide produced by the forwardmost portion of the detonation in {\ttfamily FLASH}.

\begin{figure}
	\plotone{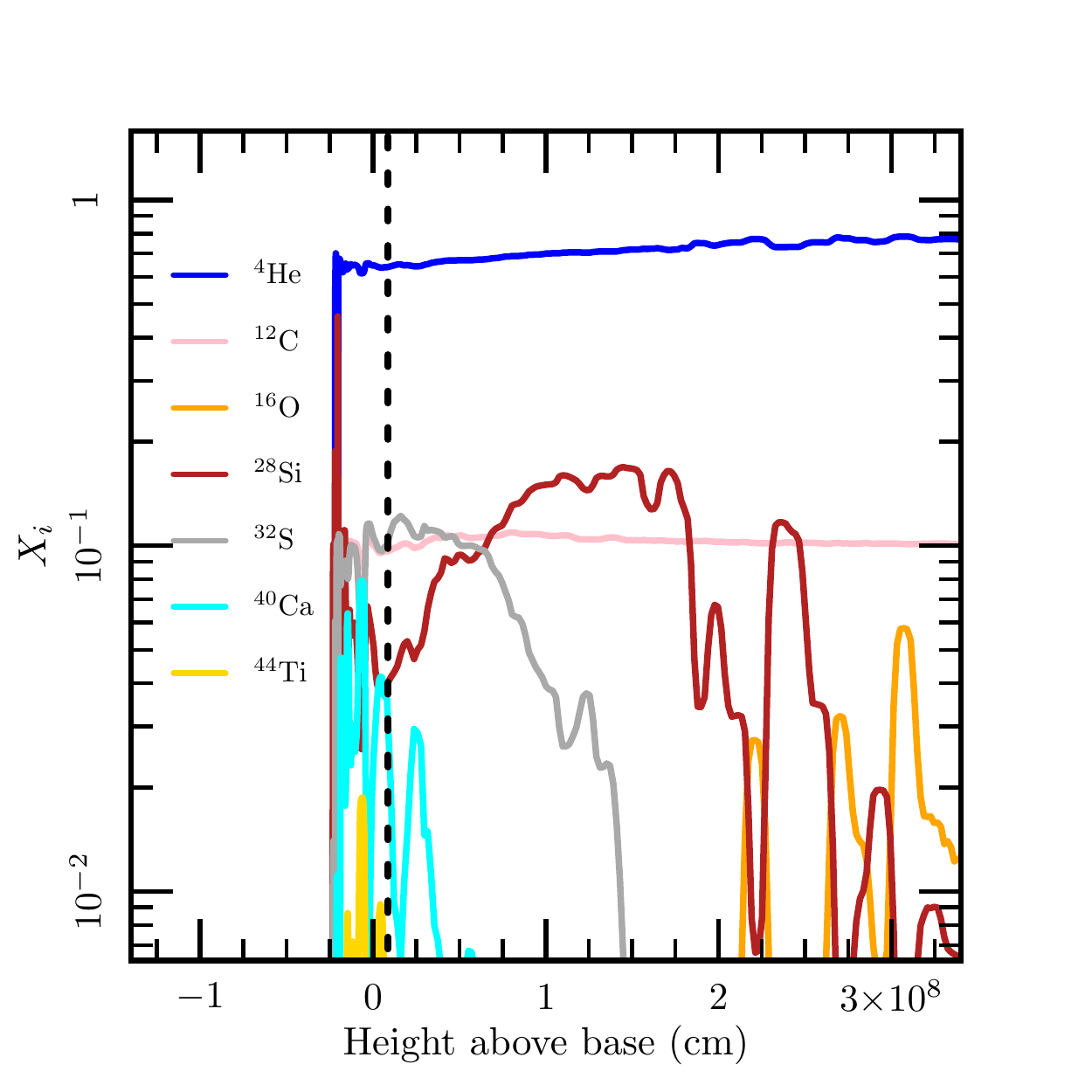}
	\caption{\label{fig:vertical_abundances} Vertical abundance profile from a {\ttfamily FLASH} simulation of a detonation with initial composition $X_4 = 0.8, X_{12} = 0.1, X_{16} = 0.1$, $\rho_b = 2\times 10^5$ g/cm$^3$, $T_b = 10^8$ K, $g_b = 2\times 10^8$ cm/s$^2$, and $H_s = 6.4\times 10^7$ cm. This led to a detonation with velocity $v_{\rm gcj} = 0.68\times 10^9$ cm/s, corresponding to the grey dot in Figure \ref{fig:final_abundance_regions_wd_he80_c10_o10}. This abundance profile is taken $10^8$ cm behind the detonation front, where burning has ceased. The zero point of the $x$-axis is taken at the initial position of the base of the helium layer, $1.5\times 10^8$ cm. Consistent with our assumptions, the most abundant isotopes lie near the centerline of the detonation - indicated by a dashed line a height $H = H_s/2$ above the current base of the layer. Elements above $^{44}$Ti do not appear on this plot because they are not produced in abundances greater than $10^{-2}$.}
\end{figure}

While detonations in constant density strips have some variation in nucleosynthesis in the direction perpendicular to the centerline (see Figure \ref{fig:strip_diagram}), detonations in finite gravity strips also have an initial density gradient in the vertical direction. This produces an even stronger variation in burning products as a function of height. When we refer to final abundances of such detonations, we mean abundances along the centerline of the detonation, a distance $H_s/2$ above the base of the layer. Detonation products as a function of height are shown in Figure \ref{fig:vertical_abundances}. As in the constant density strips, burning progresses furthest along the centerline, while material higher above burns even less completely. A series of peaks of lighter products as one moves up higher in the layer is characteristic of such detonations with finite gravity. Therefore, Figures \ref{fig:final_abundance_regions_wd_he100} and \ref{fig:final_abundance_regions_wd_he80_c10_o10} show limits on how far the burning can progress up the $\alpha$ chain. For example, a point that lies in the region where $^{48}$Cr is the dominant product corresponds to a detonation that produces isotopes up to $^{48}$Cr, with regions above $H_s/2$ burning less completely. We would thus expect to see a stratification of lighter elements outside of heavier elements in the ejecta corresponding to a laterally propagating helium detonation. We plan a more detailed characterization of ejected abundances with full-star detonation simulations in an upcoming paper.

%We consider compositions of pure helium (Fig. \ref{fig:final_abundance_regions_wd_he100}), and $90\%$ $^4$He + $10\%$ $^{12}$C (Fig. \ref{fig:final_abundance_regions_wd_he90_c10}) - motivated by pre-detonation convective burning as well as cases motivated by dredge up of C/O from the WD core -  $80\%$ $^4$He + $10\%$ $^{12}$C + $10\%$ $^{12}$O (Fig. \ref{fig:final_abundance_regions_wd_he80_c10_o10}) and $50\%$ $^4$He + $25\%$ $^{12}$C + $25\%$ $^{12}$O (Fig. \ref{fig:final_abundance_regions_wd_he50_c25_o25}).

%\begin{figure}
%	\plotone{figures/final_abundance_regions_wd_he90_c10.pdf}
%	\caption{\label{fig:final_abundance_regions_wd_he90_c10} Same as Figure \ref{fig:final_abundance_regions_wd_he100}, but with an envelope composed of $90\%$ $^4$He and $10\%$ $^{12}$C by mass.}
%\end{figure}

%\begin{figure}
%	\plotone{figures/final_abundance_regions_wd_he50_c25_o25.pdf}
%	\caption{\label{fig:final_abundance_regions_wd_he50_c25_o25} Regions where laterally propagating detonations in material composed of $50\%$ $^4$He, $25\%$ $^{12}$C, and $25\%$ $^{16}$O are allowed as a function of white dwarf mass $M_{\rm WD}$ and envelope mass $\Delta M$. Colors indicate the dominant burning product but are unshaded in this plot to enhance readability. Significant amounts of initial helium remain - the dotted blue line shows where the mass fraction of unburned helium is $50\%$ and the dashed blue line shows where the mass fraction of unburned helium is $25\%$. Regions below the $^{40}$Ca line do not allow detonation propagation because expansion happens too quickly for a post-shock sonic locus to form. Black lines show calculations guiding where dynamical shells are expected by comparing the local heating timescale $t_{\rm heat}$ to the dynamical timescale $t_{\rm dyn}$ from \citet{Shen09}.}
%\end{figure}

\subsection{Implications for explosion scenarios}
We note that nearly all of the parameter space shown is predicted to contain significant amounts ($> 25\%$) of unburned helium, and a range of radioactive elements. Detonations near the propagation cutoff line are predicted to produce very little mass in $^{56}$Ni, with most of the products being intermediate-mass elements (IMEs) such as $^{40}$Ca, $^{44}$Ti, etc. This motivates the possibility of extremely faint events that have very low yields of radioactive elements. More detailed investigations of helium detonation ignition in such low densities are necessary to determine whether detonations are likely to form in such environments.

\begin{figure}
	\plotone{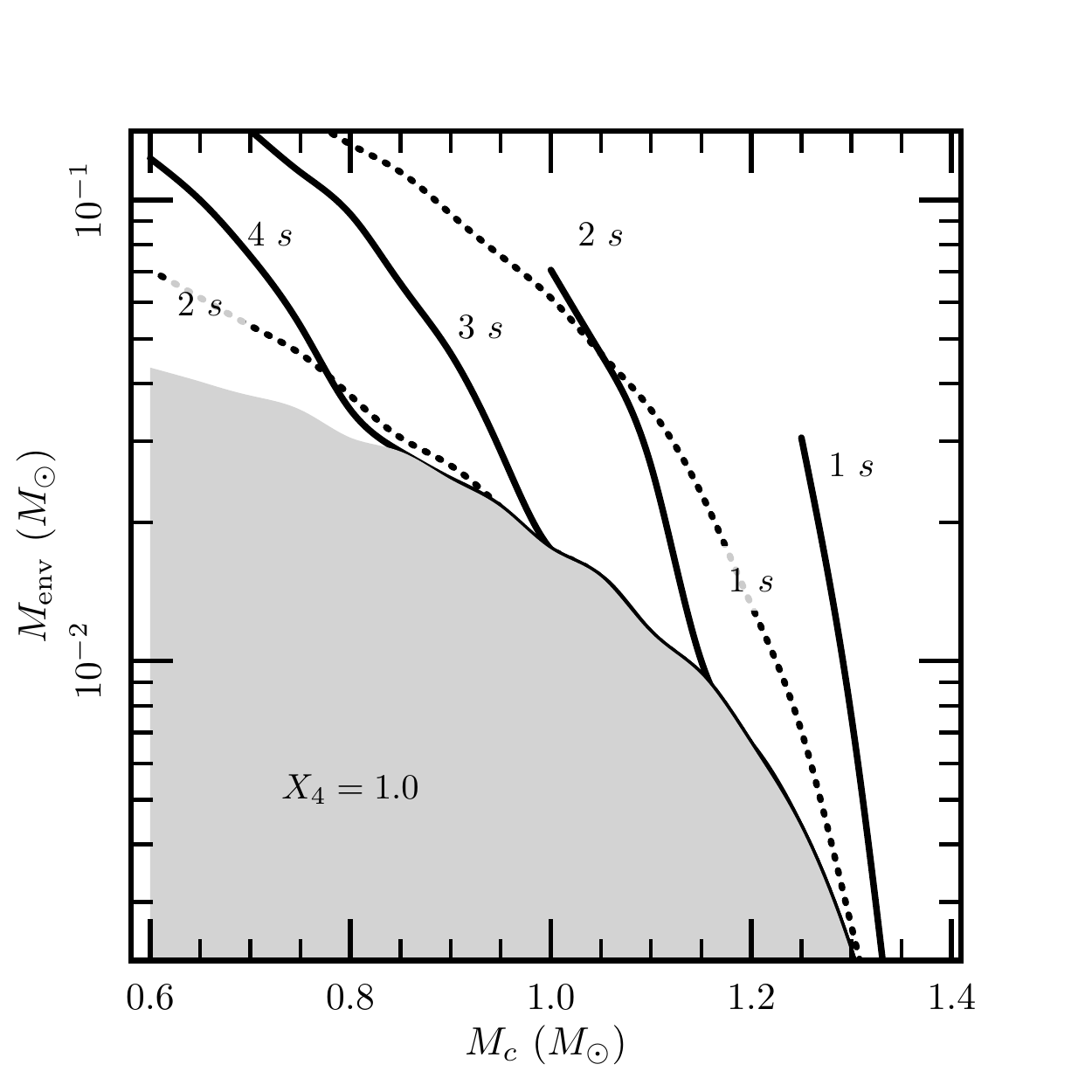}
	\caption{\label{fig:wd_timescales_he100} Lines of constant $t_{\rm sound}$ (solid lines) and $t_{\rm det}$ (dotted lines) for detonations in pure helium envelopes with $T_b = 10^8$ K on WDs in $M_c - M_{\rm env}$ space, corresponding to Figure \ref{fig:final_abundance_regions_wd_he100}. The shaded region indicates where the envelopes are too thin to support steady detonations.}
\end{figure}

\begin{figure}
	\plotone{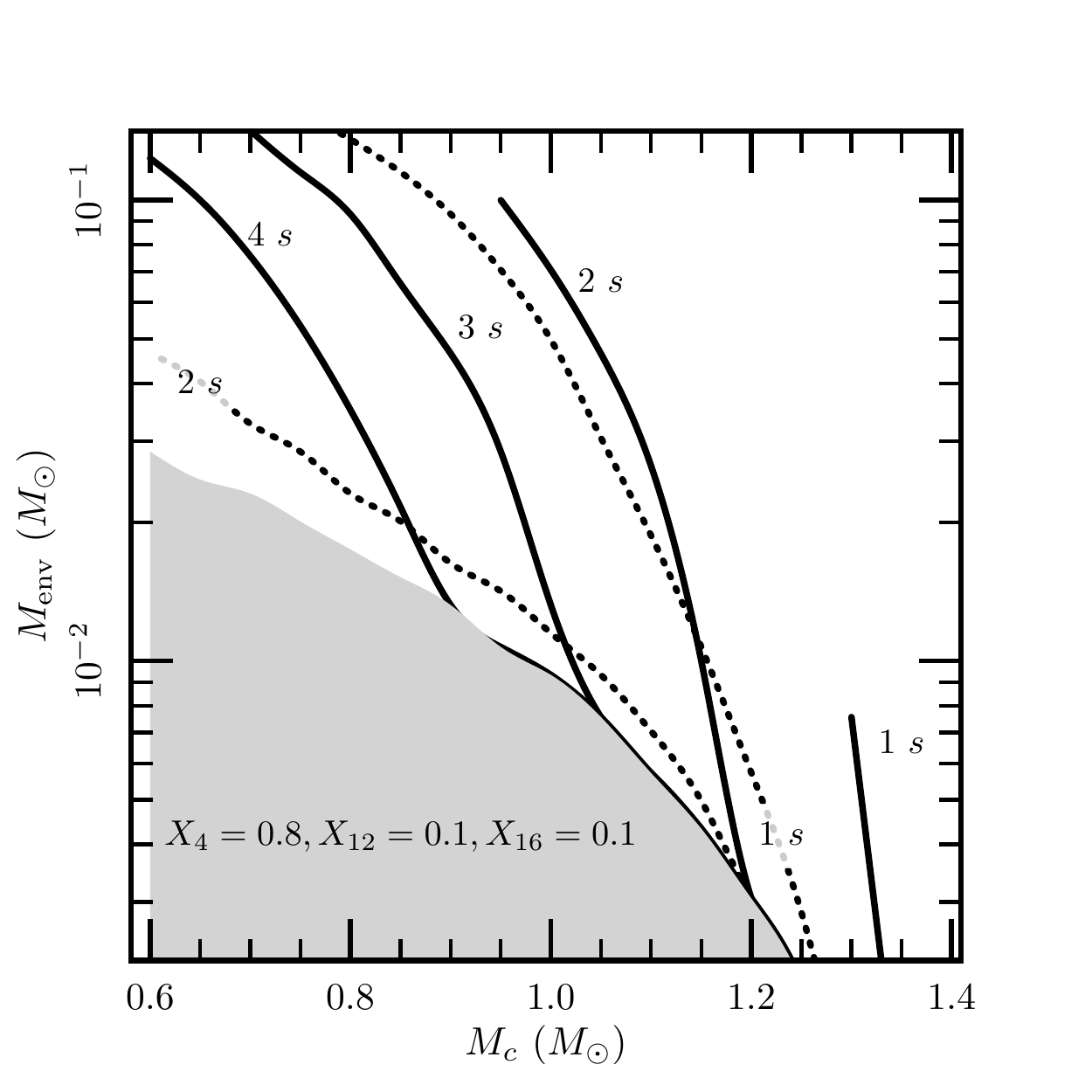}
	\caption{\label{fig:wd_timescales_he80_c10_o10} Same as Figure \ref{fig:wd_timescales_he100}, but with an envelope composition of $80\%$ $^4$He,  $10\%$ $^{12}$C, and $10\%$ $^{16}$O by mass, corresponding to Figure \ref{fig:final_abundance_regions_wd_he80_c10_o10}.}
\end{figure}

We also examine whether these surface detonations are fast enough to cause inwardly-propagating shock waves to focus in the interior of the C/O core - possibly detonating the core and making a Type Ia SN in a  double-detonation scenario. For each successful detonation in $M_c - M_{\rm env}$ space, we compute the time it takes the detonation to travel around the WD to the antipode of its initiation point,
\begin{equation}
t_{\rm det} = \frac{\pi R_c}{v_{\rm gcj}},
\end{equation}
where $R_c$ is the radius of the WD core. A lower limit on the time it takes the inwardly-propagating shock waves to traverse the interior of the core is given by the sound travel time through the core,
\begin{equation}
t_{\rm sound} = 2\int_0^{R_c} \frac{dr}{c_s}.
\end{equation}
We show lines of constant $t_{\rm det}$ and $t_{\rm sound}$ in Figures \ref{fig:wd_timescales_he100} and \ref{fig:wd_timescales_he80_c10_o10}. Virtually all detonations in these cases are fast enough to allow the inwardly propagating shock waves to focus in the interior of the WD. 
%Multiple ignition points in the envelope will decrease $t_{\rm det}$, making the outer detonation even more likely to send shock wave into the WD that may focus. 
It remains uncertain whether such focusing allows for a secondary detonation in a C/O or O/Ne/Mg core. Multidimensional simulations typically favor igniting the core \citep{Moll13, Sim12, Fink10} - see also \citet{Guillochon10}, but have a resolution much coarser than the extremely short carbon burning length scale $(\sim 1\ {\rm cm})$. High-resolution 1D calculations indicate that ignition in C/O cores may be possible if the shocks focus within a small enough critical region $(\sim 10^3\ {\rm cm})$, but ignition in carbon-deficient environments such as O/Ne/Mg is more difficult and perhaps not realizable with converging shocks in the double-detonation scenario \citep{Shen13, Seitenzahl09}.

\section{Conclusions}
\label{sec:conclusions}

Standard CJ detonations are not realizable in the low-density environments of thin helium shells on accreting WDs. Instead, a class of solutions, the generalized CJ solution \citep{He94} with $v_{\rm gcj} < v_{\rm CJ}$, determined by the expansive effects of curvature and blowout in the post-shock burning regions, is allowed in helium envelopes with large enough thicknesses. We constructed a 1D model of such detonations by adding these expansive effects into the ZND equations describing the evolution of shocked material as it burns behind the detonation front. Comparisons to 2D detonation simulations with {\ttfamily FLASH} indicate that we are capturing the relevant physics in the post-shock material with our 1D model. We find both minimum thicknesses allowing for detonation propagation and important reductions in burning length scales when isotopes such as $^{16}$O, $^{20}$Ne, and $^{24}$Mg are added to the fuel. Dredge-up from O/Ne/Mg WDs is thus expected to have more of an impact on nucleosynthesis than in C/O WDs.

We can map our 1D models to {\ttfamily FLASH} models in finite gravity, and predict which WD core + envelope configurations will support steady detonations. We can calculate the composition of the most burned section of the envelope with our 1D model, but specific vertical profiles of composition require multidimensional simulations. While detonations in pure helium generate enough energy to become unbound from the WD, the ashes from the very slowest $^{16}$O-enabled detonations are only barely gravitationally unbound, something we intend to more fully consider in future work. Our 1D model also neglects how the global curvature of the WD affects the detonation velocity - weakening as it approaches the equator and then strengthening towards the other pole due to divergence/convergence of the detonation front. We plan on addressing this and related issues in a future paper with full-star axisymmetric simulations in {\ttfamily FLASH}.

We have examined criteria allowing for the propagation of surface detonations, but have not touched on the important problem of detonation initiation in helium. Detailed investigation is necessary to determine if detonations can be ignited in some of the low-density environments considered here. The reduction in burning length scales in detonations with sufficient $^{16}$O abundances may allow detonations in helium shells to occur in a wider range of envelopes than originally considered in pure helium \citep{Holcomb13}.
%Independent of whether a detonation forms, we can address whether an existing steady detonation can propagate in helium envelopes with a variety of initial thermodynamic conditions and compositions.

We thank Ryan Foley, Dan Kasen, Bill Paxton, and Ken Shen for useful discussions and the anonymous referee for helpful comments. This work was supported by the National Science Foundation under grants PHY 11-25915 and AST 11-09174. Some of the software used in this work was in part developed by the DOE-supported ASC/Alliances Center for Astrophysical Thermonuclear Flashes at the University of Chicago. Some of the simulations for this work were made possible by the Triton Resource, a high performance research computing system operated by San Diego Supercomputer Center at UC San Diego.

%\begin{center}
%	{\bf Appendix}
%\end{center}

\appendix
\numberwithin{equation}{section}

\section{Modeling curvature in 1D}
\label{app:curvature}
%\label{subsec:curvature}
The curvature of the detonation front affects the generalized CJ velocity due to the divergence in velocity immediately behind the shock front. The theory of detonation shock dynamics (DSD) \citep{Bdzil12}, allows us to relate the curvature of the shock front to the detonation velocity through source terms in the hydrodynamic equations, similar to how curvature was modeled in 1D by \citet{Dursi06}. \citet{He94} also investigated the effects of curvature on detonation ignition and propagation speed, assuming a complete burn and a single-step reaction. We can adapt their treatment to the case of partial burning with blowout using our reaction networks. We work in the weak curvature limit here, $l_{95} \ll R_c$, where $R_c$ is the radius of curvature of the detonation front, but will justify a unified approach that allows for integration past this limit when we discuss radial expansion in the next section. To derive the source terms we
start with a simple 1D geometry with planar, cylindrical or spherical symmetry, with the detonation traveling away from the origin.

We derive the equations from the hydrodynamic equations, (\ref{eq:hydro_mass}) - (\ref{eq:hydro_energy}), and include the effects of curvature in one-dimensional, symmetric flow using
\begin{equation}
{\bf \nabla} \cdot {\bf u} = \frac{1}{r^j}\pdernp{(r^j u_{\rm lab})}{r} = \pdernp{u_{\rm lab}}{r} + \frac{j u_{\rm lab}}{r},
\end{equation}
where $j$ is a integer indicating the symmetry ($0$: plane-parallel, $1$: cylindrical, $2$: spherical) and $u_{\rm lab}$ is the lab frame velocity in the radial direction. Substituting this back into the hydrodynamic equations gives us
\begin{align}
\pdernp{\rho}{t} + \pdernp{}{r} \left(\rho u_{\rm lab} \right) + \frac{j \rho u_{\rm lab}}{r} &= 0 \\
\pdernp{}{t} \left(\rho u_{\rm lab} \right) + \pdernp{}{r}\left(P + \rho u_{\rm lab}^2 \right) + \frac{j \rho u_{\rm lab}^2}{r} &=0 \\
\pdernp{}{t} \left( \rho e \right) + \pdernp{}{r}\left[(\rho e + P) u_{\rm lab} \right] + \frac{ju_{\rm lab}(\rho e + P)}{r} &= 0.
\end{align}

In order to combine these equations with our blowout treatment, we need to rewrite them in the shock frame and find source terms to the 1D hydrodynamic equations. We define a new set of variables $(\xi, \tau)$ that are connected to the lab frame variables $(x, t)$ via $\tau = t$ and $\xi = R_s - r$, where $R_s(t) = \int_0^t v_{\rm det}(t') dt'$ is the location of the shock front (and thus also the instantaneous radius of curvature). Similarly, the lab-frame velocity $u_{\rm lab}$ is related to the shock-frame velocity $u$ via $u_{\rm lab} = v_{\rm det}-u$. When translating partial derivatives from one coordinate system to another, we use (for an arbitrary function $f$)
\begin{align}
df = \pder{f}{t}_r dt + \pder{f}{r}_t dr = \pder{f}{\tau}_\xi d\tau + \pder{f}{\xi}_\tau d\xi.
\end{align}
This implies that derivatives at constant $r$ ($dr=0$) are mapped to
\begin{align}
\pder{f}{t}_r &= \pder{f}{\tau}_\xi + \pder{f}{\xi}_\tau \pder{\xi}{\tau}_r \nonumber \\
&= \pder{f}{\tau}_\xi + \pder{f}{\xi}_\tau v_{\rm det}(\tau)
\end{align}
in the shock frame, while derivatives at constant $t$ ($dt = d\tau = 0$) are mapped to
\begin{align}
\pder{f}{r}_t &= \pder{f}{\xi}_\tau \pder{\xi}{r}_t \nonumber \\
&= -\pder{f}{\xi}_\tau
\end{align}
in the shock frame.
Putting this all together, we can write the 1D hydrodynamic equations in the shock frame as
\begin{align}
\pdernp{\rho}{\tau} + \pdernp{}{\xi}\left(\rho u \right) + \frac{j\rho (v_{\rm det} - u)}{R_s - \xi} &= 0, \\
\pdernp{}{\tau}\left(\rho u \right) + \pdernp{}{\xi}\left(P + \rho u^2 \right) - \rho \pdernp{v_{\rm det}}{\tau} + \frac{j\rho u(v_{\rm det} - u)}{R_s - \xi} &= 0, \\
\pdernp{}{\tau} \left(e + \frac{P}{\rho} + \frac{u^2}{2}\right) + u \pdernp{}{\xi} \left(e + \frac{P}{\rho} + \frac{u^2}{2}\right) - \frac{1}{\rho}\pdernp{P}{\tau} - v_{\rm det}\pdernp{v_{\rm det}}{\tau} &= 0.
\end{align}
%In the limit that the reaction zone that we are integrating through is much smaller than the radius of curvature, $l_{\rm burn} \ll R_c$, then we can drop the $\xi$ in the $R_s - \xi$ denominators. 
In the steady-state limit, all the time derivatives vanish, and the shock location $R_s(t)$ becomes a constant radius of curvature, $R_c$ in the limit $\xi \ll R_c$. 
%The curvature effects in a steady, laterally-propagating detonation correspond to the limit $\xi \ll R_c$ since the curved detonation front is not originating from a point, but steadily propagating - all fluid elements hit by the detonation front encounter a shock wave with the same curvature. 
We can then write the 1D hydrodynamic equations with source terms (\ref{eq:mass_gen}) - (\ref{eq:energy_gen}), here due to curvature as
\begin{align}
f_1 &= -\frac{j \rho (v_{\rm det} - u)}{R_c}, \label{eq:curve_source_1} \\
f_2 &= -\frac{j \rho (v_{\rm det} - u)u}{R_c}, \label{eq:curve_source_2} \\
f_3 &= 0 \label{eq:curve_source_3}.
\end{align}
These are the same source terms found in \citet{He94} used for detonations expanding from a point source as well as in the discussion of first-order curvature terms in DSD in \citet{Bdzil12}. 

We employ cylindrical symmetry ($j=1$) when making comparisons to the 2D slab detonations in {\ttfamily FLASH}. Although the post-shock structure is not exactly cylindrically symmetric, in the $l_{95} \ll R_c$ limit the radius of curvature is effectively constant, justifying our use of a symmetric coordinate system.

\section{Modeling radial expansion in 1D}
\label{app:blowout}
%\label{subsec:blowout}
The other expansive effect we consider is that due to post-shock radial expansion, which we refer to as blowout. When a shock wave propagates laterally through a surface layer on a WD initially in hydrostatic balance, the post-shock material will no longer be in hydrostatic balance due to the large jump in pressure across the shock front. In order to maintain a true 1D calculation, we treat this problem like a detonation in a pipe where we allow the cross-sectional area to expand behind the shock front, although with some important differences. The rate of such expansion is calculated directly from the hydrodynamic equations. Detonations in expanding environments have been investigated in terms of post-shock expansion and curvature in previous studies. \citet{Eyring49} examined the effects of post-shock expansion due to both the finite width of the reaction zone and curvature of the detonation front, but neglected source terms in the momentum and energy equations which are necessary for describing rapid expansion as well as deceleration due to gravity. 

%\citet{He94} looked at the effects of curvature in the context of a point explosion, somewhat different than our current case of a detonation front with constant curvature.
%\citet{Wood54} investigated the curvature of the detonation front more rigorously

We first derive the equations for blowout without including gravity since we want to initially compare against our constant density models in {\ttfamily FLASH} (see Figures \ref{fig:strip_diagram} and \ref{fig:strip_det_example}). Adding gravity to the prescription is possible by including it in the momentum equation, but we will show later that we find it unnecessary for our treatment. We take the pre-shock material to have a thickness $H$.
%The pre-shock material is in hydrostatic balance with a scale height of $H_s = P/\rho g$. We approximate the vertical acceleration of the post shock material using 
%\begin{equation}
%\frac{D{\bf u}}{D t} = -\frac{1}{\rho}{\bf \nabla} P.
%\frac{D{\bf u}}{D t} = -\frac{1}{\rho}{\bf \nabla} P + {\bf g},
%\end{equation}
%where we evaluate the acceleration due to gravity at the midpoint of the expanding helium layer, $g= GM_{\rm WD}/(R_{\rm WD} + H_s/2)^2$. Note that although $g$ is allowed to vary, the equation we're using for gravitational potential energy is assuming $H_s \ll R$. 
In the steady-state assumption the equation of motion reduces to 
\begin{equation}
%\left({\bf u}\cdot {\bf \nabla} \right){\bf u} = -\frac{1}{\rho}{\bf \nabla} P + {\bf g}
\left({\bf u}\cdot {\bf \nabla} \right){\bf u} = -\frac{1}{\rho}{\bf \nabla} P.
\end{equation}
Considering the vertical ($\hat{y}$) component of the flow, we get
\begin{equation}
%\left(u_x \pdernp{}{x} + u_y \pdernp{}{y} \right) u_y = -\frac{1}{\rho}\pder{P}{y} - g.
\left(u_x \pdernp{}{x} + u_y \pdernp{}{y} \right) u_y = -\frac{1}{\rho}\pder{P}{y}.
\end{equation}
Our 1D approximation requires that the flow only vary in the $\hat{x}$ direction, but we can approximate the vertical pressure gradient as $\pderl{P}{y} = -P/H$ and the vertical velocity gradient as $\pderl{u_y}{y} = u_y/H$, giving an equation for the evolution of the rate of vertical expansion,
\begin{equation}
%\der{u_y}{x} = \frac{P}{\rho u_x H_s} - \frac{u_y^2}{u_x H_s} - \frac{g}{u_x}.
\der{u_y}{x} = \frac{P}{\rho u_x H} - \frac{u_y^2}{u_x H}.
\end{equation}
Similarly, the $\hat{y}$ velocity controls the change in layer thickness,
\begin{equation}
\der{H}{x} = \frac{1}{u_x} \der{H}{t} =  \frac{u_y}{u_x}.
\end{equation}
The other evolution equations come from the generalization of equations (\ref{eq:mass})-(\ref{eq:energy}) in 1D with a varying thickness. Mass conservation follows the familiar $d(H \rho u_x)/dx = 0$ law from flow in a pipe with variable cross-sectional area. The momentum equation is slightly more subtle, since the pressure on the top and bottom of the slab is negligible because of the sharp shock jump conditions $P_0 \ll P$ - see Figure \ref{fig:strip_diagram}. The cross-sectional area reduces to the layer thickness $H$ in a 2D slab with no variation in the third dimension. We write the force per unit length on a fluid slab in the $\hat{x}$ direction as
\begin{equation}
F_x = P(x)H(x) - P(x+\Delta x)H(x + \Delta x) = -\der{}{x}\left(PH\right) \Delta x,
\end{equation}
to first order in $\Delta x$, where $H(x)$ is the thickness of the shocked material a distance $x$ behind the shock front. 
%This is in contrast to the more familiar result we would get if the pressure above and below the atmosphere was appreciable,
%\begin{equation}
%F_x = -\der{P}{x} H(x) \Delta x.
%\end{equation}
Adding in the momentum fluxes at each side of the slab gives us
\begin{equation}
 \rho(x) u_x(x)^2 - \rho(x+\Delta x) u_x(x + \Delta x)^2  = -\der{}{x}\left(\rho u_x^2 \right) \Delta x.
\end{equation}
This leaves us with our modifications of the hydrodynamic equations due to blowout
\begin{eqnarray}
\der{}{x}\left[H \left(\rho u_x\right) \right] &=& 0, \label{eq:mass_blowout} \\
\der{}{x}\left[H \left(P + \rho u_x^2\right) \right] &=& 0, \label{eq:mom_blowout} \\
%\der{}{x}\left[E + \frac{P}{\rho} + \frac{u_x^2 + (u_y/2)^2}{2} + \frac{g H_s}{2} \right] &=& 0 
\der{}{x}\left[E + \frac{P}{\rho} + \frac{u_x^2 + (u_y/2)^2}{2} \right] &=& 0 \label{eq:energy_blowout}.
\end{eqnarray}
The equation of conservation of energy is modified to include the kinetic energy from the vertical velocity of the material (chosen to be $u_y/2$ on average). Work against gravity can be added with a similar $gH/2$ term, but it turns out to be a small effect for WDs. Equations (\ref{eq:mass_blowout}) - (\ref{eq:energy_blowout}) allow us to determine the source terms due to blowout:
\begin{align}
f_1 &= -\frac{\rho u_x}{H} \der{H}{x} = -\frac{\rho u_y}{H}, \label{eq:blowout_source_1} \\
f_2 &= -\frac{(P+\rho u_x^2)}{H} \der{H}{x} = -\frac{(P+\rho u_x^2)u_y}{H u_x}, \label{eq:blowout_source_2} \\
%f_3 &= -\frac{u_y}{4} \der{u_y}{x} - \frac{g u_y}{2 u_x} + \frac{g H_s u_y}{2u_x(R_{\rm WD} + H_s/2)}\\
%	 &=  -\frac{P u_y}{4 \rho H_s u_x} + \frac{u_y^3}{4H_s u_x} - \frac{g u_y}{4 u_x} + \frac{g H_s u_y}{2u_x(R_{\rm WD} + H_s/2)}, \label{eq:blowout_source_3}
f_3 &= -\frac{u_y}{4} \der{u_y}{x} =  -\frac{P u_y}{4 \rho H u_x} + \frac{u_y^3}{4H u_x}, \label{eq:blowout_source_3}
\end{align}
under the conditions
\begin{align}
\der{H}{x} &= \frac{u_y}{u_x}, \\
%\der{u_y}{x} &= \frac{P}{\rho H_s u_x} - \frac{u_y^2}{H_s u_x} - \frac{g}{u_x}.
\der{u_y}{x} &= \frac{P}{\rho H u_x} - \frac{u_y^2}{H u_x}.
\end{align}
These equations will be used in conjunction with those from curvature, equations (\ref{eq:curve_source_1}) - (\ref{eq:curve_source_3}), to fully characterize the surface detonations.

\newpage

\bibliographystyle{apj}
\bibliography{Blowout}

\end{document}